\documentclass[aps,prd,10pt,showpacs,amsmath,amssymb,twocolumn,nofootinbib,nobibnotes,preprintnumbers,showkeys,longbibliography]{revtex4-1}
\pdfoutput=1 % if your are submitting a pdflatex (i.e. if you have images in pdf, png or jpg format)
\usepackage{float}
\usepackage{mathrsfs,amsfonts}
\usepackage{mathtools}
\usepackage{tensor}
\usepackage[T1]{fontenc} 
\usepackage{xcolor}
\usepackage{url}
\usepackage[hyperindex,breaklinks]{hyperref}
\usepackage{graphicx}
\pdfoutput=1
\newcommand{\sig}{\ensuremath{{\hat{\sigma}}}}
\newcommand{\s}{\ensuremath{{\hat{s}}}}

\newcommand{\sigdot}{\ensuremath{{\dot{\sigma}}}}

\newcommand{\veq}{\ensuremath{\overset{\mathrm{v}}{\approx}}}

\begin{document}
\allowdisplaybreaks[1]
\title{Scalaron-Higgs inflation}

\author{Anirudh Gundhi}
\author{Christian F. Steinwachs}
\email{anirudh.gundhi@phd.units.it}
\email{christian.steinwachs@physik.uni-freiburg.de}
\affiliation{Institut f\"ur Theoretische Physik, Universit\"at zu K\"oln,\\
Z\"ulpicher Stra\ss e 77, 50937 K\"oln, Germany}

\altaffiliation[Present address: ]{Dipartimento di Fisica, Universit\`a degli Studi di Trieste, Strada Costiera 11, 34151 Miramare-Trieste, Italy}
\affiliation{Physikalisches Institut, Albert-Ludwigs-Universit\"at Freiburg,\\
Hermann-Herder-Str.~3, 79104 Freiburg, Germany}

%%********************************************%

\begin{abstract}
In scalaron-Higgs inflation the Standard Model Higgs boson is non-minimally coupled to gravity and the Einstein-Hilbert action is supplemented by the quadratic scalar curvature invariant. For the quartic Higgs self-coupling $\lambda$ fixed at the electroweak scale, we find that the resulting inflationary two-field model effectively reduces to a single field model with the same predictions as in Higgs inflation or Starobinsky inflation, including the limit of a vanishing non-minimal coupling. 
For the same model, but with the scalar field a priori not identified with the Standard Model Higgs boson, we study the inflationary consequences of an extremely small $\lambda$. 
Depending on the initial conditions for the inflationary background trajectories, we find that the two-field dynamics either again reduces to an effective single-field model with a larger tensor-to-scalar ratio than predicted in Higgs inflation and Starobinsky inflation, or involves the full two-field dynamics and leads to oscillatory features in the inflationary power spectrum. Finally, we investigate under which conditions the inflationary scenario with extremely small $\lambda$ can be realized dynamically by the Standard Model renormalization group flow and discuss how the scalaron-Higgs model can provide a natural way to stabilize the electroweak vacuum.
\end{abstract}

%\pacs{98.80.Cq; 04.50.Kd; 12.60.-i; 04.62.+v; 11.10.Hi}


%%********************************************%
%
%
\maketitle
%
%
%%********************************************%

%%%%%%%%%%%%%%%%%%%%%%%%%%%%%%%%%%%%%%%%%%%%%%%
%%%%%%%%                                %%%%%%%
%%%%%%%%    Introduction                %%%%%%%
%%%%%%%%                                %%%%%%%
%%%%%%%%%%%%%%%%%%%%%%%%%%%%%%%%%%%%%%%%%%%%%%%

\section{Introduction}

%
%------------------------------------------------------------------------------
% Document
%------------------------------------------------------------------------------
%
\label{Sec:Introduction}
%------------------------------------------------------------------------------
%
The simplest models of inflation are based on a single additional propagating scalar degree of freedom -- the inflaton. In view of the plethora of inflationary models, it is reasonable to group different models into universality classes on a purely phenomenological basis if they lead to indistinguishable inflationary predictions \cite{Mukhanov2013,Kallosh2014,Galante2015}. However, even for models belonging to the same universality class, there is still much freedom left in the theoretical description and, unfortunately, in many models the physical nature of the inflaton and the functional form of its potential remain unexplained at the fundamental level. The model of non-minimal Higgs inflation \cite{Bezrukov2008} and Starobinsky's $R+R^2$ model \cite{Starobinsky1980} provide two exceptions. Despite their seemingly different nature, they both belong to the same universality class and are representatives of scalar-tensor theories \cite{Clifton2012,Nojiri2017} with a non-minimal coupling to gravity \cite{Spokoiny1984,Futamase1989,Salopek1989,Fakir1990} and $f(R)$ modifications of General Relativity \cite{Nojiri2003, DeFelice2010}, respectively.

What makes these two models so attractive are not just their predictions, which are in perfect agreement with recent Planck data \cite{Akrami2018}, but also their theoretical motivation. The gravity-scalar sector of these models only involves the inclusion of one single operator in addition to the Einstein-Hilbert action -- a non-minimal coupling term $\xi \varphi^2 R$ in the case of Higgs inflation and a quadratic curvature invariant $\alpha R^2$ in the case of Starobinsky's model of inflation.\footnote{In contrast to fourth order gravity \cite{Stelle1977,Stelle1978}, $f(R)$ gravity only propagates the additional ``scalaron'' \cite{Starobinsky1980}.} Each of the dimensionless coupling constants $\xi$ and $\alpha$ only adds one new parameter, respectively. This makes these models highly predictive. The motivation for the inclusion of the additional operators is not only based on their phenomenological significance -- they are the only two marginal operators that can be added to the gravity-scalar sector which neither introduce a new mass scale nor lead to propagating ghost degrees of freedom. Moreover, in both models the inclusion of the additional operator leads to an asymptotic scale invariance. Inflation is realized naturally during the almost scale invariant quasi de Sitter phase and is ended by the Einstein-Hilbert term, which breaks this scale invariance. Finally, even for an interacting scalar field minimally coupled to Einstein gravity, these two operators are unavoidably induced by quantum corrections already at the one-loop level \cite{Barvinsky1993, Shapiro1995, Steinwachs2011} and therefore have to be included for the consistency of the renormalization procedure.
In addition, in view of the perturbatively non-renormalizable character of General Relativity, from a theoretical perspective, these two operators should be included in the low energy limit that defines the particle spectrum of the corresponding effective field theory. 

The similarities between non-minimal Higgs inflation and Starobinsky inflation \cite{Barvinsky2008, Bezrukov2011,Barvinsky2012,Bezrukov2012,Kehagias2014} can be traced back to a more general equivalence between $f(R)$ gravity and scalar-tensor theories. Under certain conditions on the function $f$ and its derivatives, $f(R)$ gravity can be equivalently formulated as a scalar-tensor theory at the classical level, see e.g. \cite{Sotiriou2006,Faraoni2007,DeFelice2010,Ruf2018a}. Based on the one-loop result of $f(R)$ gravity on an arbitrary background manifold \cite{Ruf2018}, recently this equivalence was shown to also hold at the quantum level \cite{Ruf2018a, Ohta2018} - a result similar to the one obtained in \cite{Kamenshchik2015} for the quantum equivalence of scalar-tensor theories formulated in the Jordan frame and the Einstein frame parametrizations. In the scalar-tensor formulation of $f(R)$ gravity, the higher derivative scalaron degree of freedom becomes manifest. 

In this article, we analyze the model of scalaron-Higgs inflation, which results from a combination of Higgs inflation with a non-minimal coupling to gravity and Starobinsky's model of inflation.
It is a special case of a more general $f(R,\varphi)$ theory, which can be equivalently formulated as a two-field scalar-tensor theory with a curved scalar field space manifold. Multifield inflationary scenarios with and without a curved scalar field space have been studied in many variants \cite{ Kodama1984,Starobinsky1985,Salopek1989,Sasaki1996,Sasaki1998,Hwang2000,GrootNibbelink2002,Gong2002, Lee2005,Langlois2008,Senatore2012,Greenwood2013,Peterson2013,Kaiser2014}. The  particular case of two-field inflation was investigated previously in \cite{Kofman1985,Polarski1992,Polarski1994,Garcia-Bellido1996,Mukhanov1998,Starobinsky2001,Gordon2001, Bartolo2001, Wands2002, DiMarco2003, Byrnes2006,Lalak2007,Ashoorioon2009, Peterson2011,Bruck2014}. In particular, $R^2$ inflation with a minimally coupled scalar field was first considered in \cite{Kofman1985}. In addition to the adiabatic perturbations present in single-field models of inflation, a characteristic feature of multifield models is the generation of isocurvature perturbations. While so far there is no direct observational evidence for isocurvature modes, they can also source the adiabatic modes and therefore indirectly contribute to the observed adiabatic power spectrum and the derived spectral observables. 
In addition, observational signatures of non-Gaussianities  in the context of multifield inflation have been studied recently in \cite{Bernardeau2002,Vernizzi2006,Byrnes2008}.

We use the field covariant formalism for multifield models of inflation \cite{Sasaki1996,Gong2011,Peterson2013,Greenwood2013,Greenwood2013,Kaiser2014,Karamitsos2018} in order to derive the dynamics of the homogeneous and isotropic flat FLRW background and the dynamics of the perturbations propagating on that background.\footnote{For the proposal to apply the field covariant (including the metric field) geometrical formalism of \cite{Vilkovisky1984} to cosmology at the quantum level see \cite{Steinwachs2013,Steinwachs2014,Kamenshchik2015} and \cite{Moss2014a,Bounakis2018} for an explicit application.}
In contrast to single-field models of inflation, there is no unique inflationary trajectory anymore and, in general, different background trajectories originating from different initial conditions lead to different scenarios with different observational consequences.
Several aspects of the scalaron-Higgs model have been already analyzed in \cite{Salvio2015,Kaneda2016,Ema2017a,Wang2017,Pi2018,He2018,Gorbunov2018,Ghilencea2018,Wang2018}. We perform a detailed study of different initial conditions for the inflationary background trajectories and a thorough analysis of the parameter space, including the regions $\xi\ll1$ and $\lambda\ll 10^{-1}$.

We first investigate the case where the quartic Higgs coupling $\lambda=M_{\mathrm{H}}^2/2v^2\approx0.1$ is fixed by the value of the Higgs mass $M_{\mathrm{H}}\approx125$ GeV at the electroweak scale $v\approx246$ GeV.
We perform a careful study of the remaining parameter space for different initial conditions, which together determine the background dynamics in the scalaron-Higgs potential. 
For $\lambda\approx10^{-1}$ a broad range of values for the parameters $(\xi,\alpha)$ lead to two prominent valleys in the scalaron-Higgs potential, which serve as natural attractors for the inflationary background dynamics. We find that for $\lambda\approx 10^{-1}$, the observable part of the inflationary dynamics always takes place in one of the valley attractors independent of the initial conditions. This confirms the results of \cite{He2018}, which were obtained in the original Jordan frame parametrization of the scalaron-Higgs model for the case of a large non-minimal coupling $\xi\gg1$ and extends it for the case of a weak non-minimal coupling $\xi\ll1$. For $\xi\ll 1$, we find that the two valleys converge to a single broad valley located at $\varphi = 0$, which serves as the sole attractor for the model in this regime. The central valley is steep enough to prevent any slow roll in the $\varphi$ direction, or significant growth of isocurvature modes until the end of inflation. This extension provides an important result as it relaxes the situation with the strong coupling present in Higgs inflation. Therefore, for values of $\lambda$ fixed at the electroweak scale, we find that scalaron\- Higgs inflation reduces to an effective single-field model of inflation with the same universal predictions as non-minimal Higgs inflation or Starobinsky inflation, not leading to any  multifield effects during the inflationary dynamics -- even for a weak non-minimal coupling.

The situation changes significantly for very small values of $\lambda\ll 10^{-1}$. It is well known, that the divergent quantum contributions of the heavy Standard Model (SM) particles, which dominate the SM renormalization group (RG) running, can drive the Higgs self-coupling $\lambda$ to very small values at inflationary energy scales. Depending on the precise values of the top mass $M_{t}$, the Higgs mass $M_{\mathrm{H}}$ and the strong coupling constant $g_{\mathrm{s}}$ at the electroweak scale, the RG flow can even drive $\lambda$ to negative values at high energies and destabilize the electroweak vacuum. The measured values for $M_t$ and $M_{\mathrm{H}}$ imply that our universe is at the borderline between stability and non-stability \cite{Degrassi2012}.
Thus, in principle it is possible that at high energies $\lambda$ is driven to values even smaller than $\lambda=10^{-2}$ considered e.g. in \cite{Ema2017a,Pi2018,He2018,Gorbunov2018}. 
Motivated by this possibility, we first study the inflationary consequences of an extremely small quartic self-coupling $\lambda\lesssim10^{-11}$ for an abstract scalar field $\varphi$, a priori not to be identified with SM Higgs boson.

We find that this scenario leads to inflationary predictions different from non-minimal Higgs-inflation or Starobinsky inflation. Depending on the initial conditions and the specific inflationary background trajectory in the landscape of the two-field potential, different scenarios are possible: if inflation takes place in one of the two valleys for $\lambda\ll10^{-1}$, the model reduces to an effective single-field model different to the effective single-field model obtained for $\lambda=10^{-1}$ and leads to a larger tensor-to-scalar ratio than predicted in non-minimal Higgs inflation or Starobinsky inflation.

For different initial conditions parametrizing different inflationary background trajectories which start on the hilltop of the two-field potential, we find that true multifield effects are possible. They lead to features in the power spectrum and might be important for explaining the anomaly observed at large scales in the temperature anisotropy spectrum of the Cosmic Microwave Background (CMB) radiation. 

We then investigate the conditions under which the inflationary scenario with such a small self-coupling $\lambda$ can be realized dynamically by the RG flow within the scalaron-Higgs model. By analyzing the RG flow of the pure SM at inflationary energy scales, we find that the scenario with $\lambda\lesssim 10^{-11}$ is unlikely to be realized within the scalaron-Higgs model of inflation. However, this conclusion only holds provided that the extended RG flow of the scalaron-Higgs model is not significantly modified by the running couplings $\xi$ and $\alpha$ as well as their influence on the SM beta functions. 
Finally, we point out how the scalaron-Higgs model could provide a natural way to stabilize the electroweak vacuum.  

The article is structured as follows. In Sec.~\ref{Sec:Reformulation}, we reformulate a general $f(R,\varphi)$ theory as a two-field scalar-tensor theory with a curved scalar field space. In Sec.~\ref{Sec:CovMultiField}, we introduce the covariant field space formalism for a general multifield scalar-tensor theory and derive the dynamics of the inflationary background as well as that of the linear perturbations propagating on this background. In Sec.~\ref{Sec:ScalaronHiggs}, we introduce the scalaron-Higgs model, formulate it as two-field scalar-tensor theory with a curved scalar field space manifold and analyze the scalaron-Higgs potential in the Einstein frame. We classify different inflationary scenarios by their initial conditions and their associated trajectories in the scalaron-Higgs potential. In Sec.~\ref{Sec:ReductionSingleField}, we discuss the valley approximation for $\lambda=10^{-1}$ and the reduction of the scalaron-Higgs model to an effective single-field model. We derive analytical expressions for the inflationary observables in the slow-roll approximation and show that they lead to predictions indistinguishable from non-minimal Higgs inflation or Starobinsky's model of inflation with no multifield effects.
In Sec.~\ref{Sec:Isocurvature}, we analyze the inflationary consequences of a quartic self-coupling $\lambda\ll10^{-1}$ for an abstract scalar field $\varphi$ and show that compared to the $\lambda=10^{-1}$ case, this can lead to an effective single-field scenario with a larger tensor-to-scalar ratio as well as to a genuine two-field dynamics with observable multifield effects. Finally, we analyze the conditions under which such small values of $\lambda$ can be attained dynamically by the RG flow at high energies and point out a mechanism by which the electroweak vacuum might be stabilized in the scalaron-Higgs model. 
 We conclude in Sec.~\ref{Sec:Conclusion} and give a brief outlook on interesting future applications and observational consequences.    
%
%------------------------------------------------------------------------------
\section{Reformulation of $f(R,\varphi)$ gravity as two-field scalar-tensor theory }\label{Sec:Reformulation}
%------------------------------------------------------------------------------
%
We consider the action
\begin{align}
S[g,\varphi]=\int{\rm d}^4 x\sqrt{-g}\left[f(R,\varphi)-\frac{1}{2}\partial_{\mu}\varphi\partial^{\mu}\varphi+\cdots\right].\label{act1}
\end{align}
Here, $f(R,\varphi)$ is an arbitrary function of the Ricci scalar $R$ and the scalar field $\varphi$.
The higher derivatives, which enter the theory due to the dependence of $f(R,\varphi)$ on the scalar curvature $R$, lead to an additional scalar propagating degree of freedom -- the scalaron $\chi$. The ellipsis indicate that the action \eqref{act1} is considered to be the low energy approximation of an effective field theory.
The action \eqref{act1} can be represented as a two-field scalar-tensor theory in which the dependence on the scalaron $\chi$ becomes manifest. We perform the transition in two steps. In the first step, we express the action \eqref{act1} as a scalar-tensor theory with the scalaron $\chi$ non-minimally coupled to gravity. In the second step, we perform a conformal transformation of the metric $g_{\mu\nu}\to \hat{g}_{\mu\nu}$ and a reparametrization of the scalaron $\chi\to\hat{\chi}$. In terms of the variables $(\hat{g}_{\mu\nu}, \hat{\chi},\varphi)$ both scalar fields $(\hat{\chi},\varphi)$ are minimally coupled to gravity but feature a curved field space metric.
\subsection{Representation as a scalar-tensor theory}
We introduce an auxiliary action with scalar field $\psi$,
\begin{align}
S[g,\varphi,\psi]=\int{\rm d}^4 x\sqrt{-g}&\Big[f(\psi,\varphi)+f_{,\psi}(R-\psi)\nonumber\\
&-\frac{1}{2}\partial_{\mu}\varphi\partial^{\mu}\varphi\Big].\label{actaux}
\end{align}
The equations of motion for $\psi$ are given by the algebraic relation
\begin{align}
f_{,\psi\psi}(R-\psi)=0.\label{RPsiIdentification}
\end{align}
For $f_{,\psi\psi}\neq0$, the equation of motion implies $\psi=R$. Therefore, on-shell, the auxiliary action \eqref{actaux} is equivalent to the original action \eqref{act1}.
Next, we define the scalaron $\chi$ implicitly by
\begin{align}
\chi^2:=f_{,\psi}(\psi,\varphi).\label{DefChi}
\end{align} 
The knowledge of the explicit function $f(\psi,\varphi)$ together with the condition $f_{,\psi\psi}\neq0$ allows to invert \eqref{DefChi} and to express $\psi=\psi(\chi,\varphi)$ as a function of $\chi$ and $\varphi$. In particular, we can write the action \eqref{actaux} in terms of the scalaron $\chi$,
\begin{align}
S[g,\varphi,\chi]=\int{\rm d}^4 x\sqrt{-g}\left[\chi^2 R-\frac{1}{2}\partial_{\mu}\varphi\partial^{\mu}\varphi-W(\chi,\varphi)\right],\label{actJF}
\end{align}
with the two-field scalar potential
\begin{align}
W(\chi,\varphi)=\chi^2\psi(\chi,\varphi)-f(\psi(\chi,\varphi),\varphi).\label{PotW}
\end{align}
The action \eqref{actJF} corresponds to a scalar-tensor theory with two interacting scalar fields $\varphi$ and $\chi$. While $\varphi$ is minimally coupled to gravity and has a canonically normalized kinetic term, $\chi$ is non-minimally coupled to gravity and has no kinetic term at all. However, the non-minimal coupling directly couples $\chi$ to derivatives of the metric field $g_{\mu\nu}$, which after integration by parts induces a derivative coupling between the metric and $\chi$.
\subsection{Transformation to the Einstein frame and curved field space}
In order to remove the non-minimal coupling, we perform a conformal transformation of $g_{\mu\nu}$ to the new metric field $\hat{g}_{\mu\nu}$ with the field dependent conformal factor $\Omega(\chi)$,
\begin{align}
g_{\mu\nu}=\Omega\hat{g}_{\mu\nu},\label{conftraf}\qquad \Omega:=\frac{1}{2}\,\frac{M_{\mathrm{P}}^2}{\chi^2}.
\end{align}
The inverse $g^{\mu\nu}$, the determinant $g=\det\left(g_{\mu\nu}\right)$ and the Ricci scalar $R$ transform under \eqref{conftraf},
\begin{align}
g^{\mu\nu}={}&\Omega^{-1}\hat{g}^{\mu\nu},\qquad \sqrt{-g}=\Omega^2\sqrt{-\hat{g}},\\
R={}&\Omega^{-1}\left[\hat{R}-3\Omega^{-1}\hat{\Box}\Omega+\frac{3}{2}\Omega^{-2}\hat{\nabla}_{\mu}\Omega\hat{\nabla}^{\mu}\Omega\right].
\end{align}
In terms of the metric $\hat{g}_{\mu\nu}$, the action \eqref{actJF} reads
\begin{align}
S[\hat{g},\varphi,\chi]={}\int&{\rm d}^4 x\sqrt{-\hat{g}}\left[\frac{M_{\mathrm{P}}^2}{2}\hat{R}-\frac{M_\mathrm{P}^4}{4\chi^4}W(\chi,\varphi)\right.\nonumber\\&\left.-\frac{1}{4}\frac{M_\mathrm{P}^2}{\chi^2}\partial_{\mu}\varphi\partial^{\mu}\varphi-3\frac{M_{\mathrm{P}}^2}{\chi^2}\partial_{\mu}\chi\partial^{\mu}\chi\right].
\end{align}
The two scalar fields $\varphi$ and $\chi$ are minimally coupled to gravity and the dynamics of the scalaron $\chi$ is manifest.
While it is in general not possible to find a transformation $(\varphi,\chi)\to(\hat{\varphi},\hat{\chi})$ such that the kinetic terms for both scalar fields $\hat{\varphi}$ and $\hat{\chi}$ become simultaneously canonical, we are free to perform an additional reparametrization of the scalaron $\chi\to\hat{\chi}$ in such a way that $\chi$ features a canonically normalized kinetic term
\begin{align}\label{ChiEFOne}
\hat{\chi}={}&\sqrt{6}M_{\mathrm{P}}\ln\left(\frac{\sqrt{2}\chi}{M_{\mathrm{P}}}\right),\\
\chi={}&\frac{M_{\mathrm{P}}}{\sqrt{2}}\exp\left(\frac{\hat{\chi}}{\sqrt{6}M_{\mathrm{P}}}\right).\label{ChiEF}
\end{align}
We further introduce the Einstein frame potential
\begin{align}
\hat{W}(\varphi,\hat{\chi}):={}&\left.\frac{M_{\mathrm{P}}^4}{4\chi^4}  W(\varphi,\chi)\right|_{\chi=\hat{\chi}}\nonumber\\
={}&e^{-2\sqrt{\frac{2}{3}}\frac{\hat{\chi}}{M_{\mathrm{P}}}}W(\hat{\chi},\varphi).\label{PotWEF}
\end{align}
In terms of the Einstein frame field variables $\hat{g}_{\mu\nu}$ $\hat{\chi}$ and $\varphi$ and the potential \eqref{PotWEF}, the original action \eqref{act1} is expressed as scalar-tensor theory for two minimally coupled scalar fields
\begin{align}
S[\hat{g},\varphi,\hat{\chi}]=\int{\rm d}^4 x\sqrt{-\hat{g}}&\left[\frac{M_{\mathrm{P}}^2}{2}\hat{R}-\frac{1}{2}e^{-\sqrt{\frac{2}{3}}\frac{\hat{\chi}}{M_{\mathrm{P}}}}\partial_{\mu}\varphi\partial^{\mu}\varphi\right.\nonumber\\ &\left. -\frac{1}{2}\partial_{\mu}\hat{\chi}\partial^{\mu}\hat{\chi}-\hat{W}(\hat{\chi},\varphi)\right].\label{TwoFieldScalaronHiggsAct}
\end{align} 
The two-field scalar-tensor theory action \eqref{TwoFieldScalaronHiggsAct} in the Einstein frame is compactly written as 
\begin{align}
S[\hat{g},\Phi]=
\int\mathrm{d}^4 x\sqrt{-\hat{g}}&\left[\frac{M_{\mathrm{P}}^2}{2}\hat{R}-\frac{1}{2}\hat{g}^{\mu\nu}G_{IJ}(\Phi)\Phi^{I}_{,\mu}\Phi^{J}_{,\nu}\right.\nonumber\\&\left.-\hat{W}(\Phi)\right].\label{ActScal}
\end{align}
The scalars $\Phi^{I}(x)$ are viewed as local coordinates of the scalar field space with metric $G_{IJ}(\Phi)$,
\begin{align}
\Phi^{I}=\left(\begin{array}{c}\hat{\chi}\\\varphi\end{array}\right),\qquad G_{IJ}(\Phi)=\left(
\begin{array}{cc}
1&0\\
0&e^{-\gamma\frac{\hat{\chi}}{M_{\mathrm{P}}}}
\end{array}
\right). \label{MetricG}
\end{align}
Here, $\gamma:=\sqrt{2/3}$ is a numerical factor.
For models such as \eqref{act1}, originally formulated in the Jordan frame, the form of the field space metric $G_{IJ}$ is entirely fixed by the transition to the multifield Einstein frame formulation \eqref{TwoFieldScalaronHiggsAct}. Therefore, independently of the explicit form of the original function $f(R,\varphi)$, all models of the form \eqref{act1} can be represented in terms of the multifield Einstein frame formulation \eqref{TwoFieldScalaronHiggsAct} with the \textit{same} metric $G_{IJ}$. The function $f(R,\varphi)$ only determines the form of the Einstein frame potential $\hat{W}$ in \eqref{TwoFieldScalaronHiggsAct} but does not affect the geometry of the scalar field space. 
In two-field models such as \eqref{TwoFieldScalaronHiggsAct}, the scalar field space is a two dimensional space of constant curvature. The explicit expressions of all geometric objects for the scalar filed space with metric \eqref{MetricG} are provided in  \ref{App:FieldSpaceGeometry}. 
%
%------------------------------------------------------------------------------
\section{Covariant multifield formalism}\label{Sec:CovMultiField}
%------------------------------------------------------------------------------
%
\subsection{Action functional and equations of motion}
We consider a general multifield action of the form \eqref{ActScal} with arbitrary $G_{IJ}$ and $\hat{W}$,
\begin{align}
S[\hat{g},\Phi]={}&S_{\mathrm{EH}}[\hat{g}]+S_{\mathrm{M}}[\hat{g},\Phi],\label{ActScalMult}\\
S_{\mathrm{EH}}[\hat{g}]={}&\frac{M_{\mathrm{P}}^2}{2}\int{\mathrm d}^4x\sqrt{-\hat{g}}\hat{R},\\
S_{\mathrm{M}}[\hat{g},\Phi]={}&\int\mathrm{d}^4x\sqrt{-\hat{g}}\left[-\frac{1}{2}\hat{g}^{\mu\nu}G_{IJ}\Phi^{I}_{,\mu}\Phi^{J}_{,\nu}-\hat{W}(\Phi)\right].
\end{align}
The covariant treatment of the field space has been advocated already in
\cite{Sasaki1996,Gong2011,Peterson2013,Greenwood2013,Greenwood2013,Kaiser2014,Karamitsos2018}.\footnote{Note that the extended covariant treatment which includes the metric field as just another field variable has been proposed in \cite{Vilkovisky1984} in the general field theoretical context and in \cite{Steinwachs2013, Steinwachs2014,Kamenshchik2015} in the context of cosmological scalar-tensor theories.}
In the field covariant formalism, it is natural to define a field covariant differential $D$ by its action on a vector $V^{I}(\Phi)$ in the field space,\footnote{Note that $\Phi^{I}$ is not a vector in scalar field space -- the $\Phi^{I}$ are local coordinates in the scalar field space i.e. $D\phi^{I}=d\Phi^{I}$. In contrast, $\Phi^{I}_{,\mu}$ is a vector which measures the change of the field space coordinates $\Phi^{I}$ by a change of the spacetime coordinates $x^{\mu}$.}
\begin{align}
DV^{I}=dV^{I}+d\Phi^{J}\tensor{\Gamma}{^{I}_{JK}}V^{K}.\label{CovDiff}
\end{align}
The field space connection is the Christoffel connection associated with the metric $G_{IJ}$, 
\begin{align}
\tensor{\Gamma}{^{K}_{IJ}}=\frac{G^{KL}}{2}\left(G_{LJ,I}+G_{IL,J}-G_{IJ,L}\right).
\end{align} 
The inverse field space metric $G^{IJ}$ is defined via $G_{IJ}G^{JK}=\delta_{I}^{K}$.
In view of \eqref{CovDiff}, the field covariant derivative $D_{I}$ is defined by its action on a field space vector $V^{I}(\Phi)$,
\begin{align}
D_{I}V^{L}={}&\frac{\partial V^{L}}{\partial \Phi^{I}}+\frac{\partial\Phi^{J}}{\partial \Phi^{I}}\tensor{\Gamma}{^{L}_{JK}}V^{K}\nonumber\\
={}&\tensor{V}{^{L}_{,I}}+\tensor{\Gamma}{^{L}_{IK}}V^{K}.
\end{align}
Since the scalar field space coordinates $\Phi^{I}(x)$ depend on the point $x$, the change of a vector $V^{I}(\Phi)$ under a change of coordinates $x^{\mu}$ defines the spacetime field covariant derivative
\begin{align}
D_{\mu}V^{I}:={}&\partial_{\mu}V^{I}+\Phi^{J}_{,\mu}\tensor{\Gamma}{^{I}_{JK}}V^{K}=\Phi^{J}_{,\mu}D_{J}V^{I}.\label{CovDField}
\end{align}
Functional differentiation of $\eqref{ActScal}$ with respect to $g_{\mu\nu}$ leads to Einstein's field equations
\begin{align}
\hat{R}^{\mu\nu}-\frac{1}{2}\hat{g}^{\mu\nu}\hat{R}=M_{\mathrm{P}}^{-2}\hat{T}^{\mu\nu}_{\Phi},\label{EEQ}
\end{align}
with the energy momentum tensor
\begin{equation}
\begin{aligned}
\hat{T}^{\mu\nu}_{\Phi}:=&{}\frac{2}{\sqrt{-\hat{g}}}\frac{\delta S_{\mathrm{M}}}{\delta g_{\mu\nu}}\\=&\left(\hat{g}^{\mu\alpha}\hat{g}^{\beta\nu}-\frac{1}{2}\hat{g}^{\mu\nu}\hat{g}^{\alpha\beta}\right)G_{IJ}\Phi^{I}_{,\alpha}\Phi^{J}_{,\beta}-g^{\mu\nu}\hat{W}(\Phi).\label{EMTMultiFiled}
\end{aligned}
\end{equation}
Functional differentiation of \eqref{ActScal} with respect to $\Phi^{I}$ yields the Klein-Gordon equation
\begin{align}
\frac{1}{\sqrt{-\hat{g}}}D_{\mu}\left(\sqrt{-\hat{g}}\hat{g}^{\mu\nu}\partial_{\nu}\right)\Phi^I - G^{IK}\hat{W},_K = 0 .\label{KKEQ}
\end{align}
\subsection{FLRW background evolution}
The flat Friedmann-Lema\^itre-Robertson-Walker (FLRW) background line element reads
\begin{align}
\mathrm{d}s^2 = -\mathrm{d}t^2+a^2 \delta_{ij}\mathrm{d}x^i\mathrm{d}x^j\,.\label{FLRW}
\end{align}
Here, $t$ is the cosmic Friedmann time, $a(t)$ is the scale factor, $\delta_{ij}=\mathrm{diag}(1,1,1)$ and $i,\,j,\ldots=1,2,3$ are spatial indices.
The homogeneous and isotropic energy momentum tensor of a perfect fluid is given by
\begin{align}
T^{\mu\nu}=\left(\rho+p\right)u^{\mu}u^{\nu}+p\hat{g}^{\mu\nu}, \label{EMTPerfectFluid}
\end{align}
with the four velocity $u^{\mu}$ normalized by $\hat{g}_{\mu\nu}u^{\mu}u^{\nu}=-1$. By comparing \eqref{EMTPerfectFluid} and \eqref{EMTMultiFiled}, we obtain the energy density $\rho(t)$ and the pressure $p(t)$ for a  homogeneous scalar multiplet $\Phi^{I}(t)$,
\begin{align}
\rho_{\Phi}(t)={}&\frac{1}{2}G_{IJ}\dot{\Phi}^{I}\dot{\Phi}^{J}+\hat{W}\label{density},\\ p_{\Phi}(t)={}&\frac{1}{2}G_{IJ}\dot{\Phi}^{I}\dot{\Phi}^{J}-\hat{W}\label{pressure}.
\end{align} 
On the FLRW background \eqref{FLRW}, the field equations \eqref{EEQ} and \eqref{KKEQ} reduce to 
\begin{align}
H^2 ={}& \frac{M^{-2}_{\mathrm{P}}}{3}\left[\frac{1}{2}G_{IJ}\dot{\Phi}^I\dot{\Phi}^J + \hat{W}(\Phi)\right],\label{Friedmann1}\\
\dot{H} ={}& -\frac{M^{-2}_{\mathrm{P}}}{2}G_{IJ}\dot{\Phi}^I\dot{\Phi}^J,\label{Friedmann2}\\
D_{t}\dot{\Phi}^I={}& - 3H\dot{\Phi}^I - G^{IJ}\hat{W},_{J}.\label{KleinGordon}
\end{align}
Here, the dots denote derivatives with respect to the cosmic time $t$. In accordance with \eqref{CovDField}, we have also introduced the Hubble parameter $H(t)$ and the covariant time derivative $D_t$, 
\begin{align}
H(t):=\frac{\dot{a}(t)}{a(t)},\qquad D_{t}V^{I}:=\dot{V}^I+\dot{\Phi}^{J}\Gamma^{I}_{JK}V^{K}.
\end{align}
The second Friedmann equation \eqref{Friedmann2} shows that the time evolution of the Hubble parameter $H(t)$ is completely determined by the length $\dot{\sigma}$ of the velocity vector $\dot{\Phi}^{I}$ in field space, which also defines the unit velocity vector $\hat{\sigma}^{I}$ tangent to the background trajectory 
\begin{align}
\label{dsigma}
\dot{\sigma}:={}& \sqrt{G_{IJ}\dot{\Phi}^I\dot{\Phi}^J},\\
\hat{\sigma}^I:={}& \frac{\dot{\Phi}^I}{\dot{\sigma}},\qquad G_{IJ}\hat{\sigma}^{I}\hat{\sigma}^{J}={}1\,.\label{dsigmaTwo}
\end{align}
Since we restrict ourselves to a two-field model, besides $\hat{\sigma}^{I}$, we only need one additional unit vector $\hat{s}^{I}$ to span the field space. This vector is chosen to be orthogonal to $\hat{\sigma}^{I}$, which implies
\begin{align}
\label{norms}
G_{IJ}\s^I\s^J = 1,\qquad G_{IJ}\s^I\sig^J = 0.
\end{align} 
It is useful to define the turn-rate vector $\omega^{I}$ as the change in direction defined by $\hat{\sigma}^{I}$,
\begin{align}
\omega^{I}:=D_{t}\hat{\sigma}^{I}.\label{TurnVector}
\end{align} 
Since $\hat{\sigma}^{I}$ is a unit velocity vector, $\omega^{I}$ can be interpreted as acceleration vector. From \eqref{TurnVector}, it can be seen that $\omega^I$ is perpendicular to ${\hat{\sigma}}^I$ and therefore proportional to $\hat{s}^I$,
\begin{align}
\omega^{I}=\omega\hat{s}^{I}.
\end{align}
The orientation of $\hat{s}$ is chosen to be the same as that of $\omega^I$ and the magnitude $\omega$ of $\omega^{I}$ defines the turn rate of the background trajectory.
Using \eqref{dsigma} and projecting \eqref{KleinGordon} along  $\hat{\sigma}^{I}$ and $\hat{s}^{I}$ yields.\footnote{The corresponding projection operators are  $\tensor{\left[\Pi_{\sigma}\right]}{^{I}_{J}}:=\hat{\sigma}^{I}\hat{\sigma}_{J}$ and $\tensor{\left[\Pi_{s}\right]}{^{I}_{J}}=\delta^{I}_{J}-\hat{\sigma}^{I}\hat{\sigma}_{J}$.}
\begin{align}
H^2 ={}& \frac{M_{\mathrm{P}}^{-2}}{3}\left(\frac{1}{2}{\dot{\sigma}}^2 + \hat{W}\right)\label{Friedmann1Sig},\\
\dot{H} ={}& -\frac{M^{-2}_{\mathrm{P}}}{2}\dot{\sigma}^2,\label{Friedmann2Sig}\\
\ddot{\sigma}={}&-3H\dot{\sigma}- \hat{W}_{,\sigma},\label{KKSig}\\
\omega  ={}& -\frac{\hat{W},_s}{{\dot{\sigma}}}\label{TurnRateIsoDerive}.
\end{align}
Equation \eqref{Friedmann2Sig} shows that the dynamics of the Hubble parameter $\dot{H}(t)$ is entirely determined by the speed $\dot{\sigma}$ along the $\hat{\sigma}^{I}$ direction tangent to the background trajectory.
Equation \eqref{TurnRateIsoDerive} relates the turn rate $\omega$ to the directional derivative of the potential $\hat{W}$ along the $\hat{s}^{I}$ direction perpendicular to the background trajectory. 
The derivatives of the potential along $\hat{\sigma}^{I}$ and $\hat{s}^{I}$ are defined by the projections
\begin{align}
\hat{W}_{,\sigma}:=\frac{\partial \hat{W}}{\partial\Phi^I}\hat{\sigma}^{J},\qquad \hat{W}_{,s}:=\frac{\partial \hat{W}}{\partial\Phi^I}\s^{J}.
\end{align}
\subsection{Slow-roll background dynamics}
The deviation from de Sitter space $\dot{H}\neq0$ in multifield inflation is quantified by the Hubble slow-roll parameters defined along the background trajectory as in the single-field case\footnote{Here, $N$ is defined as the number of efolds left until the end of inflation. Therefore, we have $\mathrm{d}N=-H\mathrm{d}t$ with a minus sign.}
\begin{align}\label{SlowRollExactOne}
\varepsilon_{H}:={}&-\frac{\dot{H}}{H^2}=\frac{\mathrm{d}\ln H}{\mathrm{d}N}\,,\\ \eta_{H}:={}&\frac{1}{H}\frac{\dot{\varepsilon}_{H}}{\varepsilon_{H}}=-\frac{\mathrm{d}\ln\varepsilon_{H}}{\mathrm{d}N}.\label{SlowRollExact}
\end{align}
Within the slow-roll approximation we have ${\varepsilon_{H}\ll1}$ and ${|\eta_{H}|\ll1}$, and the equations \eqref{Friedmann1Sig}-\eqref{KKSig} reduce to 
\begin{align}\label{SlowRollCond}
H^2&{}\approx \frac{M^{-2}_{\mathrm{P}}}{3}\hat{W},\\
\dot{H}&{} \approx -\frac{M^{-2}_{\mathrm{P}}}{2}\sigdot^2,\\
3H\sigdot&{} \approx -\hat{W},_{\sigma}.\label{SlowRollCondTwo}
\end{align}
The slow-roll parameters \eqref{SlowRollExactOne} and \eqref{SlowRollExact} can be expressed in terms of the derivatives of the multifield potential $\hat{W}$ along the inflationary trajectory in an analogous way as for the single-field case 
\begin{align}
\varepsilon_{\sigma}:=\frac{M_{\mathrm{P}}^2}{2}\left(\frac{\hat{W}_{,\sigma}}{\hat{W}}\right)^2,\qquad \eta_{\sigma}:=M_{\mathrm{P}}^2\frac{\nabla_{\sigma}\nabla_{\sigma}\hat{W}}{\hat{W}}.\label{SlowRollPot}
\end{align}
Here $\nabla_{\sigma}:=\hat{\sigma}^{I}\nabla_{I}$ denotes the covariant directional derivative along $\hat{\sigma}$. Within the slow-roll approximation, the Hubble slow-roll parameters \eqref{SlowRollExact} are related to the potential slow-roll parameters \eqref{SlowRollPot} by $\varepsilon_{H}\approx\varepsilon_{\sigma}$ and $\eta_{H}\approx4\varepsilon_{\sigma}-2\eta_{\sigma}$.
\subsection{Cosmological perturbations} 
The scalar metric perturbations are incorporated in the perturbed FLRW line element 
\begin{align}
\mathrm{d}s^2=&-\left(1+2A\right)\mathrm{d}t^2+2a B_{,i}\mathrm{d}x^i\mathrm{d}t\nonumber
\\&+a^2\left(\delta_{ij}+2E_{ij}\right)\mathrm{d}x^i\mathrm{d}x^{j},
\end{align}
with the spatial part of the scalar metric perturbation
\begin{align}
E_{ij}:=\psi\delta_{ij}+E_{,ij}.
\end{align}
The four scalar metric perturbations $A(t,\mathbf{x})$, $B(t,\mathbf{x})$, $\psi(t,\mathbf{x})$ and $E(t,\mathbf{x})$ combine with the perturbation of the scalar multiplet $\delta\Phi^{I}(t,\mathbf{x})$.
Instead of $\delta\Phi^{I}(t,\mathbf{x})$, it is convenient to work with the gauge-invariant multifield Mukhanov-Sasaki variable \cite{Mukhanov1988,Sasaki1986,Greenwood2013},
\begin{align}
\delta\Phi^{I}_{\mathrm{g}}:=\delta\Phi^I+\frac{\dot{\Phi}^{I}}{H}\psi.\label{MukSasPhi}
\end{align}
The equation for the Fourier modes of the perturbation $\delta\Phi^{I}_{\mathrm{g}}$ is found to be \cite{Sasaki1996,Nakamura1996,Greenwood2013}\footnote{We denote the Fourier modes $\delta\Phi^{I}_{\mathbf{k}}(t)$ with wave vector $\mathbf{k}$ simply by $\delta\Phi^{I}$ and likewise for the tensor modes.},
\begin{align}
D^2_t\delta\Phi^I_\mathrm{g}+ 3HD_t \delta\Phi^I_\mathrm{g}
 +\left(\frac{k^2}{a^2}\delta^I_J+\tensor{\Omega}{^{I}_{J}}\right)\delta\Phi^J_\mathrm{g}=0.\label{DynEQPertPhi}
\end{align}
Here, $\tensor{\Omega}{^{I}_{J}}$ and the effective mass tensor $\tensor{M}{^{I}_{J}}$ are defined by
\begin{equation}
\begin{aligned}
\tensor{\Omega}{^{I}_{J}}&:={}\tensor{M}{^I_J}-M^{-2}_\mathrm{P}a^{-3}D_t\left(\frac{a^3}{H}{\dot{\Phi}}^I{\dot{\Phi}}_J\right),\\
M_{IJ} &:={} \nabla_I\nabla_J\hat{W} + R_{IKJL}\dot{\Phi}^K\dot{\Phi}^L.\label{EffMassMatrix}
\end{aligned}
\end{equation}
The effective mass tensor $M_{IJ}$ includes the Riemannian curvature tensor $R_{IJKL}$ associated with the curved field space, as well as the curvature of the multifield potential $\hat{W}$. 
The tensor modes $h$ (suppressing tensor indices) satisfy the simple mode equation,
\begin{align}
\ddot{h}+3H\dot{h}+\frac{k^2}{a^2}h=0.\label{Tensor equation}
\end{align} 
Projecting \eqref{MukSasPhi} along $\hat{\sigma}^{I}$ and $\hat{s}^{I}$ defines the adiabatic and isocurvature perturbations
\begin{align}
Q_{\sigma}:=\hat{\sigma}^{J}G_{IJ}\delta\Phi^{I}_{\mathrm{g}},\qquad Q_{\mathrm{s}}:=\hat{s}^{J}G_{IJ}\delta\Phi^{I}_{\mathrm{g}}.
\end{align}
Inserting the decomposition $\delta\Phi_{\mathrm{g}}^{I}=Q_{\sigma}\hat{\sigma}^{I}+Q_{\mathrm{s}}\hat{s}^{I}$
into \eqref{DynEQPertPhi}, we obtain the dynamical equations for the perturbations $Q_{\sigma}$, $Q_{\mathrm{s}}$ and $h$ in the large wavelength limit $k\ll aH$,
\begin{align}
\ddot{Q}_{\sigma} + 3H\dot{Q}_{\sigma} + \Omega_{\sigma\sigma}Q_{\sigma}={}& f(\mathrm{d}/\mathrm{d}t)(\omega Q_{\mathrm{s}}),\label{EQPertQsig} \\
\ddot{Q}_\mathrm{s} + 3H\dot{Q}_\mathrm{s} + m_\mathrm{s}^2Q_\mathrm{s}={}&0\label{EQPertQs},\\
\ddot{h}+3H\dot{h}={}&0.\label{TensorPert}
\end{align}
Here, we have defined the projections of \eqref{EffMassMatrix}, which include the effective isocurvature mass 
\begin{align}
\Omega_{\sigma\sigma}:=\hat{\sigma}^I\hat{\sigma}^J\Omega_{IJ}-w^2,\qquad
m^2_{\mathrm{s}}:= \s^I\s^J M_{IJ} + 3w^2\label{IsoMassDef}.
\end{align}
The operator $f(\mathrm{d}/\mathrm{d}t)$ acting on the source $\omega Q_{\mathrm{s}}$ on the right hand side of \eqref{EQPertQsig} is defined as
\begin{align}
f(\mathrm{d}/\mathrm{d}t):=2\left[\frac{\mathrm{d}}{\mathrm{d}t}-\left(\frac{W,_{\sigma}}{\dot{\sigma}} + \frac{\dot{H}}{H}\right)\right].
\end{align}
The adiabatic mode $Q_{\sigma}$ in \eqref{EQPertQsig} is sourced by the product $\omega Q_{\mathrm{s}}$ of the turn rate $\omega$ and the isocurvature mode $Q_{\mathrm{s}}$. The turn rate is determined by the background dynamics \eqref{TurnRateIsoDerive} and the isocurvature mode is obtained by solving the homogeneous equation \eqref{EQPertQs}. Only if the combination of the turn rate $\omega$ and the isocurvature mode $Q_{\mathrm{s}}$ is sufficiently large, the adiabatic mode $Q_{\sigma}$ is sourced by the isocurvature mode. Such a sourcing of the adiabatic mode by the isocurvature mode might be called ``isocurvature pumping'' and leads to potentially observable effects in the adiabatic power spectrum. 
\subsection{Inflationary observables}
As in the single-field case, we define the gauge invariant comoving curvature perturbation
\begin{align}
\mathcal{R}:=\psi-\frac{H}{\rho+p}\delta q\label{CurvPert},\qquad \delta q=\dot{\sigma}\hat{\sigma}_I\delta\phi^{I},
\end{align}
where $\delta q$ is obtained from \eqref{EMTMultiFiled} by $\delta q_{,i}=\delta T^{0}_{i}$.
Combining \eqref{CurvPert} with \eqref{density} and \eqref{pressure}, $\mathcal{R}$ is found to be proportional to the adiabatic perturbation $Q_{\sigma}$,
\begin{align}
\mathcal{R}=\psi+\frac{H}{\dot{\sigma}}\hat{\sigma}_I\delta\Phi^I=\frac{H}{\dot{\sigma}}Q_{\sigma}\label{CurvPertFin}.
\end{align}
In the single-field case, $\mathcal{R}$ is conserved on superhorizon scales $k\ll a H$.
In analogy to the curvature perturbation \eqref{CurvPertFin}, we define the isocurvature perturbation
\begin{align}
\mathcal{S}:=\frac{H}{\dot{\sigma}}Q_{\mathrm{s}}.\label{IsoCurvPert}
\end{align} 
The power spectra $\mathcal{P}_{\mathcal{R}}$ and $\mathcal{P}_{\mathcal{S}}$ for the adiabatic and isocurvature perturbations read\footnote{The power spectrum for the cross correlation $\mathcal{P}_{\mathcal{R}\mathcal{S}}$ can be obtained similarly, but is of no direct relevance for the analysis presented in this article.}
\begin{equation}
\begin{aligned}
\mathcal{P}_{\mathcal{R}}(t;k)&:=\frac{k^3}{2\pi^2}\left(\frac{H}{\dot{\sigma}}\right)^2\left|Q_{\sigma}\right|^2,\\ \mathcal{P}_{\mathcal{S}}(t,k)&:=\frac{k^3}{2\pi^2}\left(\frac{H}{\dot{\sigma}}\right)^2\left|Q_{\mathrm{s}}\right|^2.\label{PowerSpectrumadR}
\end{aligned} 
\end{equation}
Inserting the solutions $\delta\Phi^{I}$  of the mode equations \eqref{DynEQPertPhi} and the background quantities $\hat{\sigma}^{I}$, $\hat{s}^I$, $H$, $\epsilon$ obtained from the solution of the background equations in \eqref{Friedmann1Sig}--\eqref{TurnRateIsoDerive}, the power spectra $\mathcal{P}_{\mathcal{R}}(t;k)$ and $\mathcal{P}_{\mathcal{S}}(t;k)$ can be evaluated numerically.
In addition to the scalar perturbations, tensor perturbations are amplified during inflation and lead to the power spectrum for the tensor modes $h$, which are solutions of \eqref{Tensor equation},
\begin{align}
\mathcal{P}_{h}(t;k):=8\frac{k^3}{2\pi^2}\left|h\right|^2.\label{PowerSpectrumadTensor}
\end{align}
For a given Fourier mode $k$, we evaluate the power spectra at $t_{\mathrm{end}}$ and obtain a single number for $\mathcal{P}_{\mathcal{R}}(t_{\mathrm{end}};k)$, $\mathcal{P}_{\mathcal{S}}(t_{\mathrm{end}};k)$ and $\mathcal{P}_{h}(t_{\mathrm{end}};k)$.\footnote{We do not evaluate the power spectrum numerically at the $k$-dependent moment of first horizon crossing $t_{*}$ but at the end of inflation $t_{\mathrm{end}}$. This has the advantage that the potential superhorizon dynamics of the powers spectra in the multifield scenario can be taken into account.} Repeating this procedure for different Fourier modes $k$, we obtain the power spectra numerically as a function of $k$,
\begin{equation}
\begin{aligned}
\mathcal{P}_{\mathcal{R}}(k)&:=\mathcal{P}_{\mathcal{R}}(t_{\mathrm{end}},k),\\
\mathcal{P}_{\mathcal{S}}(k)&:=\mathcal{P}_{\mathcal{S}}(t_{\mathrm{end}},k),\\ \mathcal{P}_{h}(k)&:=\mathcal{P}_{h}(t_{\mathrm{end}},k).\label{PowerSpectrak}
\end{aligned}
\end{equation} 
The weak scale dependence of the power spectra motivates to fit the numerical solution obtained for \eqref{PowerSpectrak} by a power law ansatz with a reference scale $k_{*}$,
\begin{equation}
\begin{aligned}
\mathcal{P}_{\mathcal{R}}(k)&={}A_{\mathcal{R}}\left(\frac{k}{k_{*}}\right)^{n_{\mathcal{R}}-1},\\ \mathcal{P}_{\mathcal{S}}(k)&={}A_{\mathcal{S}}\left(\frac{k}{k_{*}}\right)^{n_{\mathcal{S}}-1},\\\mathcal{P}_{h}(k)&={}A_{h}\left(\frac{k}{k_{*}}\right)^{n_{h}}.\label{PowerLawPS}
\end{aligned}
\end{equation}
The constant amplitudes $A_{\mathcal{R}}$, $A_{\mathrm{S}}$ and $A_{h}$ as well as the spectral indices $n_{\mathcal{R}}$,  $n_{\mathcal{S}}$ and $n_{h}$, which we assume to be constant, are defined by \eqref{PowerLawPS}. It is convenient to define the inflationary isocurvature fraction $\beta_\mathrm{iso}$ and the tensor-to-scalar ratio $r$,
\begin{align}
\beta_{\mathrm{iso}}:=\frac{\mathcal{P}_{\mathcal{S}}(k)}{\mathcal{P}_{\mathcal{R}}(k)+\mathcal{P}_{\mathcal{S}}(k)},\qquad r:=\frac{\mathcal{P}_{h}(k)}{\mathcal{P}_{\mathcal{R}}(k)}.\label{biso}
\end{align}
The scalar amplitude $A_{\mathcal{R}}$ and the spectral index are determined at $68\%$ CL by the \textit{Planck} TT,TE,EE+lowE+lensing data at a pivot scale ${k_{*}= 0.05\; \text{Mpc}^{-1}}$ \cite{Akrami2018},
\begin{align}\label{PlanckAnsOne}
\text{ln}\left(10^{10}\,A_{\mathcal{R}}^{*}\right)={}&3.044^{+0.014}_{-0.014},\\ n_{\mathcal{R}}^{*} ={}& 0.9649\pm0.0042.\label{PlanckAns}
\end{align}
The tensor-to-scalar ratio \eqref{biso} has not been measured. For a pivot scale\footnote{The comoving reference scale $k_{*}$ corresponds to a physical wavelength $\lambda_{\mathrm{today}}=a_{\mathrm{today}}/k_{*}$ accessible to measurements today. In our conventions the scale factor today is normalized such that $a_{\mathrm{today}}=1$. We choose $k_{*}$ to correspond to that scale which exited the horizon at $N_{*}=60$ efolds before the end of inflation.} of $k_{*}= 0.002\; \text{Mpc}^{-1}$, \textit{Planck} TT,TE,EE+lowE\- +lensing+BK14 data provide an upper bound at $95\%$ CL \cite{Akrami2018},
\begin{align}
r^{*} < 0.064.\label{Planckr}
\end{align}
%
%------------------------------------------------------------------------------
\section{Scalaron-Higgs inflation}\label{Sec:ScalaronHiggs}
%------------------------------------------------------------------------------
%
In the following, we focus on the scalaron-Higgs model by identifying $\varphi$ with the Standard Model Higgs boson and by fixing the general function $f(R,\varphi)$ in \eqref{act1} to be
\begin{align}
f(R,\varphi)=\alpha R^2+\frac{1}{2}\left(M_{\mathrm{P}}^2+\xi\varphi^2\right)R-\frac{\lambda}{4}\left(\varphi^2-v^2\right)^2.\label{fHiggsStarobinsky}
\end{align}
Here, $\xi$ is the non-minimal coupling, $\lambda$ is the quartic self-coupling of the Higgs field and $v$ is the vacuum expectation value of $\varphi$, associated with the symmetry breaking scale $v\approx 246$ GeV. In addition to the non-minimal coupling term in Higgs-inflation \cite{Bezrukov2008}, the Einstein-Hilbert term is augmented by the quadratic curvature invariant $R^2$ with the dimensionless scalaron coupling constant $\alpha$. Like many other models such as Higgs-dilaton models \cite{Shaposhnikov2009, Blas2011b, Bezrukov2013,Casas2018}, or ``no-scale'' models such as \cite{Salvio2014,Rubio2017,Ferreira2018}, this model is inspired by an asymptotic classical scale invariance, which is approximately realized for large field values and curvatures where the two mass scales $M_{\mathrm{P}}\approx10^{18}$ GeV and $v\approx246$ GeV become negligible,
\begin{align}
\varphi\gg\frac{M_{\mathrm{P}}}{\sqrt{\xi}}\gg\nu,\qquad R\gg\frac{M_{\mathrm{P}}^2}{2\alpha },
\end{align}
It would also be interesting to study the consequences of an exact classical scale-invariance $(M_{\mathrm{P}},v)\to0$, where both mass scales $M_{\mathrm{P}}$ and $v$ are effectively induced via radiative corrections, such as in models of induced gravity \cite{Sakharov1968, Coleman1973, Zee1979, Adler1982, Kannike2015}.
One of the appealing features of the scalaron-Higgs model \eqref{fHiggsStarobinsky} is that the shape of the scalaron-Higgs potential $\hat{W}(\hat{\chi},\varphi)$ is completely fixed by the Standard Model Higgs potential and the two additional operators $\varphi^2R$ and $R^2$. 
Compared to the Standard-Model embedded in curved spacetime with a Higgs boson minimally coupled to Einstein gravity, the scalaron-Higgs model has two additional free parameters: the non-minimal coupling $\xi$ and the scalaron coupling $\alpha$. 
\subsection{Scalaron-Higgs inflation: two-field formulation in the Einstein frame}
We formulate the scalaron-Higgs model \eqref{fHiggsStarobinsky} as a two-field scalar-tensor theory.
Using \eqref{RPsiIdentification} and inserting \eqref{fHiggsStarobinsky} into \eqref{DefChi}, we obtain 
\begin{align}
\chi^2(R,\varphi)=2\alpha R+\frac{1}{2}\left(M_{\mathrm{P}}^2+\xi\varphi^2\right).
\end{align}
This can be inverted and solved for
\begin{align}
R(\chi,\varphi)=\frac{1}{4\alpha}\left(2\chi^2-\xi\varphi^2-M_{\mathrm{P}}^2\right).
\end{align}
Inserting this into \eqref{PotW}, 
the explicit form of the two-field scalar potential reads
\begin{align}
W(\chi,\varphi)=\frac{1}{4\alpha}\left[\frac{1}{2}\left(M_{\mathrm{P}}^2+\xi\varphi^2\right)-\chi^2\right]^2+\frac{\lambda}{4}\left(\varphi^2-v^2\right)^2.
\end{align}
In terms of the Einstein frame field $\hat{\chi}$, defined in \eqref{ChiEFOne}, the action for the scalaron-Higgs model has the form \eqref{ActScal}
with the Einstein frame potential \eqref{PotWEF} given explicitly by
\begin{align}
\hat{W}(\hat{\chi},\varphi)=\frac{e^{-2\gamma\frac{\hat{\chi}}{M_{\mathrm{P}}}}}{16\alpha}&\Big\{\left[\xi\varphi^2+M_{\mathrm{P}}^2\left(1-e^{\gamma\frac{\hat{\chi}}{M_{\mathrm{P}}}}\right)\right]^2\nonumber\\
&+4\alpha\lambda\left(\varphi^2-v^2\right)^2\Big\}.\label{Pot2F}
\end{align}
%
%------------------------------------------------------------------------------
\subsection{Properties of the scalaron-Higgs potential in the Einstein frame}
\begin{figure}
	\centering
	\begin{tabular}{cc}
		\includegraphics[width=0.45\linewidth]{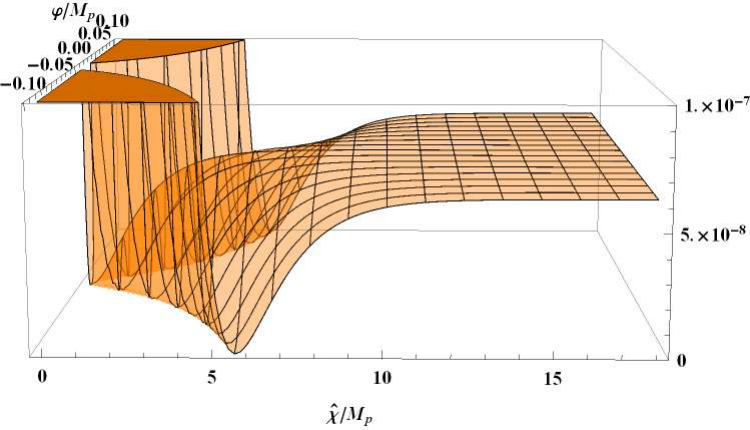}
		&
		\includegraphics[width=0.45\linewidth]{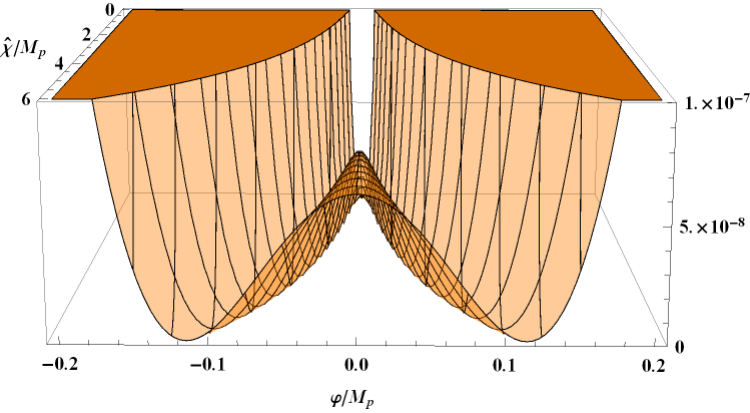}	
	\end{tabular}
	\caption{The two-field Einstein frame scalaron-Higgs potential $\hat{W}(\hat{\chi},\varphi)$ viewed from different angles. The $\varphi=\text{const}.$ profile has the form of the Starobinsky potential  (left). The $\hat{\chi}=\text{const.}$ profile has the form of the Standard Model Higgs potential (right).}
	\label{HiggsAndStarobinsky}
\end{figure}

%------------------------------------------------------------------------------
%
The scalaron-Higgs model is a combination of the Starobinsky model and the model of non-minimal Higgs inflation. This is naturally reflected in the shape of the scalaron-Higgs potential \eqref{Pot2F}, as depicted in Fig. \ref{HiggsAndStarobinsky}. Along the $\hat{\chi}$ direction (scalaron), it has the same profile as the Starobinsky potential. Along the $\varphi$ direction (Higgs), its functional form is that of the $\varphi^4$ Higgs potential. 

\noindent
For \mbox{$v/M_{\mathrm{P}}\ll\varphi/M_\mathrm{P}\ll1$}, the scalaron-Higgs potential \eqref{Pot2F} reduces to the Starobinsky potential \cite{Starobinsky1980},
\begin{align}
\hat{W}(\hat{\chi},\varphi)\approx\hat{W}(\hat{\chi})=\frac{M_{\mathrm{P}}^4}{16\alpha}\left(1-e^{-\gamma\frac{\hat{\chi}}{M_{\mathrm{P}}}}\right)^2.\label{PotStaroLimit}
\end{align}
For \mbox{$\hat{\chi}/M_{\mathrm{P}}\ll1$}, the scalaron-Higgs potential \eqref{Pot2F} reduces to the quartic Higgs potential, which for $\varphi/v\gg1$ has the form
\begin{align}
\hat{W}(\hat{\chi},\varphi)\approx\hat{W}(\varphi)=\frac{\tilde{\lambda}}{4}\varphi^4,\label{Wphi4}
\end{align}
with an effective quartic self-coupling
\begin{align}
\tilde{\lambda}:=\lambda+\frac{\xi^2}{4\alpha}.\label{tildelambda}
\end{align} 
The potential \eqref{Pot2F} has a saddle point (a minimum for $\hat{\chi}$ and a maximum for $\varphi$) at 
\begin{align}
\left(\hat{\chi}_{ \mathrm{sad}},\varphi_{\mathrm{sad}}\right)=\left(\frac{M_{\mathrm{P}}}{\gamma}\ln\left[1+4\alpha\lambda \left(\frac{v}{M_{\mathrm{P}}}\right)^4\right],0\right),\label{saddle}
\end{align}
and a two-fold degenerated minimum at
\begin{align}
\left(\hat{\chi}_{ \mathrm{min}},\varphi_{\mathrm{min}}\right)=\left(\frac{M_{\mathrm{P}}}{\gamma}\ln\left[1+\xi \frac{v}{M_{\mathrm{P}}}\right],\pm v\right).\label{min}
\end{align}
At the extrema, the potential acquires the values
\begin{align}
\hat{W}(\hat{\chi}_{ \mathrm{min}},\varphi_{\mathrm{min}})&={}0,\\
 \hat{W}(\hat{\chi}_{ \mathrm{sad}},\varphi_{\mathrm{sad}})&={}\frac{\lambda}{4}\frac{M_{\mathrm{P}}^4}{\left(M_{\mathrm{P}}/v\right)^4+4\alpha\lambda}.
\end{align}
For \mbox{$v/M_{\mathrm{P}}\ll1$}, all extrema \eqref{saddle} and \eqref{min} degenerate to a single global minimum at
\begin{align}
\left(\hat{\chi}_{\mathrm{min}},\varphi_{\mathrm{min}}\right)=\left(0,0\right).
\end{align}
For a broad range of the parameters $(\lambda,\xi,\alpha)$, the potential features two sharp valleys symmetrically aligned along the $\varphi$ axis (the potential \eqref{Pot2F} is invariant under reflections $\varphi\to-\varphi$).
In the limit \mbox{$\varphi/v\gg1$}, the location $\varphi_{\mathrm{v}}(\hat{\chi})$ of the two valleys  is determined by 
\begin{align}
\frac{\partial \hat{W}}{\partial \varphi}\Big|_{\varphi={}\varphi_\mathrm{v}}={}&0, \qquad\frac{\varphi_{\mathrm{v}}(\hat{\chi})}{M_{\mathrm{P}}}=\pm \left[\frac{\xi}{\lambda}\zeta\left(e^{\gamma\frac{\hat{\chi}}{M_{\mathrm{P}}}}-1\right)\right]^{1/2},\label{ValleyEQ}
\end{align}
where we have defined the combination of parameters
\begin{align}
\zeta:={}&\frac{\lambda}{\xi^2+4\alpha\lambda}.
\end{align}
The reason why the valley $\varphi_{v}(\hat{\chi})$, determined by the condition $\hat{W}_{,\varphi}=0$, serves as an attractor can be seen easily from the background equations of motion, which can be written as a system of first order equations
\begin{align}
\dot{\varphi}={}& v_{\varphi},\quad {\dot{v}}_{\varphi} = -\left[\left(3H - \frac{\gamma}{M_p}v_{\hat{\chi}}\right)v_{\varphi} + e^{\frac{\gamma\hat{\chi}}{M_{\mathrm{P}}}}\hat{W}_{,\varphi}\right],\\
\dot{\hat{\chi}}={}& v_{\hat{\chi}},\quad {\dot{v}}_{\hat{\chi}} = -\left[3H\,v_{\hat{\chi}} - \frac{\gamma}{M_\mathrm{P}}e^{-\frac{\gamma\hat{\chi}}{M_{\mathrm{P}}}}v_{\varphi}^2 +\hat{W}_{,\hat{\chi}}\right].
\end{align}
Clearly, a stationary point in the $(\varphi,v_{\varphi})$ plane requires $\hat{W},_{\varphi} = 0$. As the scalaron $\hat{\chi}$ rolls down the inflationary two-field potential \eqref{Pot2F}, the stationary point in the $(\varphi,v_{\varphi})$ plane moves to a different point given by $\varphi_{\mathrm{v}}(\hat{\chi})$ in \eqref{ValleyEQ}. In particular, this procedure of locating the valley is different to the procedure advocated in \cite{Wang2017}, which, unlike the simple closed expression \eqref{ValleyEQ}, leads to a rather lengthy and different expression for the valley equation.

A measure for the separation of the two valleys is provided by \eqref{ValleyEQ} and controlled by the parameter combination $\xi\zeta/\lambda$. The height difference between the hill and the valley is 
\begin{align}
H(\hat{\chi}):=&\frac{\left[\hat{W}(\hat{\chi},0)-\hat{W}(\hat{\chi},\varphi_{\mathrm{v}})\right]}{M_{\mathrm{P}}^4}\nonumber\\
={}&\frac{1}{16}\frac{\xi^2}{\lambda\alpha}\zeta\left(1-e^{-\gamma\frac{\hat{\chi}}{M_{\mathrm{P}}}}\right)^2.\label{ValleYFormation}
\end{align}
For \mbox{$\hat{\chi}/M_{\mathrm{P}}\gg1$}, the potential flattens out and asymptotically approaches the constant plateau\footnote{Avoiding trans-Planckian energy densities, the asymptotic value of $\hat{W}$ implies a lower bound on the scalaron coupling $\alpha\gtrsim10^{-1}$.}
\begin{align}
\hat{W}(\hat{\chi},\varphi)\approx\frac{M_{\mathrm{P}}^4}{16\alpha}.\label{Plateau}
\end{align} 

\begin{figure}
	\centering
	\begin{tabular}{cc}
		&\includegraphics[width=0.45\linewidth]{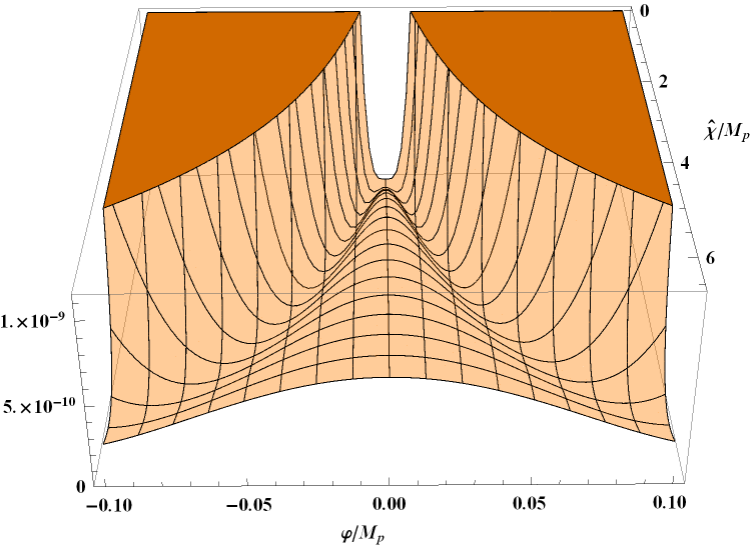}
		\includegraphics[width=0.45\linewidth]{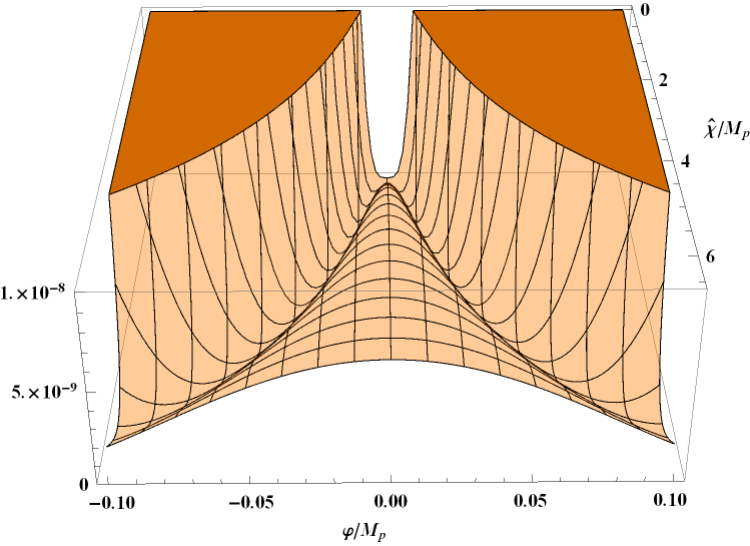}\\
		&\includegraphics[width=0.45\linewidth]{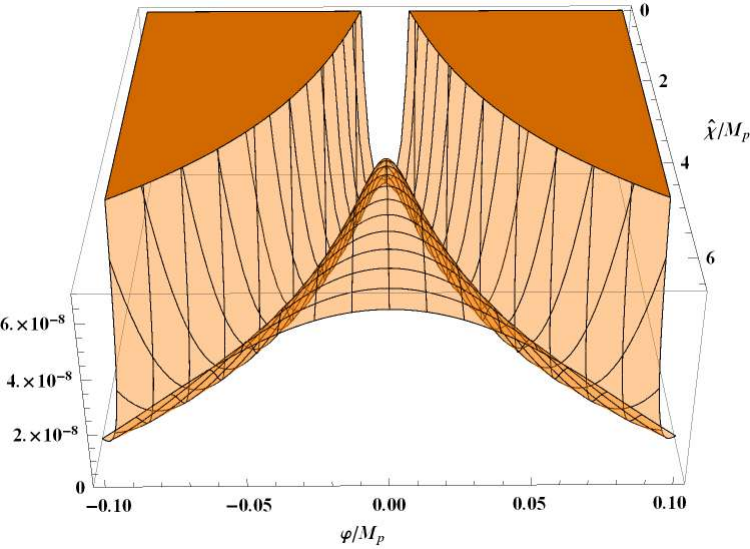}
		\includegraphics[width=0.45\linewidth]{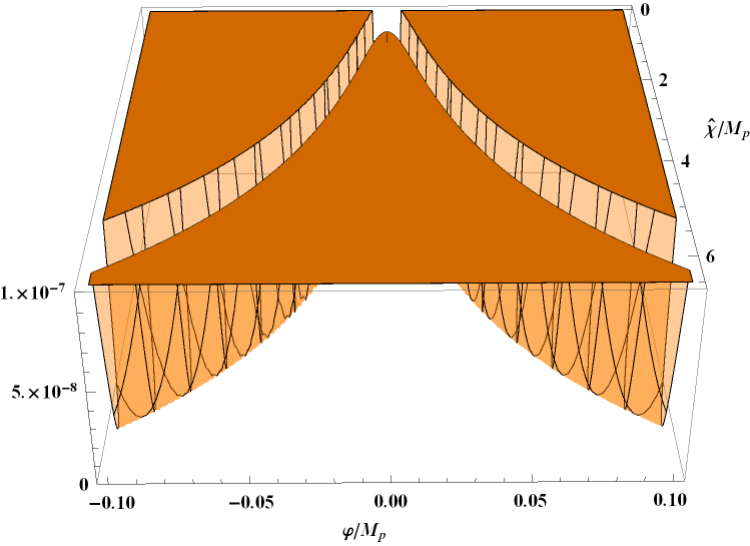}
	\end{tabular}
	\caption{Formation of the two valleys in $\hat{W}/M_{\mathrm{P}}^4$ for fixed values of $\xi=10^4$ and $\lambda=10^{-1}$ and different values for $\alpha$. From top-left to bottom-right: $\alpha=10^{8}$, $\alpha=10^{7}$, $\alpha=10^6$, $\alpha=10^5$. The lower the value for $\alpha$, the more prominent and steep are the valleys.}
\end{figure}

\noindent The two valleys located at $\pm\varphi_{\mathrm{v}}$ serve as natural attractors for the inflationary trajectory. Independent of the initial conditions, the solutions $\hat{\chi}(t)$ and $\varphi(t)$ of the background equations \eqref{KleinGordon} approach the valley at a certain moment of time $t$ before the end of their inflationary evolution. However, the history of the trajectories before entering the valley attractor can be different and might lead to observable multifield effects.
Once inside the valley, the two-field inflationary model effectively reduces to a single-field model. We perform a detailed analysis of different inflationary scenarios parametrized by different initial conditions and study their observational consequences. We classify the inflationary background trajectories into four different classes and discuss for each class how the transition to the effective single-field model takes place. We first focus on the case where $\lambda=M_{\mathrm{H}}^2/2v^2\approx10^{-1}$ is fixed by $M_{\mathrm{H}}\approx 125$ GeV and $v\approx 246$ GeV at the electroweak scale. In Sec. \ref{Sec:Isocurvature}, we explore the inflationary consequences of an extremely small coupling $\lambda$ for an abstract scalar field $\varphi$, a priori not identified with the SM Higgs boson. In Sec. \ref{Sec:Ident}, we investigate under which conditions this scenario can be dynamically realized within the RG improved scalaron-Higgs model for a running coupling $\lambda$.
%
%
%------------------------------------------------------------------------------
\subsection{Initial conditions and classification of background trajectories}
\label{ClassificationTrajectories}
%------------------------------------------------------------------------------
The inflationary background dynamics for the homogeneous background fields $a(t)$, $\hat{\chi}(t)$, and $\varphi(t)$ in the scalaron-Higgs model are determined by the system of background equations \eqref{Friedmann1}-\eqref{KleinGordon} with the scalaron-Higgs potential \eqref{Pot2F}. The time evolution of the two scalar fields $\hat{\chi}(t)$ and $\varphi(t)$ is obtained by solving the two second-order differential equations \eqref{KleinGordon}. Each equation requires two initial conditions \- $(\hat{\chi}_0,\varphi_0)$ and $(\dot{\hat{\chi}}_0,\dot{\varphi}_0)$ at some moment of time $t_0$. In the slow-roll approximation \eqref{SlowRollCond}-\eqref{SlowRollCondTwo}, equation \eqref{KleinGordon} reduces to two first-order differential equations and hence requires only two initial values $(\hat{\chi}_0,\varphi_0)$.  For slow-roll models of inflation with a single scalar field $\phi$, the inflationary trajectory is uniquely fixed and the initial condition $\phi_0$ can be expressed in terms of the value $\phi_{\mathrm{end}}$ at the end of inflation and the number of efolds $N$. For multifield inflation, in general there is no unique trajectory in the field space. Moreover, since the background solutions enter the evolution equations \eqref{DynEQPertPhi} of the perturbations $\left(\delta\hat{\chi},\delta\varphi\right)$, the inflationary observables inherit this dependence on the initial conditions.

Within an exact numerical treatment of the multifield scenario, the initial conditions which most closely resemble the slow-roll scenario are $\dot{\varphi}_0=0$ and $\dot{\hat{\chi}}_0=0$. We therefore restrict the following analysis to this case. At the end of this section, we briefly comment on the implications of more general initial conditions with non-zero ${\dot{\hat{\chi}}}_0$ and ${\dot{\varphi}}_0$. However, even for zero ``velocities'' $\dot{\varphi}_0=0$ and  $\dot{\hat{\chi}}_0=0$, the dependence on the two initial ``positions'' $\left(\hat{\chi}_0,\varphi_0\right)$ remains and various choices for $(\hat{\chi}_0,\varphi_0)$ in general lead to a plethora of trajectories in the potential landscape with different observational consequences. For more details on the impact of initial conditons in inflation see the review article \cite{Goldwirth1992}.
\begin{figure}
	\centering
	\begin{tabular}{cc}
		\includegraphics[width=0.45\linewidth]{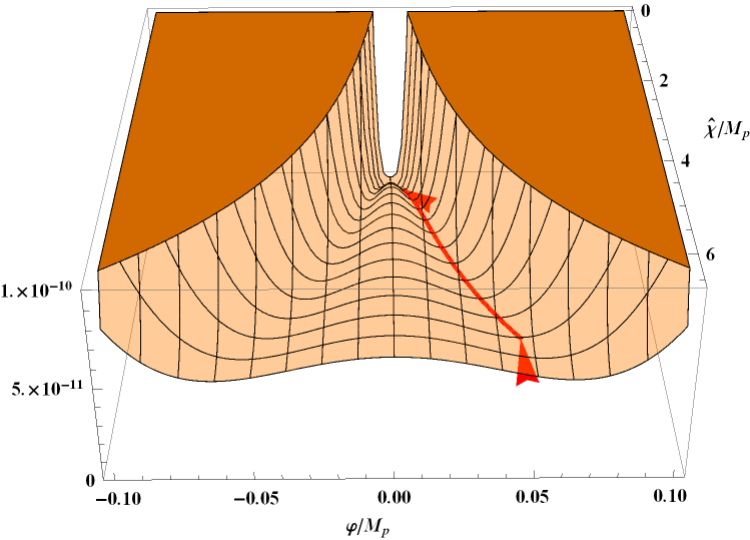}
		&
		\includegraphics[width=0.45\linewidth]{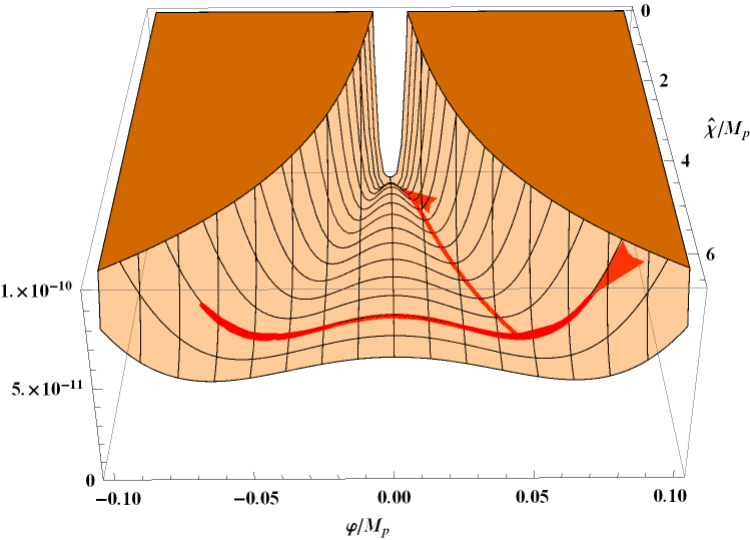}\\
		\\
		\includegraphics[width=0.45\linewidth]{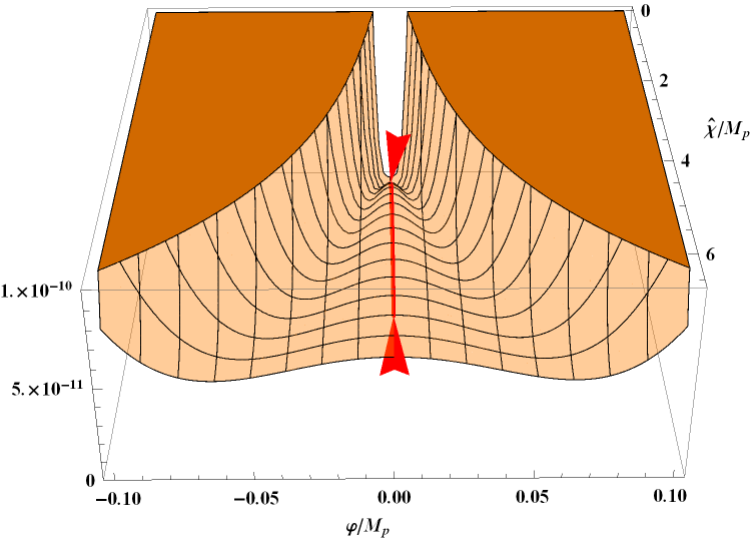}
		&
		\includegraphics[width=0.45\linewidth]{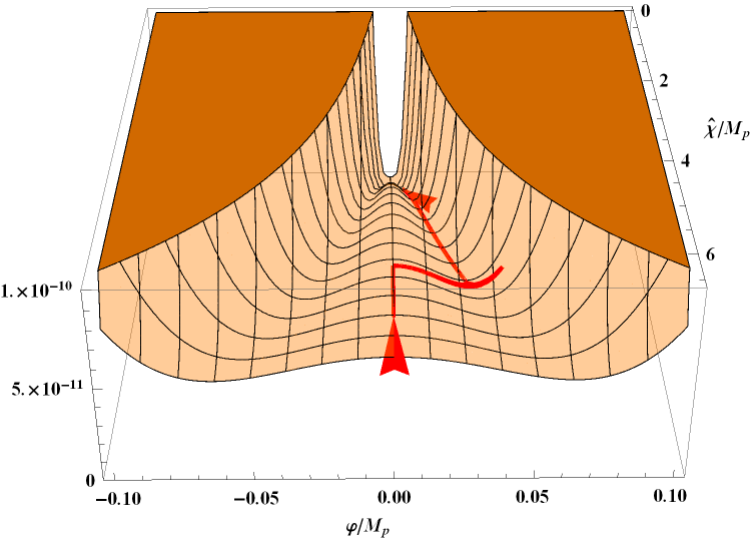}\\
	\end{tabular}
	\caption{The different inflationary background trajectories (red line in the direction of the arrows) in the scalaron-Higgs potential \eqref{Pot2F}, correspond to the four different classes of initial conditions $\varphi_0$. \mbox{(Class 1)} $\varphi_0/M_{\mathrm{P}}=\varphi_{\mathrm{v}}(\hat{\chi}_0)/M_{\mathrm{P}}=0.046$ (upper left), (Class 2) $\varphi_0/M_{\mathrm{P}}=0.07$ (upper right), (Class 3) $\varphi_0/M_{\mathrm{P}}=0$ (lower left) and (Class 4) $\varphi_0/M_{\mathrm{P}}=10^{-60}$ (lower right). In all plots we used the parameters $\alpha=10^9$, $\xi=10^4$, the initial conditions $\dot{\hat{\chi}}_0=0$, $\dot{\varphi}_0=0$ and $\hat{\chi}_0/M_{\mathrm{P}}=5.7$.}
	\label{BackgroundClassified}
\end{figure}

\noindent In previous work on the scalaron-Higgs model \cite{Wang2017, He2018} only the particular initial condition where the inflationary trajectory directly starts in the valley has been considered. We generalize the analysis and systematically study the impact of different initial conditions on the inflationary predictions. In the scalaron-Higgs model, the potential \eqref{Pot2F} has a particular structure with two prominent valleys, which allows to classify the inflationary background trajectories into four distinct classes of initial conditions $(\hat{\chi}_0,\varphi_0)$ for a broad range of parameters.
The functional form of the scalaron-Higgs potential suggests that inflationary trajectories with $\dot{\varphi}_0=0$, $\dot{\hat{\chi}}_0=0$ have to start at some \mbox{$\hat{\chi}_0/M_{\mathrm{P}}>1$}, as illustrated in Fig.~\ref{HiggsAndStarobinsky}. For a fixed value of $\varphi_0$, the value of $\hat{\chi}_{0}$ is connected to the number of efolds almost like in a single-field model of inflation. Different background trajectories can therefore be classified by different choices of $\varphi_0$. The particular shape of the scalaron-Higgs potential suggests that the space of initial conditions $\varphi_0$ is naturally divided into four classes:

For $|\varphi_{0}|=\varphi_{\mathrm{v}}$ the inflationary trajectory starts in one of the two valleys (Class $1$). For $|\varphi_{0}|>\varphi_{\mathrm{v}}$, the trajectory starts in one of the outer arms of the $\varphi^4$ part of the potential (Class $2$). For $\varphi_{0}=0$, the trajectory starts exactly on the hilltop (Class $3$). For values $\varphi_0=\pm\delta\ll1$, the trajectory starts on the hilltop but slightly displaced from the symmetrical point $\varphi=0$ (Class $4$). The different background trajectories for these four scenarios are depicted in Fig.~\ref{BackgroundClassified}.

The classification on the basis of the background dynamics is also reflected in the equations for the perturbations \eqref{EQPertQsig} and \eqref{EQPertQs}. The adiabatic mode $Q_{\sigma}$ is sourced by $\omega Q_{s}$ -- the product of the turn rate $\omega$ and the isocurvature mode $Q_{s}$. The dynamics of the adiabatic mode $Q_{\sigma}$ is only affected by the isocurvature mode $Q_{s}$ if both, $\omega$ and $Q_{s}$ are sufficiently large for a sufficient amount of time during inflation. The classification into four scenarios based on the initial conditions for the background evolution is also reflected by the classification according to the four different possible combinations of the source term $\omega Q_{s}$ in the equation for the adiabatic mode: 
\begin{enumerate}
	\item ($\omega=0, Q_{s}=0$): The turn rate $\omega$ and the isocurvature mode $Q_s$ are both negligible over the course of the inflationary evolution. The negligible turn rate can be seen from the straight line trajectory inside the valley (Class $1$).
	\item ($\omega\neq0, Q_{s}=0$): The turn rate $\omega$ is non-zero during part of the inflationary trajectory, but the isocurvature mode $Q_s$ is negligible over the course of the inflationary evolution. This scenario corresponds to the damped oscillations and the subsequent evolution in the valley (Class $2$).
	\item ($\omega=0, Q_{s}\neq0$): The turn rate $\omega$ is zero over the course of the inflationary evolution but the isocurvature perturbation $Q_s$ grows. This scenario corresponds to the trajectory that stays at the hilltop (Class $3$).
	\item ($\omega\neq0, Q_{s}\neq0$): The turn rate $\omega$ as well as the isocurvature perturbation $Q_s$ are non-zero during part of the inflationary dynamics. This scenario corresponds to the trajectory that first runs on the hilltop along the $\hat{\chi}$ direction for a certain number of e-folds and subsequently falls into the valley (Class $4$). 
\end{enumerate}
In Sections \ref{Class1}, \ref{Class2}, \ref{Class3}, and \ref{Class4}, we analyze the characteristic features of the four different scenarios and their observables consequences. 
\subsubsection{Class $1$: Effective single-field inflation inside the valley}
\label{Class1}
\begin{figure}
\centering
\begin{tabular}{cc}
\includegraphics[width=0.44\linewidth]{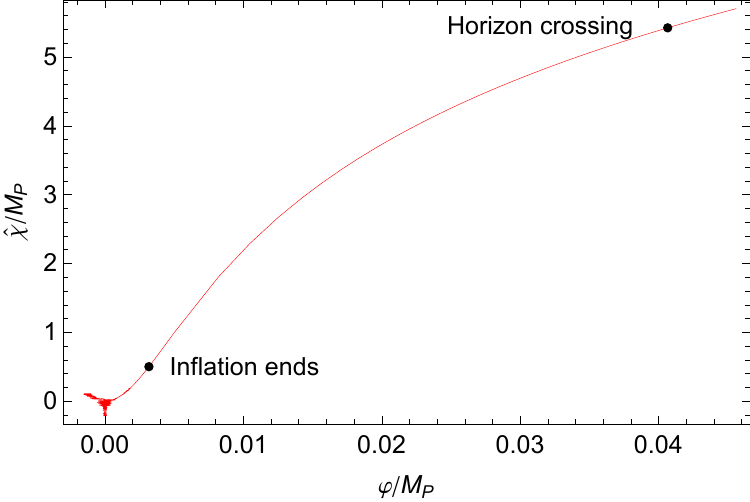}&$\;$
\includegraphics[width=0.455\linewidth]{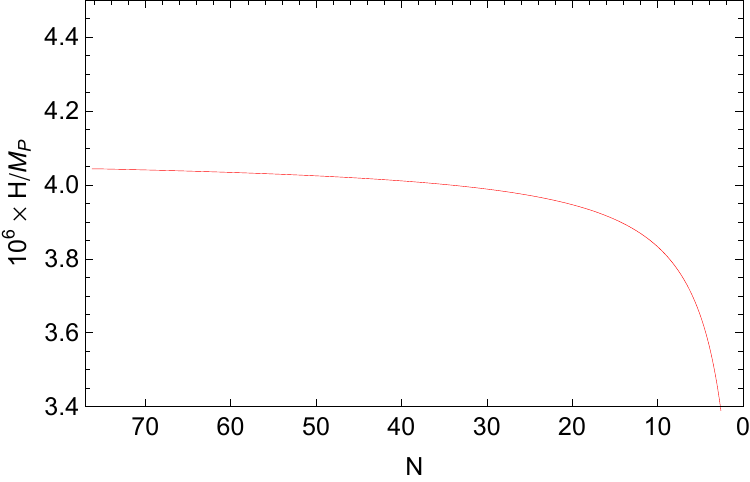}\\	
\includegraphics[width=0.44\linewidth]{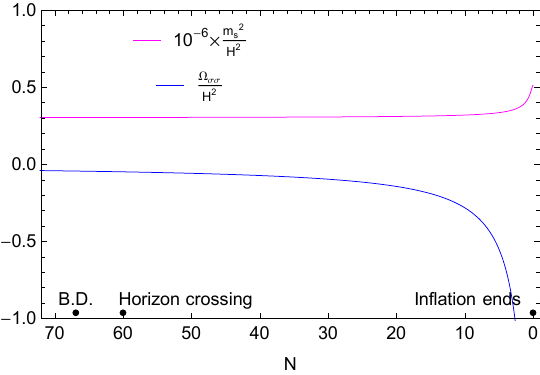}&$\;$
\includegraphics[width=0.455\linewidth]{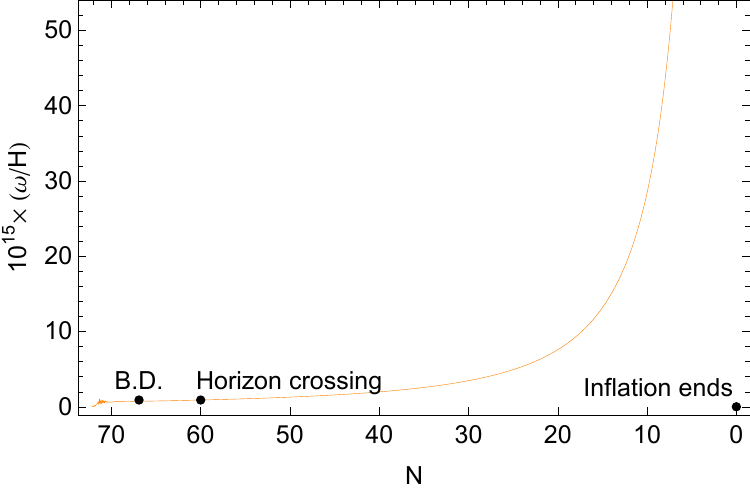}
\end{tabular}
\caption{Numerical solutions to the background equations \eqref{Friedmann1}-\eqref{KleinGordon} for the same parameters and initial conditions as in Fig.~\ref{BackgroundClassified} with $\varphi_0=\varphi_{\mathrm{v}}$ (Class 1). Upper left: Parametric plot which shows the inflationary trajectory along the valley and the subsequent oscillations around the global minimum $(\hat{\chi},\varphi)=(0,0)$. Upper right: The Hubble parameter as a function of the number of efolds is almost constant during inflation and smoothly decays at the end of inflation. Lower left: The damping ratios $m_{\rm{s}}^2/H^2$ and $\Omega_{\sigma\sigma}/H^2$ for the isocurvature mode (pink line) and the adiabatic mode (blue line) as a function of the number of efolds. In the valley, $m_{\rm{s}}^2$ is positive and large, while $\Omega_{\sigma\sigma}$ is negative and small $|\Omega_{\sigma\sigma}/m_{\rm{s}}^2|\approx 10^{-6}$. Lower right: Along the valley trajectory. the ratio of turn rate and Hubble parameter is almost constant and negligibly small $\omega/H\approx 10^{-15}$.}
\label{BackgroundClassified1}
\end{figure}

The trajectory along one of the valleys provides an attractor for the inflationary background dynamics. For generic initial conditions $\varphi_0$, inflation ultimately ends in one of the two valleys before it approaches the global minimum at $(\hat{\chi},\varphi)=(0,0)$. Therefore, we first focus on the scenario where the trajectory directly starts inside one of the valleys by setting $\varphi_0=\pm\varphi_{\mathrm{v}}$. This scenario is 
illustrated in Fig.~\ref{BackgroundClassified1}.
Inside a sharp valley, the inflationary dynamics reduces to an effective single-field model.
Inside the valley, the shape of the scalaron-Higgs potential in the $\varphi$ direction is convex and the effective isocurvature mass $m^2_{\rm{s}}$ is positive. According to \eqref{EQPertQs}, this leads to an exponential damping of the isocurvature mode $Q_{\rm{s}}$. Moreover, the turn rate $\omega$ is negligible along the trajectory inside the valley, which can be seen in the lower right plot of Fig.~\ref{BackgroundClassified1}. Therefore, along the trajectory inside the valley, the isocurvature mode can neither grow nor source the adiabatic mode and the predictions for the spectral observables have the same universal form as in the single-field models of non-minimal Higgs inflation and Starobinsky inflation.  
We discuss the reduction to the effective single-field model of inflation and the quality of the valley approximation in more detail in Sec.~\ref{Sec:ReductionSingleField}.
\subsubsection{Class $2$: Damped oscillations and subsequent inflation inside the valley}
\label{Class2}
\begin{figure}
	\centering
	\begin{tabular}{cc}
		\includegraphics[width=0.44\linewidth]{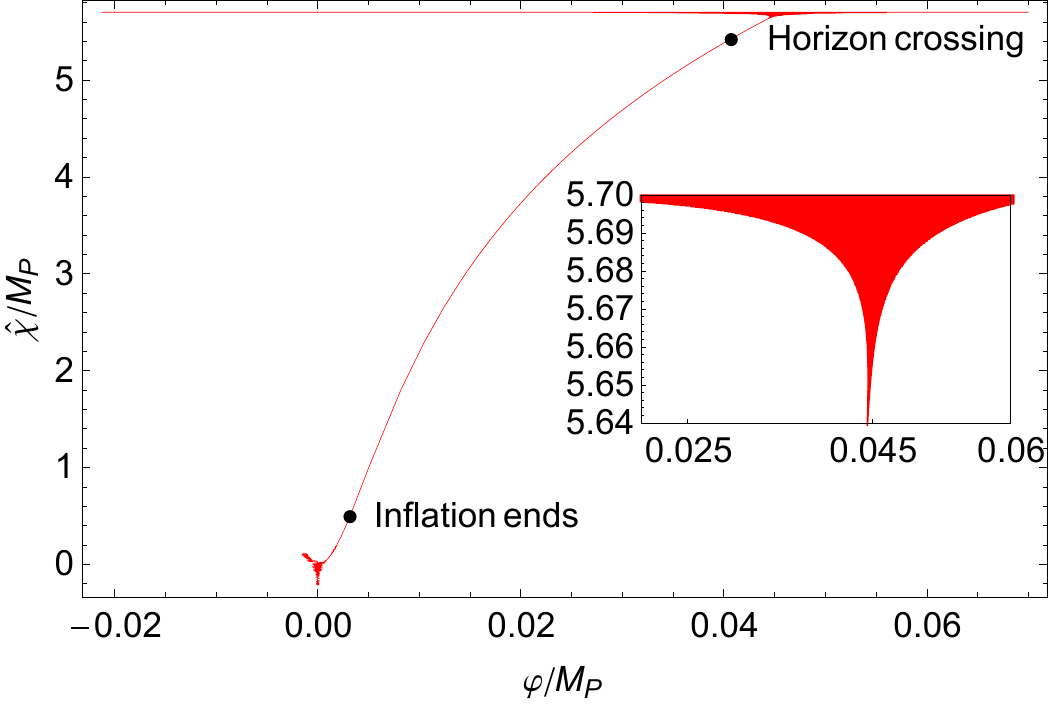}
		&$\;$
		\includegraphics[width=0.455\linewidth]{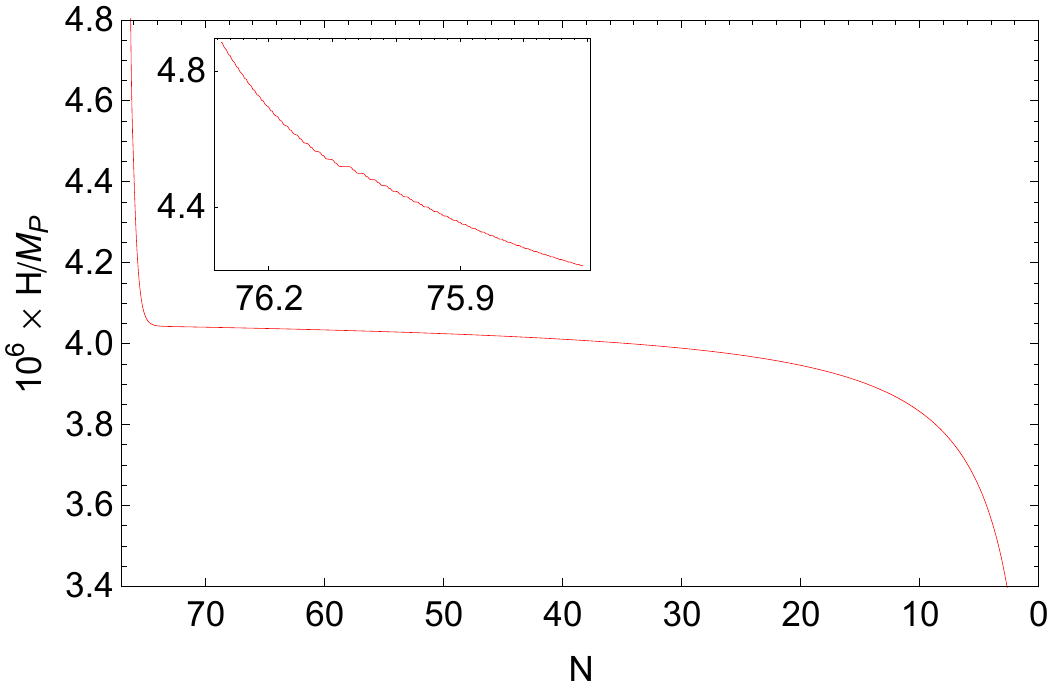}\\
		\includegraphics[width=0.44\linewidth]{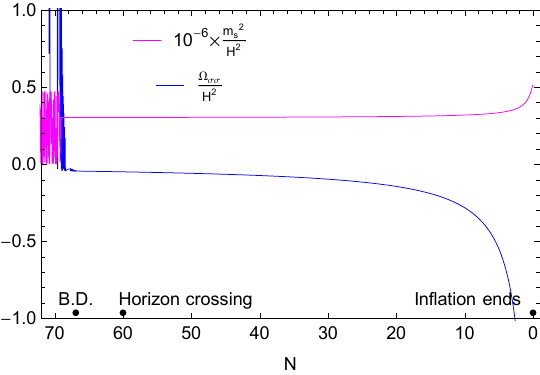}
		&$\;$
		\includegraphics[width=0.455\linewidth]{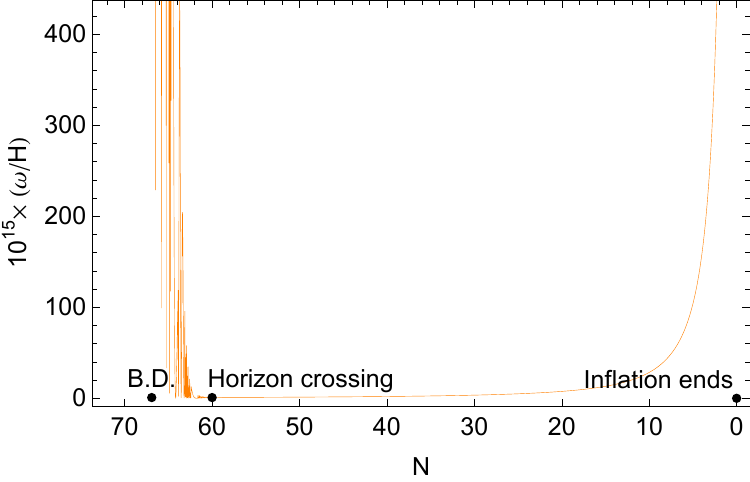}
	\end{tabular}
	\caption{Numerical solutions to the background equations \eqref{Friedmann1}-\eqref{KleinGordon} for the same parameters and initial conditions as in Fig.~\ref{BackgroundClassified} with $\varphi_0>\varphi_{\mathrm{v}}$ (Class 2). Upper left: The parametric plot shows how the oscillations in the $\varphi$ direction (highlighted in the inlay plot) are quickly damped out before horizon crossing and how the subsequent evolution of the inflationary trajectory obeys the attractor solution (Class 1). Upper right: The Hubble parameter as a function of the number of efolds starts at high values $H^2\propto\hat{W}(\hat{\chi},\varphi_0)>\hat{W}(\hat{\chi},\varphi_{v})$, drops during the first few (unobservable) efolds when the field is oscillating in $\varphi$ direction, is almost constant during inflation in the valley and smoothly decays at the end of inflation. Lower left: Inside the valley, the damping ratio $m_{s}^2/H^2$ for the isocurvature mode (pink line) is positive and large, while $\Omega_{\sigma\sigma}/H^2$ for adiabatic mode (blue line) is negative and small  $|\Omega_{\sigma\sigma}/m_{s}^2|\approx 10^{-6}$. During the first efolds, $\Omega_{\sigma\sigma}$ mainly measures the curvature of the potential along the oscillatory trajectory in $\varphi$ direction and oscillates due to the sign changes associated with the convex (in the valley) and concave (on the hill) shapes of the potential. Lower right: The ratio of the turn rate and the Hubble parameter $\omega/H$ is wildly oscillating during the first efolds and becomes almost constant once the trajectory tracks the valley solution.}
\label{BackgroundClassified2}
\end{figure}

The inflationary trajectory starts somewhere in the outer $\varphi^4$ arm of the scalaron-Higgs potential with \mbox{$\varphi_0>\varphi_{\mathrm{v}}$}. As illustrated in Fig.~\ref{BackgroundClassified2}, the trajectory is a superposition of fast oscillations in the $\varphi$ direction and a slow drift along the $\hat{\chi}$ direction. Depending on the height of the initial starting point $\hat{W}(\hat{\chi}_0,\varphi_0)$, the trajectory might perform multiple hill-crossings.

The friction term $3H\dot{\Phi}^I$ in \eqref{KleinGordon} sets the typical time scale $t_{\mathrm{osc}}\sim1/H$, after which the oscillations in $\varphi$ direction are expected to fade out. Suppose inflation lasts for a total number of $N_{\mathrm{tot}}$ efolds.\footnote{The number of total efolds  $N_{\mathrm{tot}}$ is unknown and might be chosen to be higher than the observationally required $N_{*}=50-60$. In a numerical treatment including perturbations, $N_{\mathrm{tot}}>N_*$ is required to ensure that the Bunch-Davies condition for the perturbations is imposed in the deep subhorizon regime at $N_{\mathrm{tot}}>N_{*}$.} During inflation $H\approx\text{const.}$ and $\left|\Delta N_{\rm{osc}}\right|\sim H\Delta t_{\rm{osc}}\sim \mathcal{O}(1)$. Thus, the oscillations in the $\varphi$ direction are damped out quickly after a few efolds $\Delta N_{\mathrm{osc}}\sim\mathcal{O}\left(1\right)$ and for $N_{\mathrm{tot}}>60$ cannot affect the observationally accessible scales of interest which cross the horizon between $N_{*}=50-60$.
Thus, the (Class 2) scenario quickly reduces to the effective single-field scenario in the valley (Class $1$) with no observable multifield features.
\subsubsection{Class $3$: Unstable inflation along the hilltop}
\label{Class3}
\begin{figure}
	\centering
	\begin{tabular}{cc}
		\includegraphics[width=0.44\linewidth]{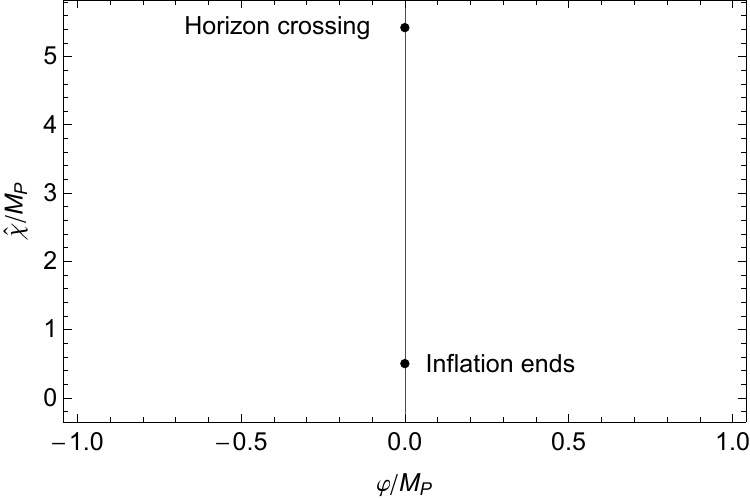}
		&$\;$
		\includegraphics[width=0.455\linewidth]{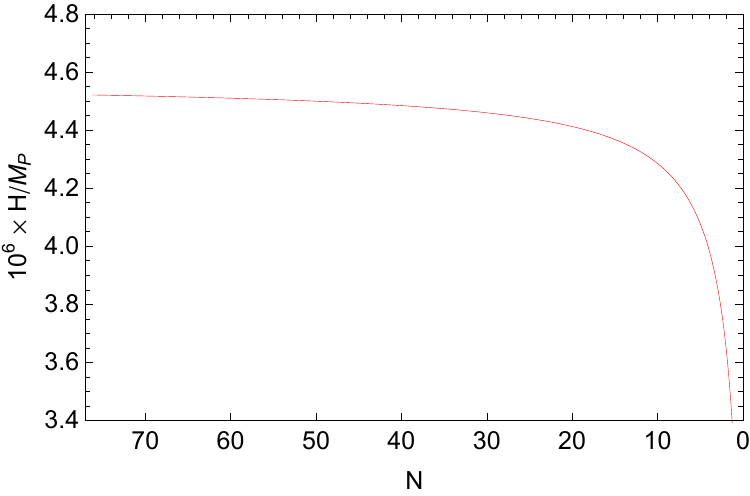}\\
		\includegraphics[width=0.44\linewidth]{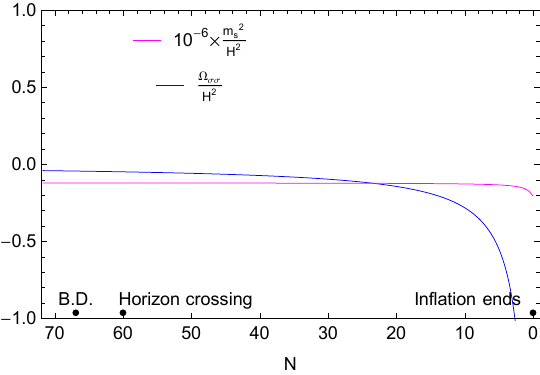}
		&$\;$
		\includegraphics[width=0.455\linewidth]{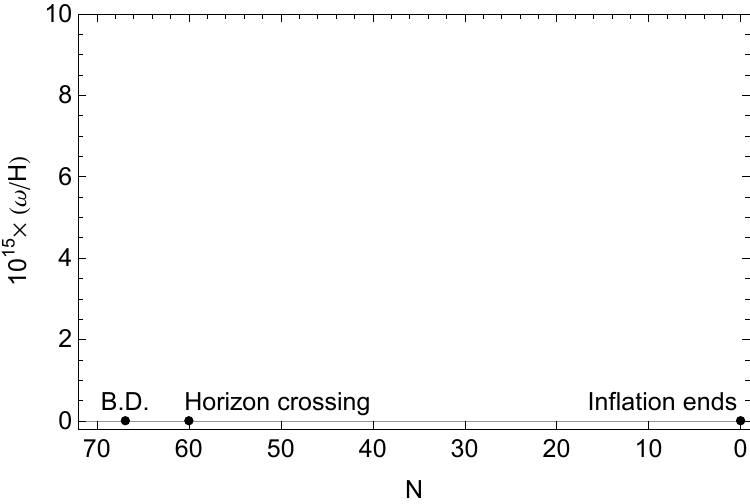}
	\end{tabular}
	\caption{Numerical solutions to the background equations \eqref{Friedmann1}-\eqref{KleinGordon} for the same parameters and initial conditions as in Fig.~\ref{BackgroundClassified} and $\varphi_0=0$ (Class 3). Upper left: The parametric plot shows the highly unstable background trajectory on the hilltop along the $\hat{\varphi}$ direction. For this plot the (strong) backreaction effects of the fluctuations on the background trajectory have been neglected. Upper right: The Hubble parameter as a function of the number of efolds is almost constant during inflation and smoothly decays at the end of inflation. Lower left: The damping ratios $m_{s}^2/H^2$ and $\Omega_{\sigma\sigma}/H^2$ for the isocurvature mode (pink line) and the adiabatic mode (blue line) as a function of the number of efolds. On the hilltop, $m_{s}^2<0$ is negative and large. This leads to an exponential growth of $Q_{s}\approx \delta\varphi$ which almost instantaneously would drive the background trajectory into one of the two valleys, but has been neglected for this plot. Lower right: The ratio of turn rate and Hubble rate $\omega/H$ is exactly zero along the (unperturbed) background trajectory.}
\label{BackgroundClassified3}
\end{figure}

The trajectory starts exactly on the hilltop at $\varphi_0=0$ (Class 3) and stays on the hilltop for the whole inflationary dynamics. 
On the hill, the concave shape of the potential in the $\varphi$ direction implies a tachyonic effective isocurvature mass $m^2_{\rm{s}}<0$, shown in the lower left diagram of Fig.~\ref{BackgroundClassified3}. According to \eqref{EQPertQs}, this leads to an exponential growth of the isocurvature modes $Q_{\rm{s}}$ and the linear split between background trajectory and perturbation $\varphi(t,\mathbf{x})=\bar{\varphi}(t)+\delta\varphi(t,\mathbf{x})$ becomes invalid once the perturbations $\delta\varphi$ grow too large (along the hilltop, the inflationary trajectory points along decreasing $\hat{\chi}$ such that $Q_{\sigma}\propto \delta\hat{\chi}$ and $Q_{\rm{s}}\propto\delta\varphi$). 
The steeper the curvature on the hilltop in $\varphi$ direction, the heavier the tachyonic mass $m_{s}^2<0$, the stronger the exponential growth of $Q_s\propto\delta\varphi$, the earlier the background trajectory will be pushed into one of the valleys.
For the initial background value $\bar{\varphi}_{0}=0$, any ever so tiny fluctuation $\delta\varphi$ will almost instantaneously push the classically highly unstable trajectory down in one of the two valleys.
Therefore also (Class 3) quickly reduces to the an effective single-field model in the valley (Class 1) with no observable multifield effects.
\noindent The scenario illustrated in Fig.~\ref{BackgroundClassified3}, in which the trajectory runs on the hilltop along the $\hat{\chi}$ direction, can therefore practically only be realized if the curvature of the potential in $\varphi$ direction on the hilltop is sufficiently small. This is realized in the small $\xi$ and small $\lambda$ limits as discussed in more detail in the Sections \ref{Sec:smallxi} and \ref{Sec:Isocurvature}. 
Nevertheless, in such a scenario, the large isocurvature mode cannot source the adiabatic mode $Q_{\sigma}$ in \eqref{EQPertQsig} because the turn rate vanishes exactly for the straight line trajectory along the hilltop.
Thus, the predictions for the adiabatic power spectrum would not be affected by isocurvature effects and would lead to the same predictions as in Starobinsky's model of inflation.
\subsubsection{Class $4$: Two-field inflation and multifield effects}
\label{Class4}
In all previous cases (Class 1)-(Class 3), the observable part of inflation ultimately takes place in one of the valleys. We show in more detail in Sec.~\ref{Sec:ReductionSingleField} that inside the valley, the scalaron-Higgs model effectively reduces to a single-field model with no observable multifield effects. 

Multifield effects associated to the fall from the hilltop into one of the valleys require the trajectory to stay sufficiently long on the hilltop such that the fall happens in the window when the observable modes cross the horizon. In addition, from the discussion of (Class 3), it is also clear that isocurvature effects can only have an observational impact if the background trajectory stays sufficiently long on the hilltop to allow the isocurvature mode to sufficiently grow ($m_{s}^2<0$). Finally, the trajectory must have a non-vanishing turn rate ($\omega\neq0$) in order to allow the sourcing of adiabatic mode. 
\begin{figure}
\centering
\begin{tabular}{cc}
\includegraphics[width=0.44\linewidth]{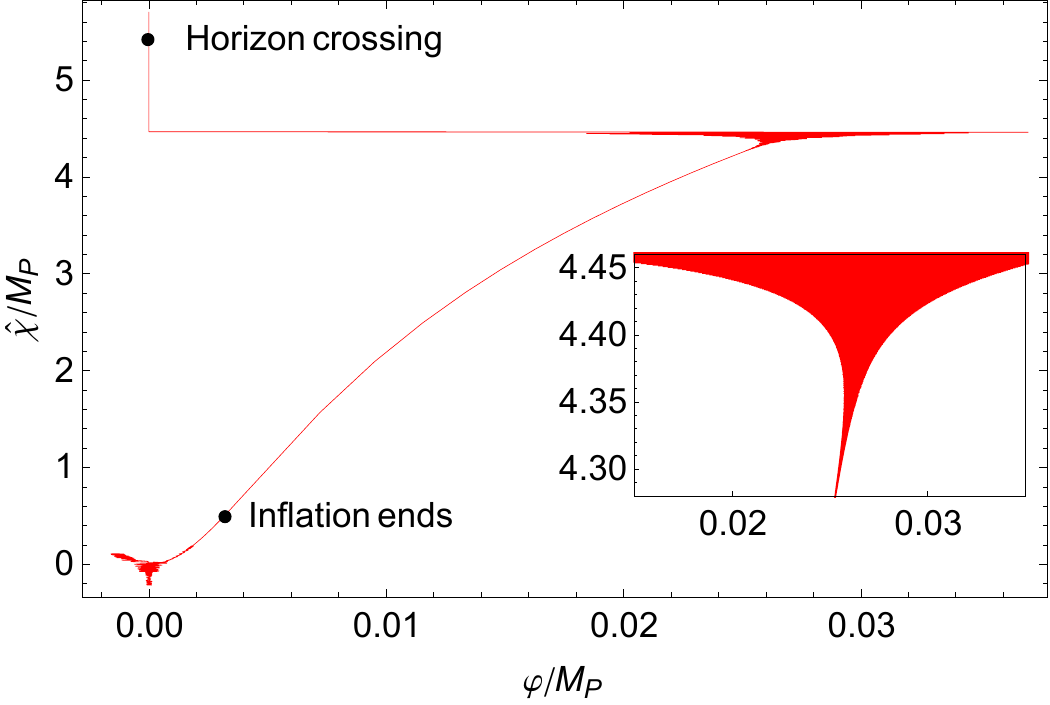}&$\;$
\includegraphics[width=0.455\linewidth]{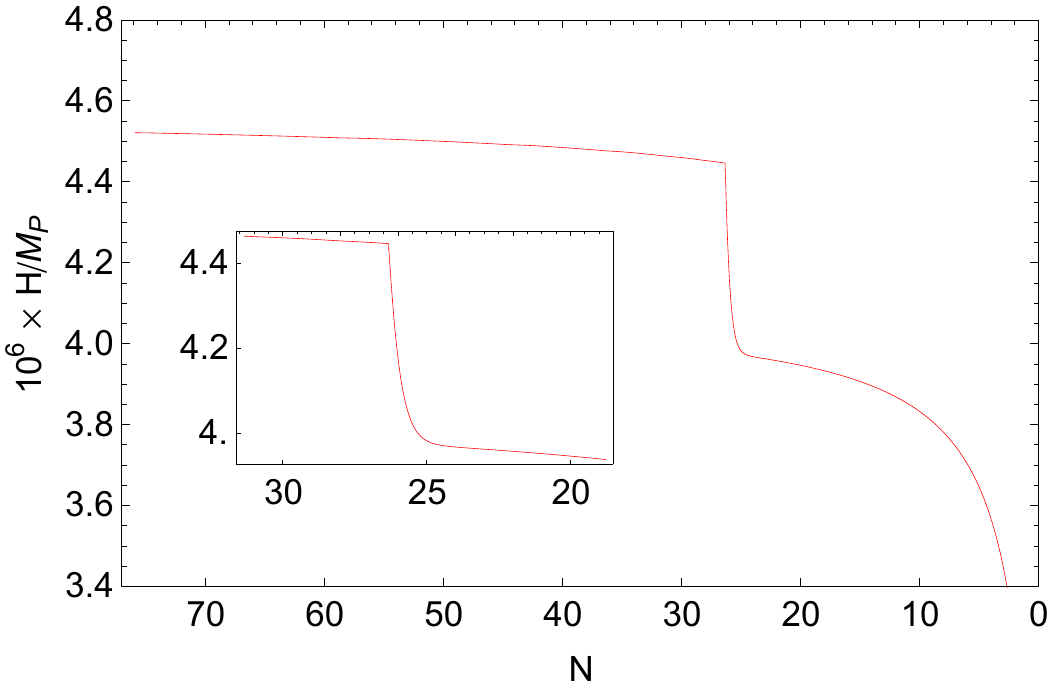}\\
\includegraphics[width=0.44\linewidth]{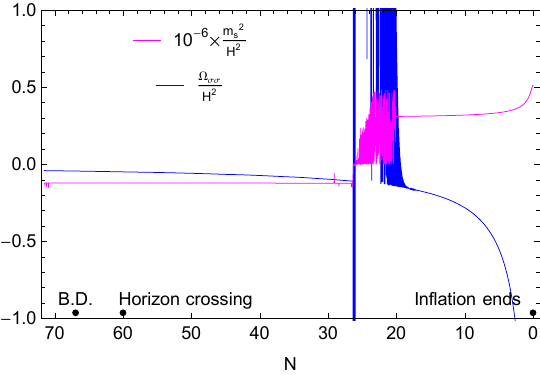}&$\;$
\includegraphics[width=0.455\linewidth]{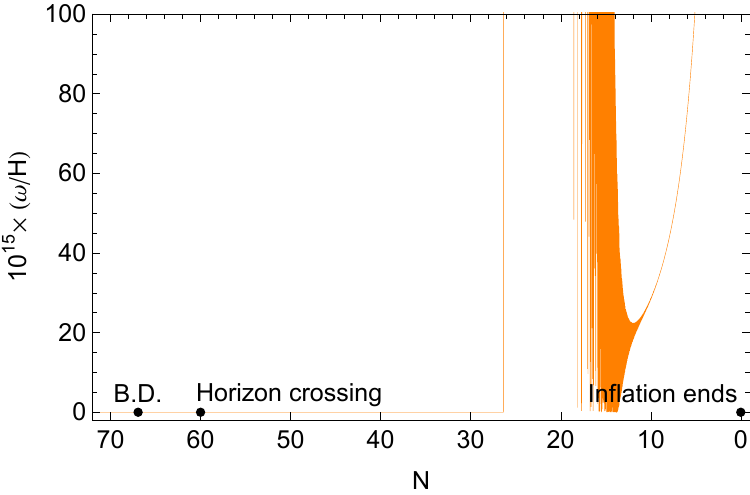}
\end{tabular}
\caption{Numerical solutions to the background equations \eqref{Friedmann1}-\eqref{KleinGordon} for the same parameters and initial conditions as in Fig.~\ref{BackgroundClassified} and $\varphi_0=\delta\ll1$ (Class 4). For this plot the (strong) backreaction effects of the fluctuations on the background trajectory have been neglected. Upper left: The parametric plot shows a short evolution along the hilltop, the subsequent fall into the valley, followed by fast oscillations around the valley (illustrated in the inlay graphic) and the tracking of the valley solution. Upper right: The Hubble parameter as a function of the number of efolds is almost constant during inflation on the hilltop as well as in the valley but shows a small dip when the background trajectory falls into the valley (illustrated in the inlay graphic). Lower left: The damping ratios $m_{s}^2/H^2$ and $\Omega_{\sigma\sigma}/H^2$ for the isocurvature mode (pink line) and the adiabatic mode (blue line) as a function of the number of efolds. During the evolution on the hilltop, both $m_{s}^2$ and $\Omega_{\sigma\sigma}$ are negative. Since $m_{s}^2$ is large and negative, isocurvature mode grows exponentially. The moment the trajectory falls into the valley, both $m_{s}^2$ and $\Omega_{\sigma\sigma}$ oscillate. Once the trajectory tracks the valley solution, $\Omega_{\sigma\sigma}$ is still negative and continues to slowly decay while $m_{s}^2$ has changed its sign and remains constant during inflation inside the valley. Lower right: The  ratio of turn and Hubble rate vanishes along the hilltop, grows when the trajectory falls into the valley, oscillates while the trajectory is oscillating around the valley and is almost constant and negligible during inflation inside the valley. }
\label{BackgroundClassified4}
\end{figure}

\noindent The case of a small deviation $\varphi_0=\delta\ll1$ from the $\varphi_0$ trajectory (Class 4) is illustrated in Fig.~\ref{BackgroundClassified4}. On the one hand, for the same reasons as in (Class 3), this is not a realistic scenario as it only takes into account the background dynamics but not the quantum fluctuations, which in this case would dominate and push the instable classical trajectory immediately into the valley. On the other hand, if the potential along the $\varphi$ direction is as steep as in Fig.~\ref{BackgroundClassified4}, initial values of $\varphi_0/M_{\mathrm{P}}>0$ (but not as small as $10^{-60}$)\footnote{This extremely small number has been chosen in the plot only for demonstration reasons, such that the trajectory does not immediately fall into the valley. Taking the feedback of quantum fluctuations into account, the purely classical evolution for such a tiny value of $\varphi_0$ becomes meaningless.} might not lie anymore on the hilltop and we effectively arrive again at the scenario described in (Class 2). On the hilltop $\hat{W},_{\varphi\varphi}/3H^2\approx \hat{W},_{\varphi\varphi}/\hat{W}$ can be interpreted as the second slow-roll parameter in the $\varphi$ direction. This quantity is equal to $-4\xi$ at $\varphi = 0$. Hence for $\xi\gtrsim 1$ we have $\eta_{\varphi\varphi}\sim 1$. Thus, slow-roll is not possible in the $\varphi$ direction and the trajectory will immediately be pushed into one of the valleys. The small $\xi$ limit, in which the valley structure disappears completely is discussed in more detail in Sec.~\ref{Sec:smallxi}.
The only way to have a trajectory on the hilltop for a sufficiently long time is to make the hilltop wider and flatter in $\varphi$ direction. It turns out that this is only possible for very small values of the quartic self-coupling $\lambda$, which we discuss in more detail in Sec. \ref{Sec:Isocurvature}.

Finally, we  briefly comment on more general initial conditions. For $\dot{\varphi}_0\neq0$ the qualitative picture does not change. The impact of a large $\dot{\varphi}_0$ would lead to a degeneracy of the four classes to (Class 2). In contrast, for $\dot{\hat{\chi}}_0\neq0$ slow-roll in $\hat{\chi}$ direction would not be possible. Nevertheless, a valid inflationary scenario with non-zero $\dot{\hat{\chi}}_0$ might still be realized. 
%
%------------------------------------------------------------------------------
\section{Effective single-field reduction}\label{Sec:ReductionSingleField}
%------------------------------------------------------------------------------
%
In the previous section we classified the background dynamics into four different classes, determined by different initial values $\varphi_0$ for a given value of $\hat{\chi}_0$, and solved the background equations numerically. In all the cases, we found that for $\lambda\approx 10^{-1}$ and $\xi\gtrsim1$, the inflationary trajectory inevitably and almost instantaneously reaches the attractor solution in one of the valleys. We conclude that for fixed quartic self-coupling $\lambda\approx 10^{-1}$ and $\xi\gtrsim1$, during the observable number of efolds $N_{*}$, scalaron-Higgs inflation takes place in one of the valleys. This is in agreement with the result in \cite{He2018} obtained for $\xi\gg1$ in the Jordan frame parametrization. The case $\xi\ll 1$ has not been studied in \cite{He2018} and requires additional care. We therefore study the consequences of $\xi\ll1$ separately in Sec.~\ref{Sec:smallxi}.

In this section, we investigate the inflationary scenario inside the valley. We discuss the reduction to the effective single-field model, the quality of the valley approximation and derive the expressions for the spectral observables.
\subsection{The valley approximation}
The valley equation \eqref{ValleyEQ} relates $\varphi$ with $\hat{\chi}$. Since Fig.~\ref{BackgroundClassified} suggests that inflation inside the valley takes place predominantly along the $\hat{\chi}$ direction, we use this relation in order to express $\varphi$ as a function of $\hat{\chi}$. Using $\varphi=\varphi_{\mathrm{v}}(\hat{\chi})$, the kinetic term of \eqref{ActScal} in the valley reads
\begin{align}
\frac{1}{2}\left.G_{IJ}\hat{g}^{\mu\nu}\Phi^{I}_{,\mu}\Phi^{J}_{,\nu}\right|_{\mathrm{v}}=\frac{1}{2}\left[1+\frac{1}{6\beta^2} K(\hat{\chi})\right]\partial_{\mu}\hat{\chi}\partial^{\mu}\hat{\chi}.
\label{KineticCompare}
\end{align}
The ratio of the kinetic terms in the valley is given by
\begin{align}
\left.\left(\frac{\partial\varphi }{\partial\hat{\chi}}\right)^2\right|_{v}={}&\frac{1}{6\beta^2} K(\hat{\chi}),\\ \beta^2:={}&\frac{\lambda}{\zeta\xi}=\xi\left(1+4\frac{\lambda\alpha}{\xi^2}\right),
\end{align}
and we have defined the $\hat{\chi}$-dependent prefactor by
\begin{align}
K(\hat{\chi}):=\frac{e^{\gamma\hat{\chi}/M_{\mathrm{P}}}}{e^{\gamma\hat{\chi}/M_{\mathrm{P}}}-1}\approx
\begin{cases}
1&\text{for }\gamma\hat{\chi}/M_{\mathrm{P}}\gg 1,\\
\frac{M_{\mathrm{P}}}{\gamma\hat{\chi}}&\text{for } \gamma\hat{\chi}/M_{\mathrm{P}}\ll 1.
\end{cases}\label{Kfunction}
\end{align}
During inflation inside the valley we have \mbox{$\hat{\chi}/M_{\mathrm{P}}>1$}, which according to \eqref{Kfunction} implies \mbox{$K(\hat{\chi})\approx1$}. 
We first investigate the case in which the kinetic term for $\hat{\chi}$ dominates, and the second term in \eqref{KineticCompare} can be safely neglected during the whole inflationary trajectory.\footnote{The opposite case is analyzed in Sec.~\ref{Sec:EffectiveSingleFieldu} and requires ${\lambda\ll 10^{-1}}$.}
For this approximation to hold, we must have $\beta^2\gg 1$.\footnote{For large $\xi$ with $\xi^2/\lambda\gg4\alpha$, we have $\zeta\approx\lambda/\xi^2$ and the numerical factor in \eqref{KineticCompare} is suppressed as $1/\beta^2\approx1/\xi\ll1$.}
Inserting \eqref{ValleyEQ} into the potential \eqref{Pot2F}, we obtain the effective single-field potential
\begin{align}
\hat{V}(\hat{\chi}):=\hat{W}(\hat{\chi},\varphi_{\mathrm{v}})=\frac{}{}\zeta\frac{M_{\mathrm{P}}^4}{4}\left(1-e^{-\gamma\frac{\hat{\chi}}{M_{\mathrm{P}}}}\right)^2.\label{potValley}
\end{align}
The functional form of the potential is the same as in non-minimal Higgs inflation \cite{Bezrukov2008} or Starobinsky inflation \cite{Starobinsky1980}.
Thus, the original two-field scalaron-Higgs model effectively reduces to a model of a single minimally coupled scalar field $\hat{\chi}$ with potential $V(\hat{\chi})$,
\begin{align}
S_{\mathrm{eff}}[\hat{g},\hat{\chi}]=\int\mathrm{d}^4x\sqrt{-\hat{g}}\left[\frac{M_{\mathrm{P}}^2}{2}\hat{R}-\frac{1}{2}\partial_{\mu}\hat{\chi}\partial^{\mu}\hat{\chi}-\hat{V}(\hat{\chi})\right].
\label{EffectiveSfAction}
\end{align}
We derive the inflationary observables in the next section and show that the CMB normalization leads to the constraint $\zeta\approx10^{-9}$.
For the value of the quartic self-coupling at the electroweak scale $\lambda\approx10^{-1}$, this implies the upper bounds $\alpha\lesssim10^{9}$ and $\xi\lesssim 10^4$. 
%
%--------------------------------------------------------------------
%
\subsection{Effective single-field predictions in scalaron-Higgs inflation}\label{SubSec:Predictions}
In single-field inflation, isocurvature effects are absent and the scalar perturbations are adiabatic. 
To first order in the slow-roll approximation, the inflationary observables for the scalar and tensor power spectrum \eqref{PowerLawPS} can be expressed analytically in terms of the inflaton potential $\hat{V}(\hat{\chi})$ and the slow-roll parameters.
The first two slow-roll parameters $\epsilon_{\rm{v}}$ and $\eta_{\rm{v}}$ in turn are expressed in terms of the inflaton potential $\hat{V}$ and its first and second derivatives $\hat{V}_{1}$ and $\hat{V}_{2}$,
\begin{align}
\epsilon_{\rm{v}}:=\frac{M_{\rm{P}}^2}{2}\,\left(\frac{\hat{V}_{1}}{\hat{V}}\right)^2,\qquad\eta_{\rm{v}}:=M_{\rm{P}}^2\,\left(\frac{\hat{V}_{2}}{\hat{V}}\right).\label{SlowRollParameters}
\end{align}
The field value $\varphi_{\mathrm{end}}$ is defined by the breakdown of the slow-roll approximation
\begin{align}
\epsilon_{\mathrm{v}}(\hat{\chi})|_{\hat{\chi}=\hat{\chi}_{\mathrm{end}}}:=1.\label{EndOfInfl}
\end{align}
The power spectra \eqref{PowerLawPS} and consequently \eqref{SlowRollParameters} are to be evaluated for the value of the inflaton field $\hat{\chi}_{*}$, which can be expressed in terms of the number of e-folds $N_{*}$ by solving the integral equation for $\varphi_{*}$,
\begin{align}
N_{*}-N_{\mathrm{end}}=\int_{t_{*}}^{t_{\rm{end}}}\,\text{d}t\,H\simeq M_{\mathrm{P}}^{-2}\,\int^{\hat{\chi}_{*}}_{\hat{\chi}_{end}}\,\text{d}\hat{\chi}\,\frac{\hat{V}}{\hat{V}_{1}}.\label{NumberEfolds}
\end{align}
Here $N$ runs from \mbox{$N_{*}\simeq60$} at the beginning of inflation to \mbox{$N_{\mathrm{end}}=0$} at the end of inflation.
The amplitudes and the spectral indices at horizon crossing are given by
\begin{equation}
\begin{aligned}
A_{h}^{*}:={}&\frac{2\,\hat{V}^{*}}{3\,\pi^2\,M_{\rm{P}}^4},\qquad n_{h}^{*}=-2\,\epsilon_{\rm{v}}^{*},\\ A_{\mathcal{R}}^{*}:={}&\frac{\hat{V}^{*}}{24\,\pi^2\,M_{\rm{P}}^4\,\epsilon_{\rm{v}}^{*}},\qquad n_{\mathcal{R}}^{*}=1+2\,\eta_{\rm{v}}^{*}-6\,\epsilon_{\rm{v}}^{*}.\label{CMBAmpl}
\end{aligned}
\end{equation}
The tensor-to-scalar ratio in terms of the first slow-roll parameter $\epsilon_{\rm{v}}$ reads
\begin{align}
r^{*}= \frac{{\cal P}_{h}(k_*)}{{\cal P}_{\mathcal{R}}(k_*)}=\frac{A_{h}^{*}}{A_{\mathcal{R}}^{*}}=16\,\epsilon_{\rm{v}}^{*}=-8\,n_{h}^{*}\label{TTS}.
\end{align}
The last equation is a consistency relation of single-field models of inflation, which effectively reduces the inflationary parameters from the four observables \eqref{CMBAmpl} to the three observables $A_{\mathcal{R}}$, $n_{\mathcal{R}}$ and $r$.  

We can evaluate the expressions \eqref{CMBAmpl}-\eqref{TTS} for the effective single-field potential \eqref{potValley}.
The integral \eqref{NumberEfolds} is dominated by the upper integration bound $\hat{\chi}_{*}$. Keeping only the dominant exponential and inverting the relation we obtain $\hat{\chi}_{*}$ as a function of $N_{*}$,
\begin{align}
\hat{\chi}(N_{*})\approx\frac{M_{\mathrm{P}}}{\gamma}\ln\left(2\gamma^2 N_{*} \right).\label{HCchi}
\end{align}
Inserting this into the slow-roll parameters yields to leading order in the large $N_*$ limit
\begin{align}
\varepsilon_{\mathrm{v}}^{*}\approx\frac{1}{2\gamma^2\,N_*^2},\qquad \eta_{\mathrm{v}}^{*}\approx-\frac{1}{N_{*}}.\label{slowrollStar}
\end{align}
Finally, inserting \eqref{HCchi}-\eqref{slowrollStar} with $\gamma=\sqrt{2/3}$ into the expressions for the slow-roll observables \eqref{CMBAmpl} and \eqref{TTS}, we obtain to leading order in $N_{*}$ the well-known results
\begin{align}
A_{\mathcal{R}}^{*}=\frac{N_{*}^2}{72\pi^2}\zeta,\qquad n_{\mathcal{R}}^{*}=1-\frac{2}{N_{*}},\qquad r^{*}=\frac{12}{N_{*}^2}.\label{ObsStar}
\end{align}
Assuming that the scale $k_{*}=0.002\rm{Mpc}^{-1}$ crossed the horizon at $N_{*}= 60$, we obtain the predictions for the effective single-field model in the valley approximation from \eqref{ObsStar},
\begin{align}
A_{\mathcal{R}}^{*}\approx\frac{50}{\pi^2}\zeta,\qquad n_{\mathcal{R}}^{*}\approx 0.967,\qquad r^{*}\approx0.0033.\label{PredStar}
\end{align}
The scalar spectral index and the tensor-to-scalar ratio of the Starobinsky and Higgs-inflation scenarios are independent of the model parameters and fall into the same universal attractor regime \eqref{ObsStar}, which is in excellent agreement with the recent \textit{Planck} data \eqref{Planckr}. The only constraint on the model parameters comes from the normalization of the scalar amplitude. Combining \eqref{PlanckAns} and \eqref{PredStar} leads to
\begin{align}
\zeta=\frac{\lambda}{\xi^2+4\alpha\lambda}\approx 4.143\times 10^{-10}\label{ConstCMBStar}.
\end{align}
The constraint \eqref{ConstCMBStar} simultaneously implies upper bounds on $\alpha$ and the ratio $\xi^2/\lambda$,
\begin{align}
\alpha\lesssim 10^{9},\qquad \frac{\xi^2}{\lambda}\lesssim 10^{9}.\label{Boundsalphaxi}
\end{align}
For $\lambda\approx 10^{-1}$, \eqref{Boundsalphaxi} leads to an upper bound on $\xi$ alone $\xi\lesssim10^{4}$, resulting in two cases:
\begin{enumerate}
	\item For $\alpha\gg\xi^2$, the CMB constraint \eqref{Boundsalphaxi} is determined by $\alpha$ alone and therefore fixes $\alpha\approx 10^8$. In this case, $\xi$ is only bounded from above $\xi<10^4$. In particular the non-minimal coupling $\xi$ can take on small values which relaxes the situation with the strong coupling present in non-minimal Higgs inflation. The extreme limit $\xi\to0$ is investigated separately in Sec.~\ref{Sec:smallxi}.
	\item For $\xi^2\gg\alpha$, the CMB constraint \eqref{Boundsalphaxi} is determined by $\xi$ alone and therefore fixes $\xi\approx 10^4$.  In this case $\alpha$ is only bounded from above $\alpha<10^9$. However, in contrast to the $\xi\to0$ limit, the $\alpha\to 0$ limit is singular, which can be seen for example from the two-field potential \eqref{Pot2F} -- a manifestation of the associated discontinuous change in the propagating degrees of freedom, as setting $\alpha\to0$ in \eqref{act1} would eliminate the scalaron.
\end{enumerate}

\noindent For fixed $\lambda=10^{-1}$ the bounds \eqref{ConstCMBStar} and \eqref{Boundsalphaxi} are self-consistent with the elimination of the $\varphi$ kinetic term inside the valley \eqref{KineticCompare}, as $1/\beta^2\approx 10^{-8}\xi<10^{-4}$. This does not identically prove that the kinetic term for $\varphi$ can be neglected for any combination of the free parameters $\left(\xi,\alpha\right)$, as the results \eqref{ConstCMBStar} and \eqref{Boundsalphaxi} were derived under this assumption. But given the order of magnitude of $\zeta = \mathcal{O}\left(10^{-9}\right)$, which according to \eqref{potValley} controls the potential energy inside the valley, it is hard to imagine a situation where the kinetic term for $\varphi$ might become dominant, since this would require $1/\beta^2 = \xi\zeta/\lambda\gg 1$, which in the light of \eqref{ConstCMBStar} and \eqref{Boundsalphaxi} can hardly be realized, unless $\lambda$ is allowed to be significantly smaller than $10^{-1}$. Thus, for $\lambda=10^{-1}$ and a broad range of the parameters $\left(\xi,\alpha\right)$, the CMB constraint restricts the inflationary background trajectory to point predominantly along the $\hat{\chi}$ direction.
\subsection{Turn rate, effective mass and absence of isocurvature effects}
The observationally distinct feature of a single-field scenario is the absence of multifield and isocurvature effects. From the dynamical equation \eqref{EQPertQs}, it is clear that the effective isocurvature mass $m^2_{s}$, defined in \eqref{IsoMassDef}, decides whether the isocurvature perturbation $Q_{s}$ grows or is sufficiently suppressed. Moreover, for a sourcing of the adiabatic mode by the isocurvature mode, the turn rate must be non-vanishing for a sufficiently long time.
Since for $\lambda\approx10^{-1}$, the $\varphi$ kinetic term is strongly suppressed compared to that of $\hat{\chi}$ inside the valley, this implies that $\hat{\sigma}^{I}$ points approximately in the $\hat{\chi}$ direction.
In view of \eqref{dsigma}, the speed $\dot{\sigma}$ is determined by $\dot{\hat{\chi}}$ alone, 
\begin{align}
\dot{\sigma}\veq\left[1+\frac{1}{4\beta^2} K(\hat{\chi})\right]^{1/2}\dot{\hat{\chi}}\approx\dot{\hat{\chi}}.\label{speedvalley}
\end{align}	
Here we have introduced the symbol $\veq$, which denotes equality under the assumption that the valley approximation holds. In particular, inside the valley, the derivatives of the potential $\hat{W}$ are to be evaluated at $\varphi=\varphi_{\mathrm{v}}$ after differentiation.
Together with \eqref{dsigmaTwo} and \eqref{norms}, \eqref{speedvalley} implies for the components of $\hat{\sigma}^{I}$ and $\hat{s}^{I}$,
\begin{align}
\hat{\sigma}^{I}={}&
\left(
\begin{array}{c}
\hat{\sigma}^{\hat{\chi}}\\
\hat{\sigma}^{\varphi}
\end{array}
\right)
\veq
\left(
\begin{array}{c}
1\\
0
\end{array}
\right),\\
\hat{s}^{I}={}&
\left(
\begin{array}{c}
\hat{s}^{\hat{\chi}}\\
\hat{s}^{\varphi}
\end{array}
\right)
\veq
\left(
\begin{array}{c}
0\\
\pm e^{\frac{\gamma}{2}\frac{\hat{\chi}}{M_{\mathrm{P}}}}
\end{array}
\right).
\label{InflatonApprox}
\end{align}
Thus, the isocurvature unit vector $\hat{s}^{I}$ points along the $\hat{\varphi}$ direction and the inflaton unit vector $\hat{\sigma}^{I}$ in the $\hat{\chi}$ direction. This also implies that $\delta\chi$ can be associated with the adiabatic inflaton perturbation $Q_{\sigma}$ and $\delta\varphi$ with the isocurvature perturbation $Q_{s}$. Combining \eqref{InflatonApprox} with \eqref{ValleyEQ}, we find
\begin{align}
\hat{W}_{,s}=\hat{s}^{I}\frac{\partial \hat{W}}{\partial\Phi^{I}}\veq e^{-\frac{\gamma}{2}\frac{\hat{\chi}}{M_{\mathrm{P}}}}\hat{W}_{,\varphi}\veq 0.\label{potsValley}
\end{align}
Therefore, due to \eqref{TurnRateIsoDerive}, to first approximation the turn rate is  negligible inside the valley\footnote{Inside the valley, $\hat{s}^{I}$ is only approximately pointing in $\varphi$ direction and therefore \eqref{potsValley} is only approximately satisfied. For slow-roll inflation inside the valley, both $\dot{\sigma}\ll1$ as well as $\hat{W}_{,s}\ll1$ and $\omega$ can be non-zero.}
\begin{align}
\omega=-\frac{\hat{W},_{s}}{\dot{\sigma}}\veq 0.\label{turnvalley}
\end{align}
The vanishing of the turn rate is directly related to the accuracy of the valley approximation and implies that $Q_{\sigma}$ and $Q_{s}$ do not couple and evolve independently.
Moreover, for a vanishing turn rate, the effective isocurvature mass \eqref{IsoMassDef} only depends on the projection of the effective mass matrix $M_{IJ}$ \eqref{EffMassMatrix} in the $\hat{s}^{I}$ direction and therefore only receives contributions from the field space curvature and the curvature of the potential $\hat{W}$ along the $\hat{s}^{I}$ direction
\begin{align}
m_{s}^2\veq \hat{s}^{I}\hat{s}^{J}\left(\nabla_{I}\nabla_{J}\hat{W}+R_{IKJL}\dot{\Phi}^{K}\dot{\Phi}^{L}\right).
\end{align}
Making use of \eqref{GeoCurv} and \eqref{GeoGam}, we evaluate both contributions in the valley approximation
\begin{align}
\hat{s}^{I}\hat{s}^{J}R_{IKJL}\dot{\Phi}^{K}\dot{\Phi}^{L}\veq{}&\dot{\sigma}^2\hat{s}^{\varphi}\hat{s}^{\varphi}R_{\varphi\hat{\chi}\varphi\hat{\chi}}=-\frac{\dot{\sigma}^2}{6M_{\mathrm{P}}^2},\label{CurvValley}\\
\hat{s}^{I}\hat{s}^{J}\nabla_{I}\nabla_{J}\hat{W}\veq{}& \hat{s} ^{\varphi}\hat{s} ^{\varphi}\hat{W}_{,\varphi\varphi}-\hat{s} ^{\varphi}\hat{s} ^{\varphi}\Gamma^{\hat{\chi}}_{\varphi\varphi}\hat{W}_{,\hat{\chi}}\nonumber\\
\veq{}& \hat{s} ^{\varphi}\hat{s} ^{\varphi}\hat{W}_{,\varphi\varphi}-\frac{\gamma}{2 M_{\mathrm{P}}}\hat{W}_{,\hat{\chi}}.\label{MassValley}
\end{align}
During slow-roll in the valley, \eqref{CurvValley} and the last term in \eqref{MassValley} can be expressed in terms of $\varepsilon_{\sigma}$ via \eqref{SlowRollCond}-\eqref{SlowRollCondTwo} and \eqref{SlowRollPot},
\begin{align}
-\frac{\dot{\sigma}^2}{6M_{\mathrm{P}}^2}\veq{}&-\frac{1}{3}H^2\varepsilon_{\sigma},\\ -\frac{\gamma}{2M_{\mathrm{P}}}\hat{W}_{,\hat{\chi}}\veq{}& -H^2\sqrt{3\varepsilon_{\sigma}}.
\end{align}
Inside the valley, the dimensionless damping ratio $m_{s}^2/3H^2$ becomes
\begin{align}
\frac{m_{s}^2}{3H^2}\veq M_{\mathrm{P}}^2\frac{s ^{\varphi}s ^{\varphi}\hat{W}_{,\varphi\varphi}}{\hat{W}}-\frac{\varepsilon_{\sigma}}{9}-\frac{1}{3}\sqrt{3\varepsilon_{\sigma}}.\label{msvalley}
\end{align} 
The first term can be evaluated explicitly and yields
\begin{align}
M_{\mathrm{P}}^2\frac{{\hat{s}} ^{\varphi}{\hat{s}} ^{\varphi}\hat{W}_{,\varphi\varphi}}{\hat{W}}\veq{}& 2\frac{\xi}{\alpha\zeta}K(\hat{\chi})\nonumber\\
={}& 8\xi\left(1+\frac{\xi^2}{4\alpha\lambda}\right)K(\hat{\chi}).\label{DoubleDerivePotV}
\end{align}
The slow-roll parameter $\varepsilon_{\sigma}$ for the scalaron-Higgs potential in the valley approximation  reads
\begin{align}
\varepsilon_{\sigma}\veq \frac{4}{3}\left(1-e^{\gamma\frac{\hat{\chi}}{M_{\mathrm{P}}}}\right)^{-2}.\label{SReps}
\end{align}
Inflation in the valley takes place for $\hat{\chi}/M_{\mathrm{P}}>1$ and stops at $\hat{\chi}/M_{\mathrm{P}}\approx1$, where $\varepsilon_{\sigma}\veq 1$. During slow-roll inflation, the last two terms in \eqref{msvalley} can be neglected for $\xi\gtrsim 1$. Therefore, for a Higgs self-coupling $\lambda\approx10^{-1}$ along with a non-minimal coupling $\xi\gtrsim 1$, the isocurvature damping ratio reduces to
\begin{align}
\frac{m_{s}^2}{3H^2}\veq
\begin{cases}
8\xi K(\hat{\chi})&\text{for }\alpha\gg\xi^2/\lambda ,\\
 2\frac{\xi}{\alpha}\frac{\xi^2}{\lambda} K(\hat{\chi})\,&\text{for } \xi^2/\lambda\gg \alpha.
\end{cases}\label{mssvalley}
\end{align}
Since $K(\hat{\chi})$ is positive, the square of the isocurvature mass in the valley is always positive -- a consequence of the convex shape of $\hat{W}$ along the $\varphi$ direction inside the valley. The case $\xi\to0$ needs extra care and is treated separately in Sec.~\ref{Sec:smallxi}. Since $K(\hat{\chi})\approx 1$ during inflation, the magnitude of the damping ratio \eqref{mssvalley} grows to large positive values for both the cases in \eqref{mssvalley}. The CMB constraint \eqref{Boundsalphaxi} restricts the magnitude of the isocurvature damping
\begin{align}
\frac{m_{s}^2}{3H^2}\veq
\begin{cases}
 10\lesssim 8\xi\lesssim 10^5&\text{for }\alpha\gg\xi^2/\lambda ,\\
\frac{\sqrt{\lambda}}{\alpha}10^{14}>\sqrt{\lambda}10^{5}\,&\text{for } \xi^2/\lambda\gg \alpha.
\end{cases}\label{mssvalley2}
\end{align}
Equation \eqref{EQPertQs} shows that a positive damping ratio $m_{s}^2/3H^2>0$ leads to an exponential suppression of the isocurvature modes $Q_{s}$.
From equation \eqref{mssvalley2} it can be seen that for $\xi\gtrsim 1$, the isocurvature modes are heavily suppressed. It was argued in the previous sections that for generic initial conditions $\varphi_0$ as well as fixed $\lambda\approx10^{-1}$ and $\xi\gtrsim 1$, the trajectory falls into the valley within $\mathcal{O}(1)$ efolds. We therefore conclude that the problem reduces to an effectively single-field one for $\xi\gtrsim 1$.

\begin{figure}
 	\centering
 	\begin{tabular}{cc}
 		\includegraphics[width=0.45\linewidth]{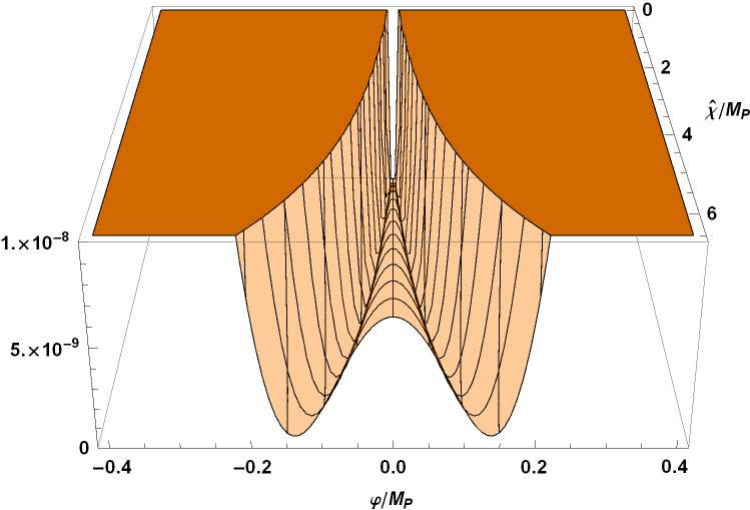}&
 		\includegraphics[width=0.45\linewidth]{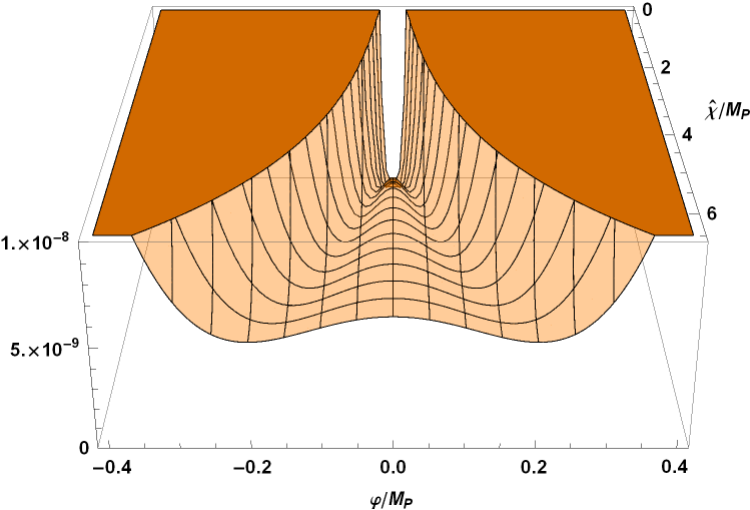}\\
 		\includegraphics[width=0.45\linewidth]{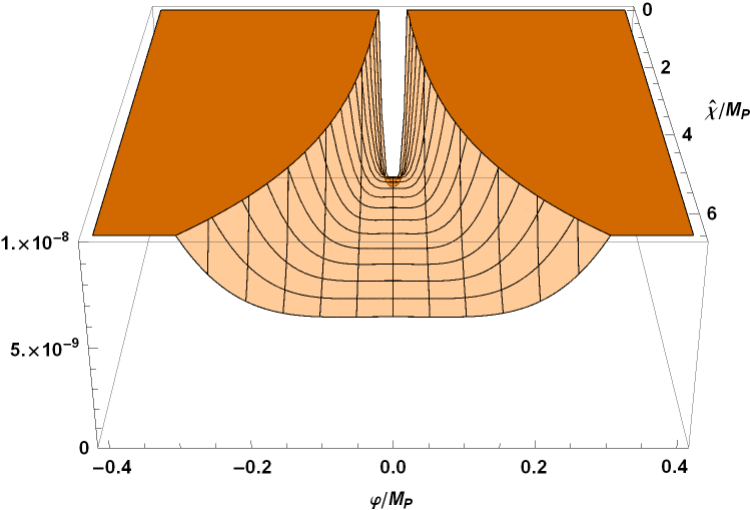}&
 		\includegraphics[width=0.45\linewidth]{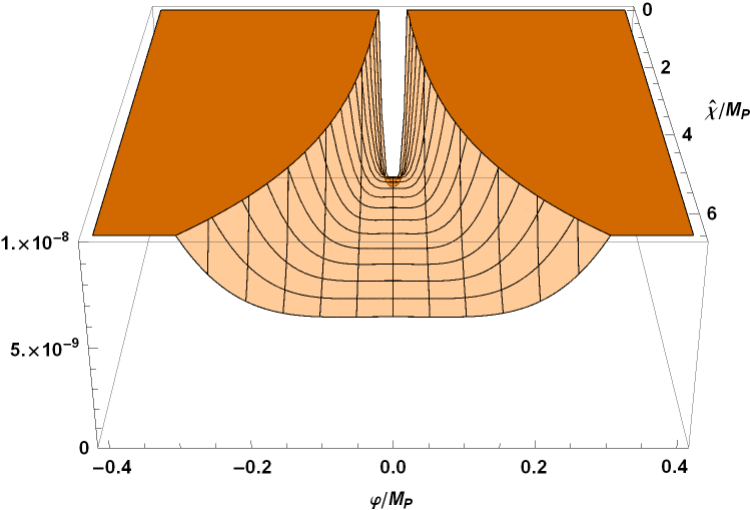}
 	\end{tabular}
 	\caption{Disappearing valleys in $\hat{W}/M_{\mathrm{P}}^4$ for fixed value of $\lambda=10^{-1}$ and $\alpha=10^{7}$ and varying values of $\xi$. From top-left to bottom-right: $\xi=10^{4}$, $\xi=10^{3}$, $\xi=10^{2}$, $\xi=10$. In the small $\xi$ limit the two valleys disappear and form a single broad flat valley. }
 	\label{smallxi}
 \end{figure}

So far, we restricted our analysis to large values of the non-minimal coupling. The conclusion that the scalaron-Higgs model reduces to an effective single-field one for $\xi\gtrsim 1$, relied upon the fact that on the hilltop at $\varphi=0$, the curvature in the $\varphi$ direction is proportional to $\xi$, and hence inflationary trajectories starting on the hill are immediately pushed into one of the valleys, in which inflation takes place predominantly in the $\hat{\chi}$ direction and the isocurvature modes are suppressed. This conclusion is also in agreement with the findings of the recent work \cite{He2018}. However, for $\xi\ll 1$ this argument is no longer valid and a more careful analysis is required -- not explored in \cite{He2018}. Moreover, from \eqref{ValleYFormation} and the CMB constraint \eqref{ConstCMBStar}, it can be seen that the height difference between the hilltop and the valley disappears in the small $\xi$ limit. Likewise, \eqref{ValleyEQ} implies that the separation between the two valleys vanishes in the low $\xi$ limit. Therefore, the two valleys degenerate and form a single broad valley for $\xi\ll1$, as can be seen in Fig.~\ref{smallxi}. In this scenario, inflation essentially takes place inside this central valley along ${\varphi=0}$ line. We perform a detailed analysis of the $\xi\ll1$ regime in Sec.~\ref{Sec:smallxi} and show that for $\lambda\approx 10^{-1}$, even in the extreme limit $\xi\to0$, the scalaron-Higgs model gives the same attractor predictions as non-minimal Higgs inflation or Starobinsky inflation with no significant production of isocurvature modes until the end of inflation. This might not be intuitively expected, since in contrast to the pure single-field Starobinsky model, in the low $\xi$ limit of the scalaron-Higgs model there are two dynamical fields, both of which become equally light in this regime.
\subsection{Small $\xi$ limit}\label{Sec:smallxi}
In this section, we investigate whether or not a small non-minimal coupling constant $\xi$, can lead to a finite isocurvature mode surviving at the end of inflation. To address this question, we analytically study the limit $\xi\to 0$. In this extreme limit, the isocurvature mode is suppressed the least during its evolution on the potential landscape. From \eqref{msvalley} and \eqref{DoubleDerivePotV}, it can be seen that a finite positive $\xi$ leads to a positive $m^2_{\rm{s}}$ and hence enhance the suppression of the isocurvature mode. Thus, if the isocurvature mode is sufficiently suppressed already in the extreme limit $\xi\to0$, it is expected to be even more strongly suppressed for higher values of $\xi$. In addition to the approximate analytic investigation, we strengthen our results by an independent numerical analysis.

\noindent The valley equation \eqref{ValleyEQ} shows that the two valleys degenerate to one broad valley around $\varphi=0$ in the limit \mbox{$\xi\to 0$}. Hence the trajectory inside the valley is almost a straight line along the $\hat{\chi}$ direction. For $\xi\ll1$, the turn rate in the valleys is even smaller and can safely be ignored. This implies that the modes $Q_{s}$ and $Q_{\sigma}$ evolve independently also in the limit $\xi\to0$. As can be seen from \eqref{EQPertQsig} and \eqref{EQPertQs}, the individual growths of the adiabatic and isocurvature modes $Q_{\sigma}$ and $Q_{\rm{s}}$ are determined by $\Omega_{\sigma\sigma}$ and $m_{s}^2$ respectively. 

\begin{figure}[H]
	\centering
	\begin{tabular}{cc}
		\includegraphics[width=0.45\linewidth]{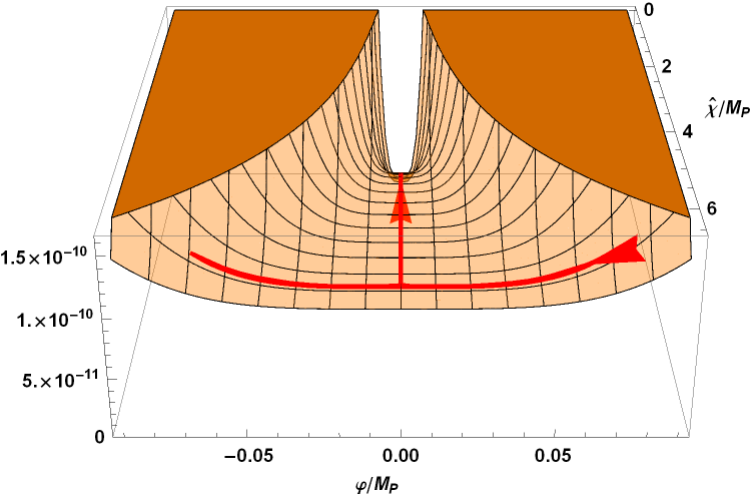}&
		\includegraphics[width=0.45\linewidth]{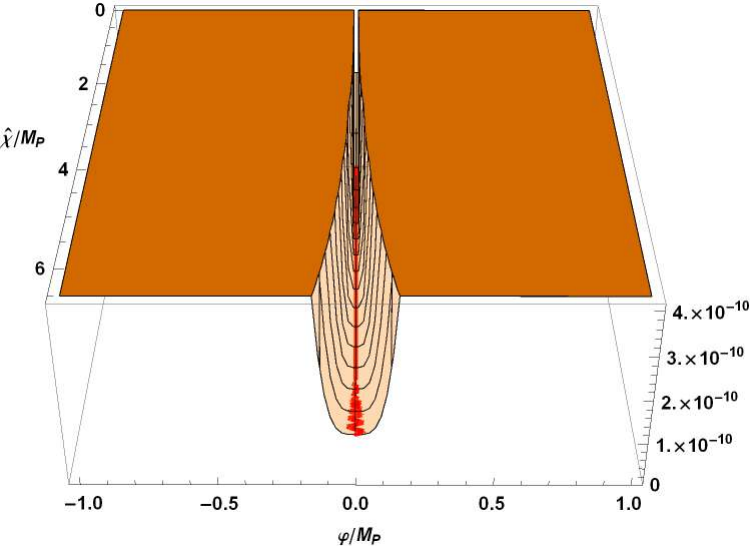}\\
		\includegraphics[width=0.45\linewidth]{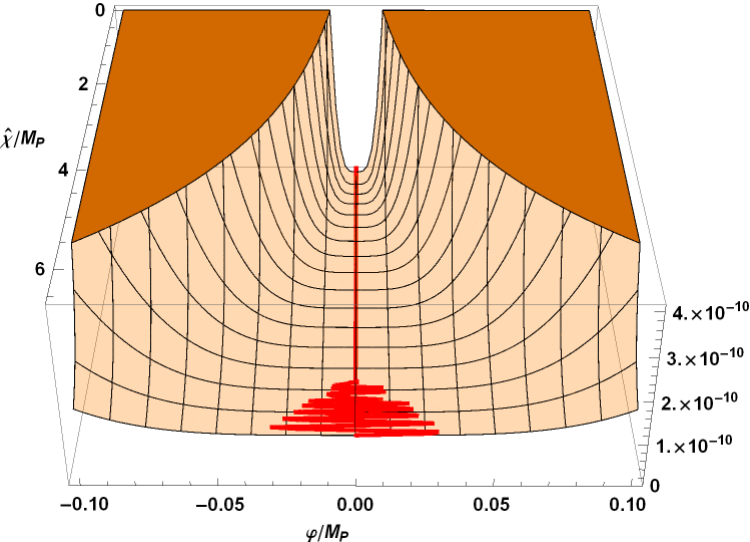}&
		\includegraphics[width=0.45\linewidth]{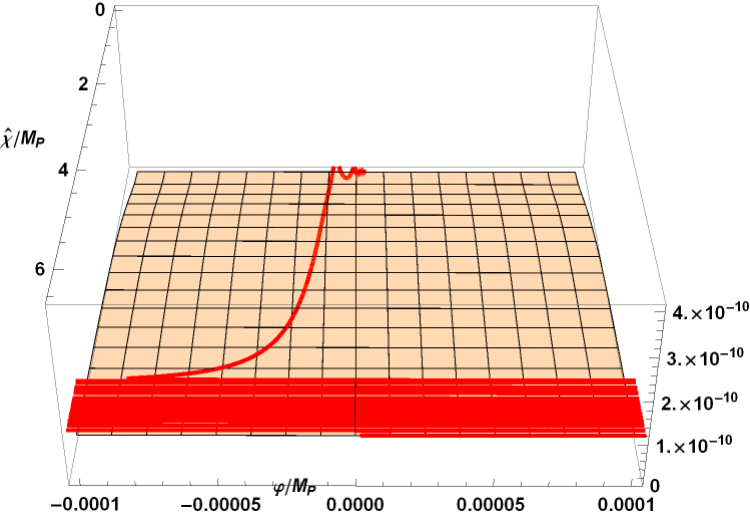}
	\end{tabular}
	\caption{Upper Left: Scalaron-Higgs potential for $\xi = 10^{-3}$, $\lambda = 10^{-1}$, and $\alpha = 6.03\times 10^8$. In the small $\xi$ limit, the valley structure disappears and the two valleys degenerate to one broad valley centered at $ \varphi=0$. The inflationary background trajectory (red line) is superimposed on the potential and shows that for generic initial conditions the trajectory quickly approaches the attractor solution along $\varphi=0$. Upper right to lower right: Scalaron-Higgs potential with identical parameters and different resolutions in $\varphi$-direction, which illustrate that the field excursion in $\varphi$-direction is negligible. }\label{ValleyDissapear}
\end{figure}

Before we compute $\Omega_{\sigma\sigma}$ and $m_{s}^2$, we first notice that in the limit $\xi\to 0$, the approximation $\hat{\sigma}^{\chi}\gg\hat{\sigma}^{\varphi}$ is satisfied with an even higher accuracy. This can be seen from \eqref{KineticCompare}, as for $\lambda=10^{-1}$ and $\zeta$ fixed by the CMB constraint \eqref{CMBAmpl}, the kinetic term for $\varphi$ is even more negligible compared to that of $\hat{\chi}$. From \eqref{msvalley} and \eqref{DoubleDerivePotV}, the expression for the effective isocurvature mass in the $\xi\to0$ limit reduces to
\begin{align}
\frac{m_{s}^2}{3H^2}\veq -\frac{\varepsilon_{\sigma}}{9}-\frac{1}{\sqrt{3}}\sqrt{\varepsilon_{\sigma}}.\label{IsoMassXiZero}
\end{align}
The two terms on the right hand side originate from the curved scalar field space geometry and are expressed in terms of slow-roll parameters. However, in contrast to the estimate \eqref{mssvalley} for $\xi\gtrsim1$, these terms cannot be neglected anymore in the $\xi\to 0$ limit, for which the effective isocurvature mass is dominated by the curvature of the scalar field space and not be the two-field potential. 
Using the Friedmann equations in the slow-roll approximation and the definition \eqref{IsoMassDef}, the valley expression for $\Omega_{\sigma\sigma}$ in the $\xi\to0$ limit reads
\begin{align}
\frac{\Omega_{\sigma\sigma}}{3H^2} = \eta_{\sigma} - \left(2\varepsilon_{\sigma}+\frac{2}{3}\varepsilon_{\sigma}^2-\frac{4}{3}\epsilon_{\sigma}\eta_{\sigma}\right).
\end{align}
Retaining only terms up to first order in the slow-roll, we obtain
\begin{align}
\frac{\Omega_{\sigma\sigma}}{3H^2} \approx \eta_{\sigma} - 2\varepsilon_{\sigma}.\label{OmegaFirstOrder}
\end{align}
With inflation predominantly along the $\hat{\chi}$ direction, the slow-roll parameters reduce to
\begin{align}
\eta_{\sigma}\approx M^2_{\rm{P}}\frac{{\hat{V}}_{,\hat{\chi}\hat{\chi}}}{\hat{V}},\qquad\varepsilon_{\sigma}\approx \frac{M^2_{\rm{P}}}{2}\left(\frac{{\hat{V}}_{,\hat{\chi}}}{\hat{V}}\right)^2.\label{OmegaSlowRoll}
\end{align}
The results for $m_{s}^2$ and $\Omega_{\sigma\sigma}$ in \eqref{IsoMassXiZero} and \eqref{OmegaFirstOrder} were obtained in the valley approximation $\varphi=\varphi_{\mathrm{v}}$ with $\varphi_{\mathrm{v}}$ given by \eqref{ValleyEQ}. For $\xi\ll 1$, the inflationary trajectory is inevitably driven towards the single attractor solution in the broad valley along $\varphi = 0$. This is in contrast to the case with $\xi\gg 1$, where a central hill located at $\varphi = 0$ pushed the trajectory away from $\varphi = 0$.  The scenario with $\xi\ll1$ is illustrated in Fig.~\ref{ValleyDissapear}.

To demonstrate that the two valleys converge and form a single attractor along the $\varphi = 0$ line, we show that the expressions for \eqref{IsoMassXiZero} and \eqref{OmegaFirstOrder}, which were obtained using the valley equation \eqref{ValleyEQ}, become identical if evaluated along the $\varphi = 0$ line for $\xi\ll1$.
\noindent As can be seen from \eqref{Pot2F}, the effective single-field potential for $\varphi=0$  on the hilltop has the same functional dependence as the effective single-field potential \eqref{potValley} inside the valley
\begin{align}
\hat{V}(\hat{\chi})\overset{\mathrm{h}}{\approx} \hat{W}(\hat{\chi},0)=\frac{M_{\mathrm{P}}^2}{16\alpha}\left(1-e^{\gamma\frac{\hat{\chi}}{M_{\mathrm{P}}}}\right)^2\label{HilltopPot}.
\end{align}
Therefore, the ratios $\hat{V}_{,\hat{\chi}}/\hat{V}$ and $\hat{V}_{,\hat{\chi}\hat{\chi}}/\hat{V}$ inside the valley and on the hilltop are identical.
Consequently, the slow-roll parameters \eqref{OmegaSlowRoll} and the expression for $\Omega_{\sigma\sigma}$ on the hilltop and inside the valley are identical in the $\xi\to0$ limit
\begin{align}
\left.\epsilon_{\sigma}\right|_{h}\overset{\xi\to 0}{=}&\left.\epsilon_{\sigma}\right|_{v},\\ \left.\eta_{\sigma}\right|_{h}\overset{\xi\to 0}{=}&\left.\eta_{\sigma}\right|_{v},\\
\left.\frac{\Omega_{\sigma\sigma}}{3H^2}\right|_{h} \overset{\xi\to 0}{=}&\left.\frac{\Omega_{\sigma\sigma}}{3H^2}\right|_{v}\approx \eta_{\sigma} - 2\varepsilon_{\sigma}. 
\end{align}
For the isocurvature mass, the contribution from the second derivative of the potential are different at the hilltop and inside the valley, 
\begin{align}
\frac{\hat{W},_{\varphi\varphi}}{3H^2}\overset{\mathrm{h}}{\approx}  -4\xi K(\hat{\chi}),\qquad\frac{\hat{W},_{\varphi\varphi}}{3H^2}\overset{\mathrm{v}}{\approx}  8\xi K(\hat{\chi}).\label{DoubleDerivativeIsoDiffer}
\end{align}
However, this difference disappears in the limit $\xi\to 0$ and from \eqref{msvalley}, we obtain
\begin{align}
\left.\frac{m^2_{\mathrm{s}}}{3H^2}\right|_{h} \overset{\xi\to 0}{=}\left.\frac{m^2_{\mathrm{s}}}{3H^2}\right|_{v} =  -\frac{\varepsilon_{\sigma}}{9}-\frac{1}{\sqrt{3}}\sqrt{\varepsilon_{\sigma}}.\label{IsomassHill}
\end{align}
Therefore, in the limit $\xi\to 0$, both the $\hat{\chi}$ and the $\varphi$ directions of the potential are almost flat and the modes $Q_{\sigma}$ and $Q_{\rm{s}}$ are light. Together with the vanishing turn rate, this implies that the slow-roll approximation is satisfied and the mode equations \eqref{EQPertQsig} and \eqref{EQPertQs} reduce to the first order equations
\begin{align}
\frac{d Q_{\sigma}}{dN} \approx \frac{\Omega_{\sigma\sigma}}{3H^2}Q_{\sigma}, \qquad\frac{d Q_{s}}{dN} \approx \frac{m^2_s}{3H^2}Q_{s}.\label{redslowrolleq}
\end{align}
Here we have expressed the differentials in terms of $\mathrm{d}N=-H\mathrm{d}t$, which means that the modes are evolved backward in time for increasing $N$. For the inflationary dynamics in the effective single-field potential \eqref{HilltopPot} along the $\varphi=0$ trajectory, we have the same inflationary predictions for $\varepsilon_{\sigma}$ and $\eta_{\sigma}$ as for the effective single-field inflation inside the valley \eqref{slowrollStar},
\begin{align}
\varepsilon_{\sigma}(N)\approx\frac{3}{4}\frac{1}{N^2},\qquad \eta_{\sigma}(N)\approx -\frac{1}{N}.
\end{align}
Thus, in view of \eqref{IsomassHill} and \eqref{OmegaFirstOrder}, we obtain to first order in slow-roll
\begin{equation}
\begin{aligned}
\frac{\Omega_{\sigma\sigma}}{3H^2} \approx{}&\eta_{\sigma}-2\varepsilon_{\sigma}\approx-\frac{1}{N}-\frac{3}{2N^2},\\ \frac{m_{s}^2}{3H^2}\approx{}&\frac{\eta_{\sigma}}{2}-\frac{\varepsilon_{\sigma}}{9}\approx-\frac{1}{2N}-\frac{1}{12N^2}.\label{mssOmsigmaslowroll}
\end{aligned}
\end{equation}
Since $\Omega_{\sigma\sigma}/3H^2$ and $m_{s}^2/3H^2$ are both negative, \eqref{redslowrolleq} implies that this leads to an exponential growth of $Q_{\sigma}$ and $Q_{s}$.
However, the slow-roll analysis \eqref{mssOmsigmaslowroll} shows that for all values of $N$, the growth rate of the isocurvature mode is always smaller than that of the adiabatic mode.
\begin{figure}
	\centering
\begin{tabular}{cc}
	\includegraphics[width=0.455\linewidth]{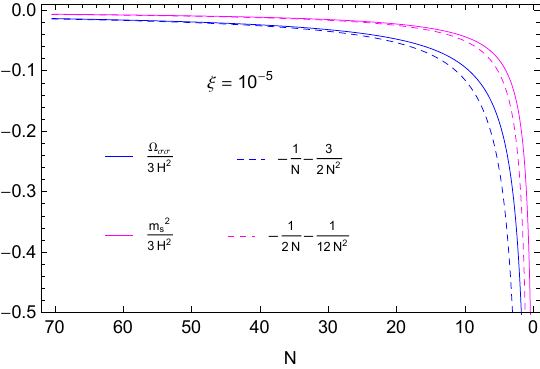} &
	\includegraphics[width=0.455\linewidth]{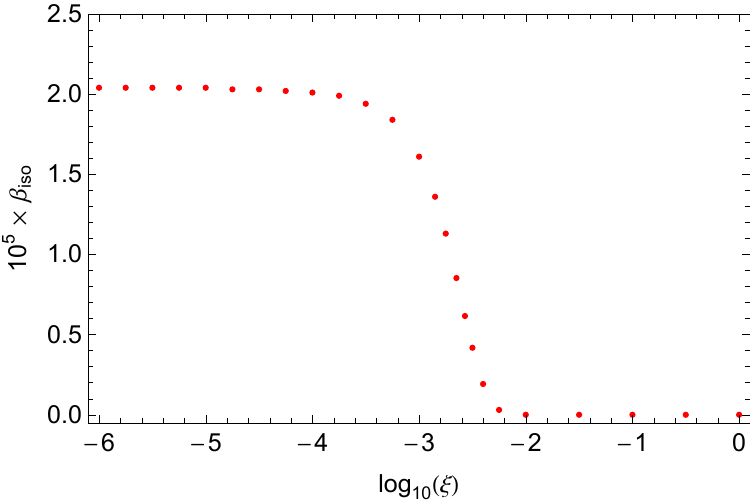}
\end{tabular}	
	\caption{Left: Exact numerical evaluation of $\Omega_{\sigma\sigma}/3H^2$ (solid blue line) and $m_{s}^2/3H^2$ (solid pink line) and the corresponding analytic expressions  \eqref{mssOmsigmaslowroll} to first order in slow-roll (dashed lines) for the inflationary trajectory, which starts inside the valley at $\chi_{0}/M_{P} = 5.7$, $\varphi_{0}/M_{\mathrm{P}} = \varphi_{\mathrm{v}}/M_{\mathrm{P}}$. The parameters are $\lambda=10^{-1}$, $\alpha=6.03\times 10^{8}$, and $\xi=10^{-5}$. 
	Right: Numerically generated values of $\beta_{\mathrm{iso}}$ for different values of $\xi$ with initial conditions $\hat{\chi}_{0}/M_{\rm{P}} = 5.7$ and $\varphi_{0}/M_{\rm{P}} = 10^{-5}$, as well as parameters $\lambda=10^{-1}$ and $\alpha = 6.03\times 10^{8}$. Here, the initial value $\varphi_0$ was chosen to be small (which leads to maximum growth of the isocurvature mode) in order to demonstrate that even for this extreme case there are no sizable isocurvature effects.}
	\label{fiso}
\end{figure}

\begin{figure}
	\centering
	\begin{tabular}{cc}
		\includegraphics[width=0.46\linewidth]{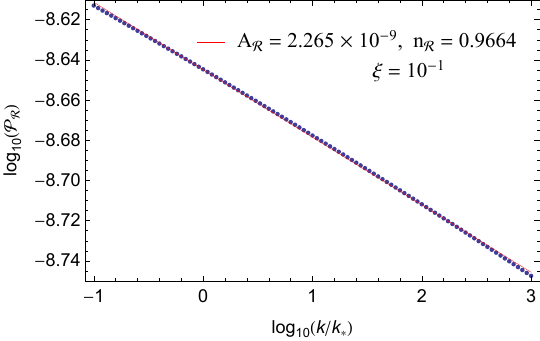}&
		\includegraphics[width=0.46\linewidth]{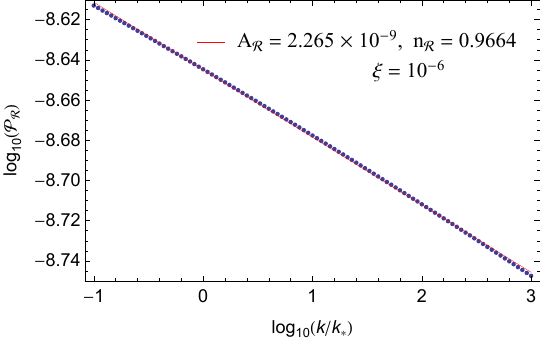}
	\end{tabular}
	\caption{The logarithmic power spectra $\log_{10}{\left(\mathcal{P_{\mathcal{R}}}\right)}$ evaluated at the end of inflation as a function of $\log_{10}\left({k/k_{*}}\right)$ are obtained numerically (blue dots) for $\varphi_0 = 10^{-5}$, $\hat{\chi}_0 = 5.7$, $\alpha = 6.03\times 10^{8}$, $\lambda = 10^{-1}$ and $\xi=10^{-1}$ (left) as well as $\xi=10^{-6}$ (right). The amplitude and the spectral index are extracted from the linear fit (red line).}
	\label{PowerSpectraLowXi}
\end{figure}

\noindent This is illustrated in Fig.~\ref{fiso}, where the exact numerical solutions of the growth rates are compared to the analytic slow-roll estimate \eqref{mssOmsigmaslowroll}. Moreover, the exact numerical evolution of the inflationary isocurvature fraction $\beta_{\mathrm{iso}}$ yields\footnote{Note that $\beta_{\mathrm{iso}}$, is defined in \eqref{biso} as the ratio of the adiabatic and isocurvature power spectra at the end of inflation and we make no statements about post-inflationary isocurvature effects. A careful analysis of these effects is certainly interesting but goes beyond the scope of this paper.}
\begin{align}
\beta_{\mathrm{iso}}\approx 10^{-5}.
\end{align}  
Summarizing we find that even in the extreme limit of $\xi\to 0$, the inflationary isocurvature fraction is found to be negligible. Although both the fields are light in this regime, the growth rate of the adiabatic mode $Q_{\sigma}$ is always greater than that of the isocurvature mode $Q_{\mathrm{s}}$. In addition, as a consequence of the straight line trajectory along $\varphi=0$, the turn rate is negligible and the curvature power spectrum is unaffected. This is illustrated in Fig.~\ref{PowerSpectraLowXi} by the exact numerical evolution of the curvature power spectrum \eqref{PowerSpectrak} for two different values of $\xi$.
 
We therefore conclude that for fixed $\lambda=10^{-1}$, an exhaustive range of the initial conditions $\left(\hat{\chi}_{0},\,\varphi_{0}\right)$ and free parameters $\left(\xi,\alpha\right)$ lead to a curvature power spectrum indistinguishable from that of Higgs inflation or Starobinsky inflation with no observable multifield effects. In particular, even for small values of $\xi\ll1$ the inflationary isocurvature fraction at the end of inflation is negligibly small and the inflationary power spectra are identical to those in non-minimal Higgs inflation and Starobinsky inflation. 

Therefore, the analysis of the $\xi\ll1$ case (with $\lambda$ fixed at $10^{-1}$) provides an important extension of the results obtained in \cite{He2018} for $\xi\gg1$, as it relaxes the situation with the strong non-minimal coupling present in Higgs inflation. Moreover, this is in contradiction to the statement made in \cite{Wang2017} that the observed primordial perturbations cannot be produced for $\xi\varphi^2\ll M^2_{\mathrm{P}}$ and $\xi\lesssim1$.
	
%
%--------------------------------------------------------------------
%
\section{Multifield effects with a tiny quartic self-coupling $\lambda$}\label{Sec:Isocurvature}
%
%--------------------------------------------------------------------
%
So far, we have investigated the scalaron-Higgs model for fixed $\lambda\approx10^{-1}$. In this section we explore several aspects of this model for a very small quartic Higgs self-coupling constant $\lambda\ll10^{-1}$.
Small values of the quartic Higgs coupling are expected at high energies due to the Standard Model RG flow.
Like in the model of non-minimal Higgs inflation, where the RG improvement turned out to be crucial for the connection between particle physics and inflationary cosmology \cite{Bezrukov2009a,DeSimone2009,Barvinsky2009,Bezrukov2009,Bezrukov2011,Barvinsky2012}, the Standard Model RG running drives the Higgs self-coupling $\lambda$ to very small values at high energies.
\begin{figure}
	\centering
	\begin{tabular}{cc}
		\includegraphics[width=0.42\linewidth]{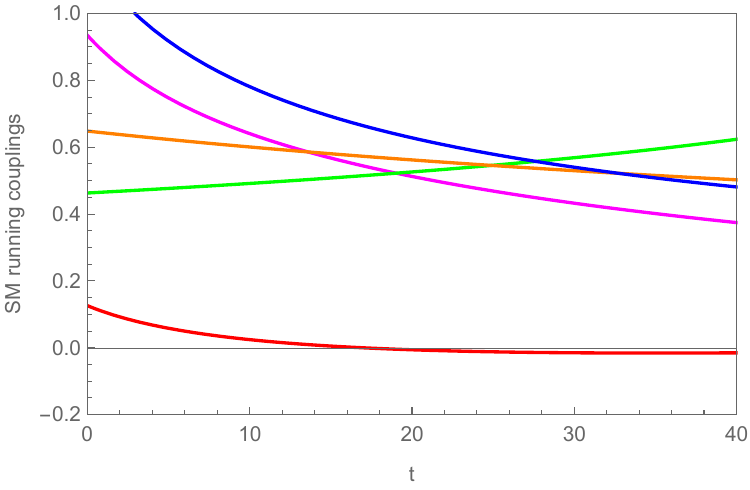}
		&$\qquad$
		\includegraphics[width=0.42\linewidth]{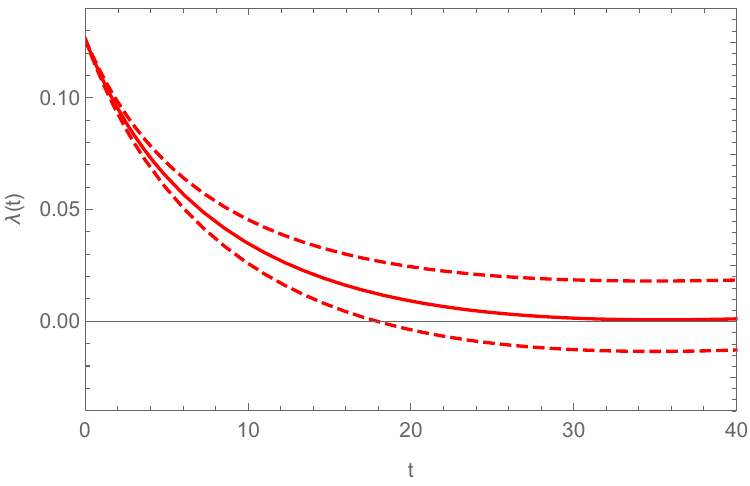}	
	\end{tabular}
	\caption{Left: The pure Standard Model running of the quartic Higgs coupling $\lambda$ (red line), the Yukawa top quark coupling $y_{t}$ (Magenta line) and the electroweak and strong gauge couplings $g_1$ (green line), $g_2$ (orange line) and $g_3$ (blue line) for a Higgs mass $M_{\mathrm{H}}=125.1\;\mathrm{GeV}$ and a top-quark mass $M_{\mathrm{t}}=172.9\;\mathrm{GeV}$. Right: The pure Standard Model running of $\lambda$ for fixed value of the Higgs mass $M_{\mathrm{H}}=125.1$ GeV and different values of the top mass $M_{t}=168$ GeV (upper red dashed line), $M_{t}=170.8$ GeV (central red solid line) and $M_{t}=173$ GeV (lower red dashed line). The two-loop SM beta functions were extracted from Appendix B of \cite{Buttazzo2013}. The logarithmic RG scale is defined as  $t=\ln(E/\mu_0)$ with the arbitrary renormalization point $\mu_0$ identified with the top-quark mass $\mu_0=M_{\mathrm{t}}$. }
	\label{SMrunning}
\end{figure}
\noindent Moreover, as can be seen from the right plot in Fig \ref{SMrunning}, the Higgs self-coupling $\lambda$ might even be driven to negative values, rendering the electroweak vacuum unstable. Since the RG flow is extremely sensitive to the conditions at the electroweak scale, the question about the stability of the vacuum strongly depends on the precise values of the Higgs mass $M_{\mathrm{H}}$ and the top mass $M_{t}$. Current measurements of $M_{\mathrm{H}}$ and $M_{t}$ seem to indicate that the electroweak vacuum is just at the borderline of being stable \cite{Degrassi2012}.

Before we investigate constraints from the RG flow of the pure SM and questions regarding the stability of the effective potential in scalaron-Higgs inflation, we first investigate the inflationary consequences of a very small $\lambda\ll10^{-1}$ for the two-field model \eqref{ActScal} with \eqref{fHiggsStarobinsky} in which $\varphi$ is considered as abstract scalar field, a priori \textit{unrelated} to the SM Higgs field. At the end of this section we analyze for which values of $\lambda\ll10^{-1}$ the abstract scalar field $\varphi$ might be identified with the SM Higgs and derive   constraints from extrapolating the RG flow of the pure SM to the model of scalaron-Higgs inflation. 

We consider two different scenarios for tiny values of the self-coupling $\lambda\ll10^{-1}$, which mainly differ by their initial conditions. Within the classification scheme of Sec.~\ref{ClassificationTrajectories} these two scenarios correspond to (Class 1) where the inflationary trajectory starts out in one of the two valleys and to (Class 4), where the inflationary trajectory starts on the hilltop and subsequently falls into one of the valleys.
\subsection{Effective single-field model with large tensor-to-scalar ratio}
\label{Sec:EffectiveSingleFieldu}
\begin{figure}
	\centering
	\begin{tabular}{cc}
		\includegraphics[width = 0.45\linewidth]{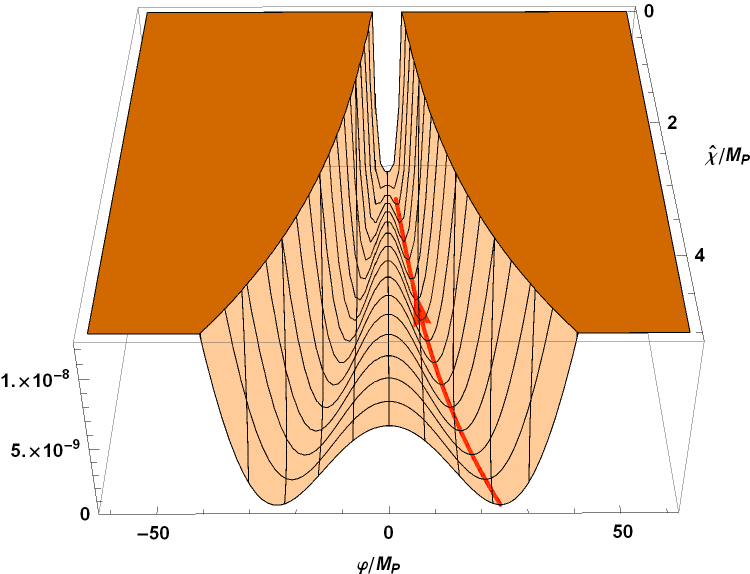}&
		\includegraphics[width = 0.45\linewidth]{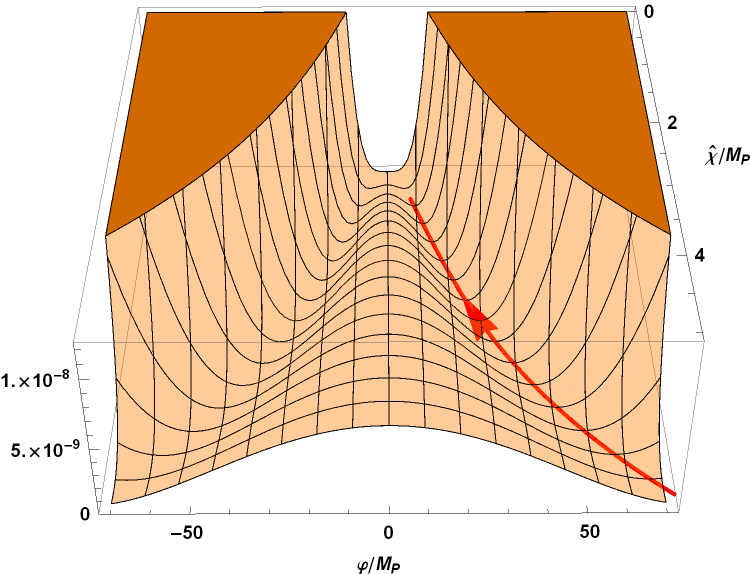}\\
		\includegraphics[width = 0.45\linewidth]{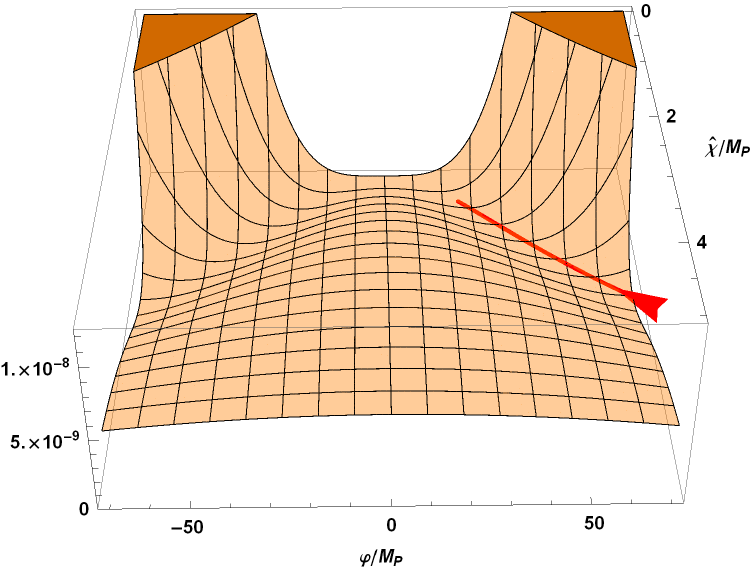}&
		\includegraphics[width = 0.45\linewidth]{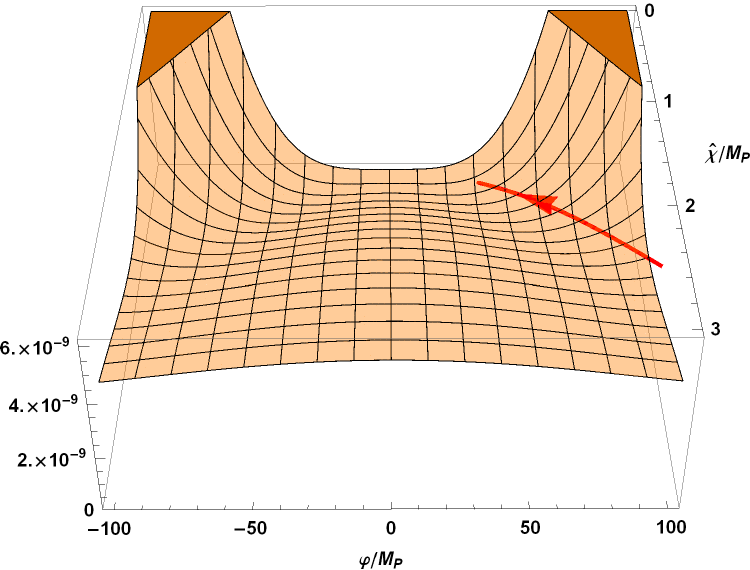}
	\end{tabular}
	\caption{In the low $\lambda$ limit, the valleys in $\hat{W}/M_{\mathrm{P}}^4$ are stretched out in the $\varphi$ direction and the angle enclosed by the two valleys increases. The plots are for fixed $\lambda=10^{-15}$ and fixed $\alpha=10^7$ with varying $\xi$. From top-left to bottom-right: $\xi=10^{-1}$, $\xi=10^{-2}$, $\xi=10^{-3}$, $\xi=10^{-4}$. Inflation inside the valley then predominately takes place in the $\varphi$ direction.}
	\label{smalllambda}
\end{figure}

Since for fixed $\zeta$, the separation between the two valleys is controlled by the ratio $\xi/\lambda$, the main effect of a tiny self-coupling $\lambda\ll 10^{-1}$ is to stretch out the valleys as well as the hilltop plateau of the potential \eqref{Pot2F} in $\varphi$ direction. This effect is illustrated in Fig.~\ref{smalllambda}.

\noindent
For the initial condition $\varphi_0=\varphi_{\mathrm{v}}$ chosen such that the dynamics start inside one of the valleys, the inflationary trajectory tracks the valley up to the global minimum at $(\hat{\chi},\varphi)=(0,0)$.
In contrast to the case for $\lambda = 10^{-1}$, where inflation takes place inside the valley (Class 1) predominantly along the $\hat{\chi}$ direction,  for $\lambda\ll 10^{-1}$ inflation inside the valley takes place predominantly in the $\varphi$ direction. This effect is visible in Fig.~\ref{smalllambda}, which clearly shows that a smaller value of $\lambda$ allows for a much wider separation of the valleys and an increased angle between them. Consequently, for $\lambda\ll 10^{-1}$ the inflaton velocity inside the valley is dominated by the velocity of the $\varphi$ field rather than $\hat{\chi}$.
Thus, compared to the valley analysis for $\lambda=10^{-1}$ in Sec.~\ref{Sec:ReductionSingleField}, the roles of $\hat{\chi}$ and $\varphi$ are interchanged for $\lambda\ll 10^{-1}$. Therefore, the valley equation \eqref{ValleyEQ} should be used to eliminate $\hat{\chi}$ in favor of $\varphi$ in order to arrive at an effective single-field description inside the valley for $\lambda\ll10^{-1}$.
The valley equation reads
\begin{align}
e^{\gamma\frac{\hat{\chi}}{M_{\mathrm{p}}}} \overset{\mathrm{v}}{=} 1+\left(\frac{\varphi}{M}\right)^2,\qquad M^2:=\frac{M_{\mathrm{P}}^2}{\beta^2},\label{valleyequ}
\end{align}
where we have defined the new mass scale $M$. However, even if \eqref{valleyequ} is used to eliminate $\hat{\chi}$ in favor of $\varphi$, we need to perform an additional field redefinition $\varphi\to u$ to the canonically normalized scalar field $u$. The transformation rules are found from the condition
\begin{align}
\dot{u}^2:={}&\left. G_{IJ}\dot{\Phi}^{I}\dot{\Phi}^{J}\right|_{\mathrm{v}}=\dot{\hat{\chi}}^2+e^{\gamma\frac{\hat{\chi}}{M_{\mathrm{P}}}}\dot{\varphi}^2\nonumber\\
={}&\left(6\beta^2\frac{\varphi^2/M^2}{1+\varphi^2/M^2}+1\right)\frac{\dot{\varphi}^2}{1+\varphi^2/M^2}\nonumber\\\veq{}&\frac{\dot{\varphi}^2}{1+\varphi^2/M^2},\label{kineticcompareu}
\end{align}
where we use the notation $\veq$ for an equality which holds in the valley for $\beta^2\ll1$.
This leads to the simple differential relation between $u$ and $\varphi$,
\begin{align}
\mathrm{d}u \veq \frac{\mathrm{d}\varphi}{\sqrt{1+(\varphi/M)^2}}.
\end{align}
Integration yields the desired transformation law to the canonical variable $u$,
\begin{align}
u \veq M\,\text{arcsinh}\left(\varphi/M\right),\qquad \varphi \veq M\sinh\left(u/M\right).\label{uphi}
\end{align}
Consequently, inflation is driven by an effective single-field potential\footnote{The new mass scale $M$ should not be confused with the effective mass $M_{\mathrm{eff}}^2:=\hat{V}_{,uu}$ of the propagating scalar degree of freedom of the effective single field reduction in the $\beta^2\ll1$ case. In particular, during slow-roll inflation, the scalar degree of freedom associated with $u$ is light as $M^2_{\mathrm{P}}M_{\mathrm{eff}}^2/\hat{V}=\eta_\sigma\ll1$ and $\hat{V}<M^4_{\mathrm{P}}$ which implies $M_{\mathrm{eff}}^2/M_{\mathrm{P}}^2\ll1$.}
\begin{align}
\hat{V}(u):=\hat{W}\left(\hat{\chi}(\varphi(u)),\varphi(u)\right)\veq\frac{M_{\mathrm{P}}^4}{4}\zeta\tanh^4(u/M).\label{Vsingleu}
\end{align}
The inflationary slow-roll dynamics in the valley for ${\beta^2\ll 1}$ can be studied by applying the formalism of Sec.~\ref{SubSec:Predictions} to the potential \eqref{Vsingleu}.
The slow-roll parameters are 
\begin{align}
\varepsilon_{\sigma}(u)={}&\frac{32\beta^2}{\sinh^2\left(\frac{2u}{M}\right)},\label{epsu}\\ \eta_{\sigma}(u)={}&-\frac{1}{2}\varepsilon_{\sigma}\left[\cosh\left(\frac{2u}{M}\right)-4\right].\label{slowrollparau}
\end{align}
At the end of inflation $u_{\mathrm{end}}$, we find from $\varepsilon_{\sigma}(u)=1$ the relation
\begin{align}
\sinh^2\left(\frac{2u_{\mathrm{end}}}{M}\right)=32\beta^2.\label{sinhend}
\end{align}
The number of inflationary efolds with $N_{\mathrm{end}}=0$ is defined as
\begin{align}
N_{*}={}&\frac{1}{M_{\mathrm{P}}^2}\int_{u_{\mathrm{end}}}^{u_{*}}\mathrm{d}u\frac{\hat{V}(u)}{\hat{V}_{,u}(u)}=\left[\frac{1}{16\beta^2}\cosh\left(\frac{2u}{M}\right)\right]_{u_{\mathrm{end}}}^{u_{*}}.\label{efoldsu}
\end{align}
From \eqref{sinhend}, we obtain for ${\beta^2\ll1}$ with ${\cosh^2(x)-\sinh^2(x)=1}$,
\begin{align}
\cosh\left(\frac{2u_{\mathrm{end}}}{M}\right)=\sqrt{1+32\beta^2}\approx1+16\beta^2.\label{coshuend}
\end{align}
Inserting this into \eqref{efoldsu}, we obtain \eqref{coshuend} for $u_{\mathrm{in}}$ in terms of $N_{*}$,
\begin{align}
\cosh\left(\frac{2u_{*}}{M}\right)\approx 1+16\beta^2(N_{*}+1)\approx1+16\beta^2 N_{*}.
\end{align}
Inserting this into the slow-roll parameters \eqref{epsu} and \eqref{slowrollparau}, we find
\begin{align}
\varepsilon_{\sigma}(u_{*})={}&\frac{1}{N_{*}+8\beta^2N_{*}^2},\\ \eta_{\sigma}(u_{*})={}&\frac{3-16\beta^2 N_{*}}{2N_{*}+16\beta^2 N_{*}^2}.
\end{align}
In this way, we obtain the inflationary observables at horizon crossing
\begin{align}
A_{\mathcal{R}}^{*}={}&\frac{2\lambda}{3\pi^2\xi}\frac{\beta^2 N_{*}^3}{\left(1+8\beta^2 N_{*}\right)},\label{Astaru}\\
n_{\mathcal{R}}^{*}={}&1-\frac{2}{N_{*}}-\frac{1}{N_{*}+8\beta^2 N_{*}^2}, \label{nsstaru}\\
r^{*}={}&\frac{16}{N_{*}+8\beta^2N_{*}^2}.\label{rstaru}
\end{align}
Evaluating these expressions at $N_{*}=60$ we can derive bounds on the parameters $\lambda$ and $\xi$.
First we combine \eqref{nsstaru} with \eqref{PlanckAns} and solve $n_{\mathcal{R}}^{*}=0.9649$ for $\beta^2$, leading to
\begin{align}
\beta^2\approx 1.76\times 10^{-2}.\label{constraintbeta2}
\end{align}
Next, we combine \eqref{Astaru} with \eqref{PlanckAnsOne} and solve $\ln(10^{10}A_{\mathcal{R}}^{*})=3$ for $\lambda$, yielding
\begin{align}
\lambda\approx 7.72\times10^{-11}\xi.\label{constraintLambda}
\end{align}
Finally, inserting \eqref{constraintLambda} into \eqref{constraintbeta2}, we find the relation
\begin{align}
\xi=1.76\times 10^{-2}-3\times10^{-10}\alpha.\label{constraintxi}
\end{align}
For $\alpha\lesssim10^{7}$, the negative $\alpha$ contribution to \eqref{constraintxi} is negligible and we obtain
\begin{align}
\xi\approx\beta^2\approx 1.76\times 10^{-2},\qquad\lambda\approx 1.36\times10^{-12}.\label{ParameterChoice}
\end{align}
By construction, for fixed parameters \eqref{ParameterChoice}, the observables  $A_{\mathcal{R}}^{*}$ and $n_{\mathcal{R}}^{*}$ are in perfect agreement with \textit{Planck} data \eqref{PlanckAnsOne} and \eqref{PlanckAns}. A non-trivial consistency check is provided by the tensor-to-scalar ratio, which for \eqref{ParameterChoice} is given by
\begin{align}
r_{*}=0.0283,\label{ttslowlambda}
\end{align}
and which is still well below the current upper bound \eqref{Planckr}. Moreover, the tensor-to-scalar ratio in this model is significantly higher than the universal attractor value \eqref{PredStar} for the Higgs-inflation or Starobinsky model.\footnote{In \cite{Wang2017} also a larger tensor-to-scalar ratio was found, however by a very different approach than ours, which make a direct comparison of the results for the inflationary observables difficult. In particular, the relations \eqref{Vsingleu}, and \eqref{Astaru}-\eqref{rstaru} were not obtained in \cite{Wang2017}.}

\begin{figure}
	\centering
	\begin{tabular}{cc}
		\includegraphics[width=0.45\linewidth]{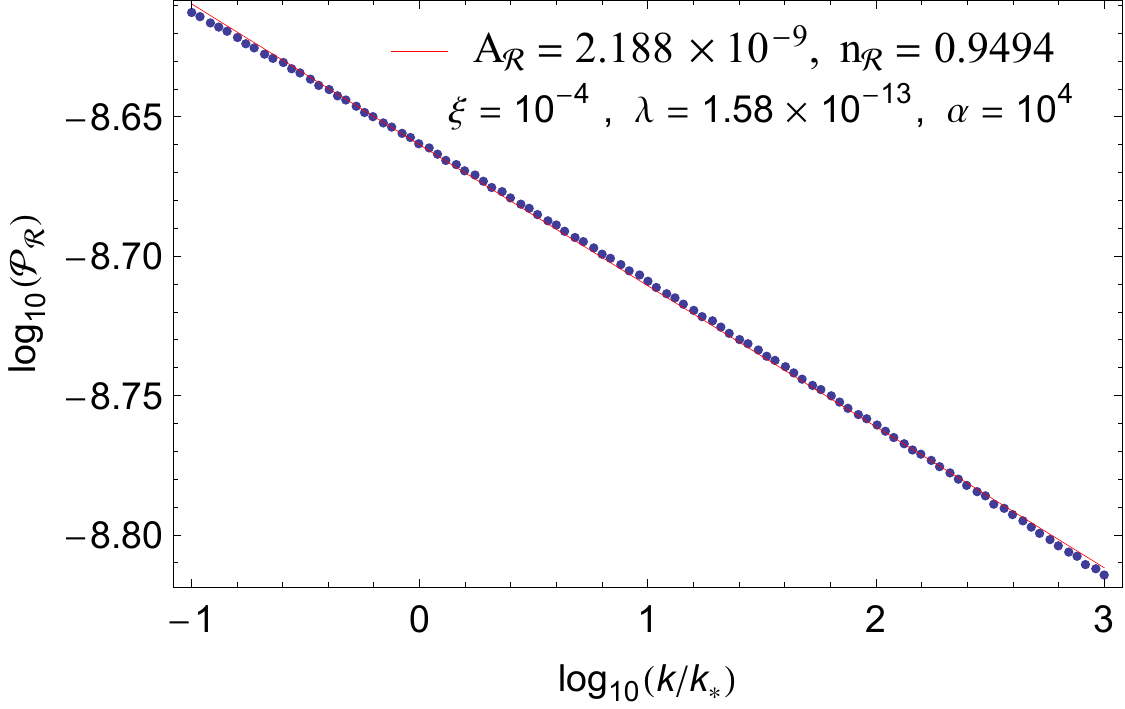}&
		\includegraphics[width=0.45\linewidth]{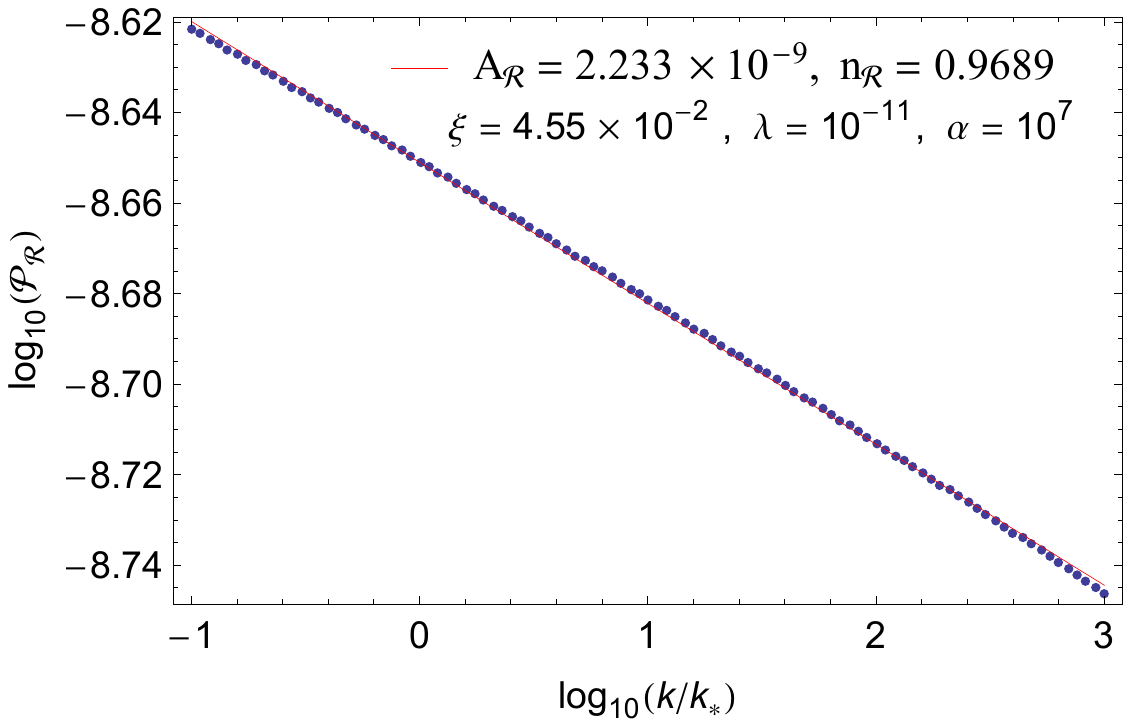}\\
		\includegraphics[width=0.45\linewidth]{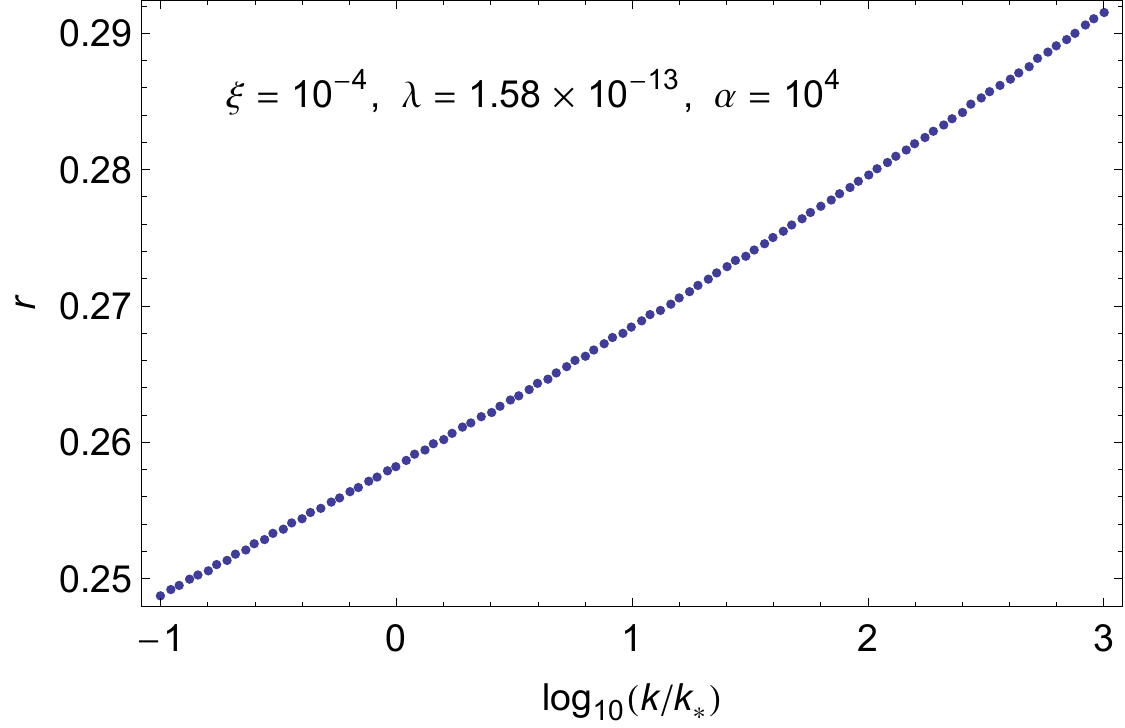}&
		\includegraphics[width=0.45\linewidth]{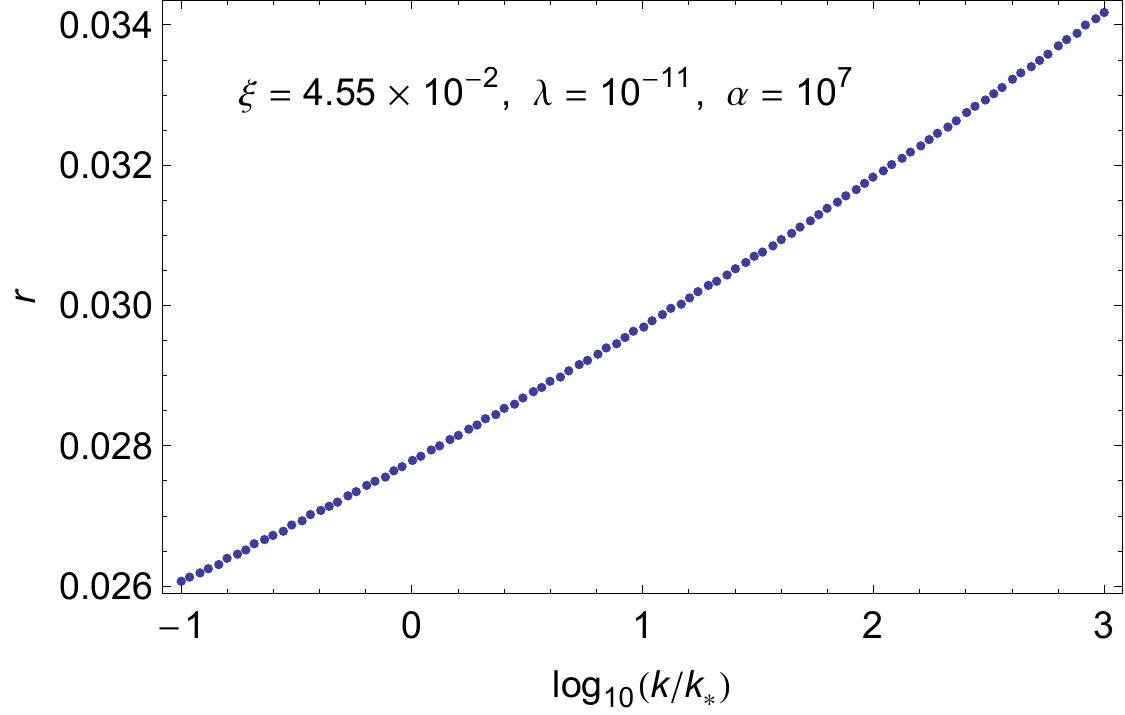}
	\end{tabular}
	\caption{The first row shows the numerical results (blue dots) of the power spectra for different choices of the parameters with $\beta^2 = 1.63\times 10^{-4}$ (left) and $\beta^2 = 5.43\times 10^{-2}$ (right). The values of $A_{\mathcal{R}}$ and $n_{\mathcal{R}}$ are obtained by a linear fit (red line). The lower row shows the numerically obtained tensor-to-scalar ratio (blue dots) for different values of parameters. The parameters in the left column correspond to the $\beta^2\to0$ limit and are in good agreement with the analytic estimate of the effective single-field scenario. The parameters in the right column corresponds to  $\beta^2=5.43\times 10^{-2}$ which is of the same order of magnitude as the value for $\beta^2$ in \eqref{constraintbeta2}, for which the effective single-field description is not a good approximation. Nevertheless, the results for the spectral observables are in good agreement with the cosmological data.}
\label{TensorToScalarRatio}
\end{figure}

Although the value for the tensor-to-scalar ratio obtained in \eqref{ttslowlambda} is consistent with the observational bound, the calculation was performed under the assumption that the approximation $6\beta^2\ll 1$ is satisfied. From the value of $\beta^2$ obtained in $\eqref{constraintbeta2}$, we find $6\beta^2 = 0.103$. Although the $\varphi$ kinetic term is still dominant, the precise numerical value of the tensor-to-scalar ratio obtained in \eqref{ttslowlambda} for the parameters \eqref{ParameterChoice} does not fit the exact numerical result. Nevertheless, the analytic expressions obtained by the single-field approximation provide a useful tool to navigate through the parameter space.
This is illustrated in the right column of Fig.~\ref{TensorToScalarRatio}. The numerical results for the amplitude of the curvature power spectrum, the spectral index of the curvature perturbation and the tensor-to-scalar ratio are found within experimental bounds for a choice of parameters which are of the same order of magnitude as those obtained in the analytic estimate \eqref{ParameterChoice}. In particular, these numerical results show that for small $\lambda\ll10^{-1}$ the scalaron-Higgs model allows for a tensor-to-scalar ratio which is one order of magnitude higher than the universal attractor value \eqref{ObsStar}, obtained  in the corresponding effective single-field model for $\beta^2\gg1$ discussed in Sec.~\ref{SubSec:Predictions}. 
\noindent As an independent check of the single-field approximation, we consider the limit $\beta^2\ll 1$ for which the spectral index and the tensor-to-scalar ratio converge to the attractor values  
\begin{align}
n_{\mathcal{R}}^{*}={}&1-\frac{3}{N_{*}}\approx 0.95, \label{nsstarb0}\\
r^{*}={}&\frac{16}{N_{*}}\approx 0.2667.\label{rstarb0}
\end{align}
Although the values \eqref{nsstarb0} and \eqref{rstarb0} lie outside the observationally allowed bounds \eqref{PlanckAns} and \eqref{Planckr}, they are nevertheless useful as an independent check of the exact numerical results, which should reproduce the same numbers in the $\beta^2\to0$ limit. The numerical results for the limit $\beta^2\to0$ are presented in the first column of Fig.~\ref{TensorToScalarRatio}. 
For the exact numerical evaluation of the power spectra, we choose an inflationary trajectory that initially starts inside the valley $\varphi_{0}=\varphi_{\mathrm{v}}$. From the linear fits \eqref{PowerLawPS} of the numerically obtained power spectra \eqref{PowerSpectrak}, the numerical values for the amplitude and the spectral index are extracted.
They are in excellent agreement with approximate analytic results \eqref{nsstarb0} and \eqref{rstarb0} at $k=k_*$.

For values of $6\beta^2\sim 1$ the analytic slow-roll single-field description breaks down since the kinetic terms for $\varphi$ and $\hat{\chi}$ contribute equally to the total inflaton velocity. Therefore, the precise numerical values obtained from the analytic expressions for the spectral observables cannot be trusted. Nevertheless, with a choice of parameters slightly different from those obtained in \eqref{ParameterChoice}, the exact numerical results still lead to viable predictions for the spectral observables, as illustrated in the right column in Fig.~\ref{TensorToScalarRatio}.
\subsection{Multifield effects: wiggles in the power spectrum}
\label{Wiggles}
Allowing for smaller values of $\lambda\ll10^{-1}$ also stretches out and flattens the hilltop in the $\varphi$ direction. Therefore, slow-roll inflation along the $\varphi$ direction on the hilltop is possible for $\lambda\ll10^{-1}$. This is in contrast to the analysis of (Class 3) and (Class 4) for $\lambda=10^{-1}$, where the narrow hilltop descends sharply in $\varphi$ direction due to the steep and prominent valley structure and the unstable inflationary trajectories on the hilltop are almost instantaneously pushed down into one of the two valleys.
\begin{figure}
	\centering
	\includegraphics[width=0.65\linewidth]{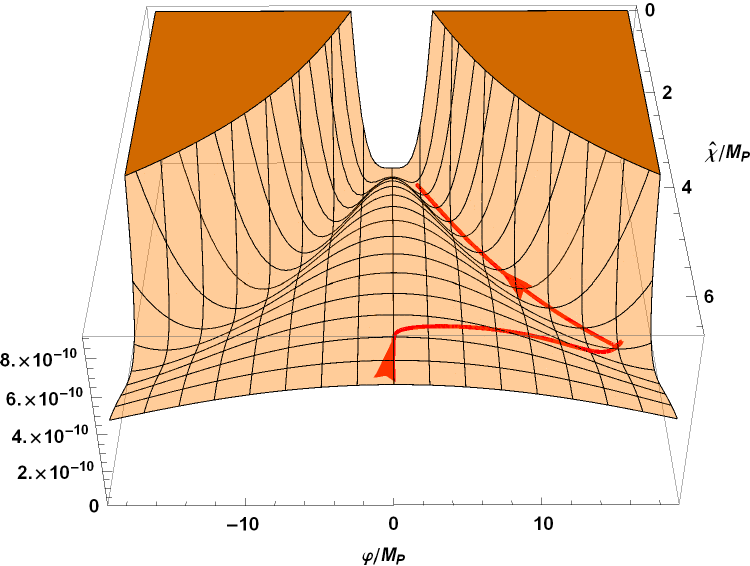}
	\caption{The plot shows an inflationary trajectory for the parameters $\lambda=10^{-11}$, $\xi=10^{-1}$ and $\alpha=10^8$, which slowly rolls on the hilltop along the $\varphi$ direction and subsequently falls into the valley.}
	\label{Fig:Multifieldtrajectory}
\end{figure}

\begin{figure}
	\centering
	\begin{tabular}{cc}
		\includegraphics[width=0.455\linewidth]{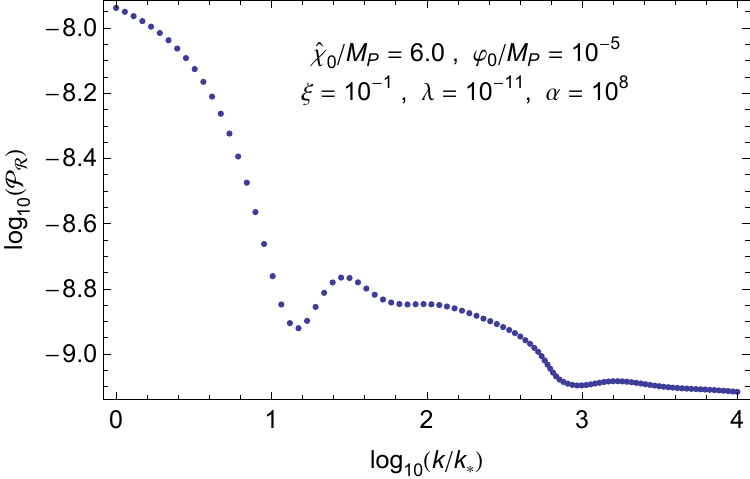}&
	\includegraphics[width=0.455\linewidth]
	{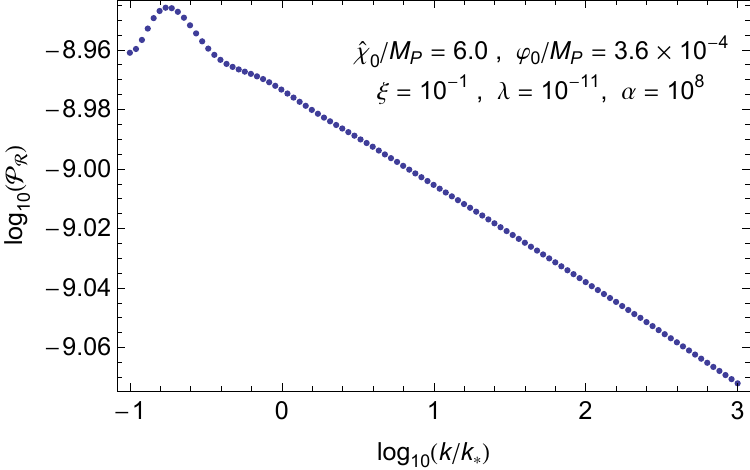}
	\end{tabular}
	\caption{The adiabatic power spectrum for the window of observable scales. Each blue dot corresponds to the numerical solution of \eqref{PowerSpectrak} at $t_{\mathrm{end}}$ for a given mode $k$.  The wiggles at small $k$ (large scales) are a clear multifield effect. For larger $k$, after the trajectory has settled down into the valley, the power spectrum approaches the linear fit. The structure of the wiggles is mainly determined by the parameter $\alpha$. Lower values of $\alpha$ lead to a higher hilltop and therefore to stronger oscillations and to more pronounced wiggles. By choosing larger values of the initial condition $\varphi_0$, the position of the wiggle structure can be shifted. For $\varphi_0/M_{\mathrm{P}}=10^{-5}$ the wiggle structure is well inside the observable window (left), while for larger $\varphi_0/M_{\mathrm{P}}=3.6\times10^{-4}$ (where the fall from hilltop to valley happens ``earlier'') is shifted partially out of the observable window towards lower $k$ (right).}
	\label{FeaturesPowerSpectrum}
\end{figure}

\noindent Thus, besides the effective single-field inflationary scenario for trajectories that start inside the valley, discussed in Sec.~\ref{Sec:EffectiveSingleFieldu}, a small $\lambda\ll10^{-1}$ also allows for an inflationary background trajectory that starts on the hilltop and slowly rolls on the hilltop in the $\varphi$ direction. If the trajectory stays for sufficiently many efolds on the hilltop, such that the subsequent fall into the valley happens during the observable number of efolds, this can lead to observable multifield effects.
The transition of the inflationary trajectory from the hilltop to the valley is followed by oscillations inside the valley. This is reflected by a sharp drop and subsequent oscillations in the potential, which ultimately lead to wiggles in the adiabatic power spectrum \cite{Polarski1992, Polarski1994,Adams2001,Hazra2014,Hazra2014a,Mori2017,LHuillier2018,Pi2018}. 
The resulting wiggles in the power spectrum arise due to a combined effect of isocurvature sourcing, the large drop in the Hubble parameter from $H^2\sim \hat{W}_{\mathrm{hill}}$ to $H^2\sim \hat{W}_{\mathrm{valley}}$, and the temporary spikes that arise in the slow-roll parameters at the point of fall into the valley.  This is exactly the scenario discussed in Section \ref{Class4}, which however was not possible for $\lambda = 10^{-1}$ as in this case, the inflationary trajectory on the hilltop is pushed almost immediately into the valley before the observable modes cross the horizon, resulting in no observable multifield effects. However, for $\lambda\ll 10^{-1}$, the trajectory stays on the hill for a sufficient amount of time, allowing isocurvature perturbations to grow for modes which exit the horizon while the trajectory is still rolling down the hill. At the point of fall, the turn rate spikes up, allowing isocurvature modes to significantly affect the super horizon evolution of the adiabatic modes via the sourcing term on the right-hand side of equation \eqref{EQPertQsig}. The modes which cross the horizon after the trajectory has settled down inside the valley, are not affected by isocurvature sourcing, as the corresponding isocurvature perturbations are quickly damped out, along with the turnrate settling down to a much lower value inside the valley. This  mode $k$ dependent sourcing mechanism therefore plays a crucial role in determining the overall shape and the nature of the wiggles in Fig. \ref{FeaturesPowerSpectrum}.
\noindent The wiggle structure in the curvature power spectrum is illustrated in Fig.~\ref{FeaturesPowerSpectrum} for different initial values $\varphi_0$.
The position of the wiggle structure is controlled by the initial condition $\varphi_0$. For smaller $\varphi_0$, closer to $\varphi_0=0$, the trajectory stays longer on the hilltop plateau and falls into the valley with the subsequent oscillations taking place within the observable number of efolds. Vice versa, by increasing $\varphi_0$ to larger values, the wiggle structure can be shifted to the lower end of the observable $k$ window corresponding to the largest scales. Oscillatory features in the power spectrum due to multifield effects were discussed already long time ago in the context of two free scalar fields \cite{Polarski1992} and recently in the context of the model \eqref{act1} for a scalar field $\varphi$ with a quadratic potential not associated with the SM Higgs boson \cite{Pi2018}. It would be interesting to investigate whether the wiggles found in Fig.~\ref{FeaturesPowerSpectrum} for the model \eqref{fHiggsStarobinsky} might explain the anomalies observed at large scales in the CMB temperature anisotropy spectrum. This requires a more detailed numerical analysis which we leave for future work. 
\subsection{Identification of $\varphi$ with the SM Higgs boson}
\label{Sec:Ident}
From the preceding analysis, we found that the quartic self-coupling needs to be very small $\lambda\ll 10^{-1}$ in order for multifield effects to become relevant and extremely small $\lambda\lesssim 10^{-11}$ in order to enter the parameter region $\beta^2\ll1$ without violating observational constraints. For the realization of this scenario within the scalaron-Higgs model $\varphi$ must be identified with the SM Higgs boson. Consequently, such tiny values of $\lambda$ would have to be generated dynamically by the extended RG flow of the SM, including the $\xi$ and $\alpha$ dependent contributions to the SM beta functions, as well as their own running. Since the beta functions for this extended RG flow are not known explicitly, we investigate this question based on the pure RG flow of the SM (i.e. neglecting any effects due to $\alpha$ and $\xi$). While this might provide a rough estimate of whether such a scenario has a chance to be realized or not, we emphasize that their might be essential modifications due to $\xi$ and $\alpha$ dependent effects.

It is well established that the RG flow of the SM drives the quartic Higgs self-coupling $\lambda$ to very tiny values at high energies. 
Moreover, the RG flow of the SM is highly sensitive to the ``initial condition'' at the electroweak scale -- in particular to the Higgs mass $M_{\mathrm{H}}$ and the top mass $M_{\mathrm{t}}$. Restricting these masses to lie within the experimentally allowed ranges, different scenarios are possible, depending on whether $\lambda$ becomes negative at some high energy scale or not. In particular, varying $M_{\mathrm{H}}$ and $M_{\mathrm{t}}$ (or $\lambda$ and $y_\mathrm{t}$, respectively) at the EW scale leads to a continuous change of $\lambda(t)$ from a strictly positive RG trajectory to one which touches zero or one which even becomes negative. Here $t=\ln E/\mu_0$ is the logarithmic RG scale, $E$ is the energy scale of the process under consideration and $\mu_0$ an arbitrary renormalization point, here chosen to coincide with the top-quark mass $\mu_0=M_{\mathrm{t}}$. 
For the $\lambda\lesssim10^{-11}$ scenario to be dynamically realized within the scalaron-Higgs model two conditions must be satisfied:
\begin{enumerate}
	\item The Higgs self-coupling $\lambda$ must become very small $\lambda(t_{\mathrm{inf}})$ $\lesssim 10^{-11}$ at the energy scale of inflation $E_{\mathrm{inf}}$.\label{cond1}
	\item The Higgs self-coupling $\lambda$ must sustain this small value during the $N\approx60$ efolds of inflation.\label{cond2}
\end{enumerate}   
Both conditions are non-trivial. While the initial conditions at the EW scale might be easily tuned such that $\lambda$ becomes sufficiently small at \textit{some} energy scale, it might happen that this scale is not compatible with the energy scale of inflation.
The second condition requires not only a small value of $\lambda$ at a given point $t_{\mathrm{inf}}$ but in an interval $\Delta t_{\mathrm{inf}}$ chosen to correspond to $N=60$ efolds,
\begin{align}
\Delta t_{\mathrm{inf}}:=|t_{\mathrm{in}}-t_{\mathrm{end}}|=\ln |E_{\mathrm{in}}/E_{\mathrm{end}}|.
\end{align}
The energy scale during inflation can be estimated from the value of the two-field potential $E\simeq\hat{W}^{1/4}$. We numerically determine the moment of time corresponding to the end of inflation by solving $\varepsilon_{\sigma}(t_{\mathrm{end}})=1$ for $t_{\mathrm{end}}$. The moment of time corresponding to the beginning of inflation is then obtained by numerically solving the equation $N(t_{\mathrm{in}})=60$ for $t_{\mathrm{in}}$. We then numerically evaluate the two-field potential by inserting the scalar field values $\varphi(t_{\mathrm{in/end}})$, $\hat{\chi}(t_{\mathrm{in/end}})$ and obtain $\hat{W}_{\mathrm{in/end}}$. In this way, we obtain $E_{\mathrm{in/end}}$ and ultimately the logarithmic scales $t_{\mathrm{in/end}}$ and the inflationary RG interval $\Delta t_{\mathrm{inf}}$. 

We perform this numerical analysis for the two scenarios with different initial conditions (``valley'' vs. ``hill'') considered in Sec. \ref{Sec:Isocurvature} with the parameters $\lambda$, $\xi$, $\alpha$ and initial conditions $\hat{\chi}_0$, $\varphi_0$ chosen as in the right column of Fig. \ref{TensorToScalarRatio} (valley) and the right column of Fig. \ref{FeaturesPowerSpectrum} (hilltop). The corresponding energy scales and inflationary RG intervals for both scenarios are summarized in Table \ref{TablePar}.
\begin{table}[h!]
	\begin{tabular}{cccc}
	scenario & $\quad E_{\mathrm{in}}\,[\mathrm{GeV}]\quad$ & $\quad E_{\mathrm{end}}\,[\mathrm{GeV}]\quad$ & $\quad \Delta t_{\mathrm{inf}}$\\
	\hline
	valley & $1.34\times 10^{16}$ & $8.67\times 10^{15}$ & $0.44$\\
	hilltop & $8.79\times 10^{15}$ & $5.66\times 10^{15}$& $0.44$\\
	\end{tabular}
\caption{Energy scales and inflationary RG interval for the two inflationary scenarios with extremely small $\lambda$ with different initial conditions for the background dynamics starting inside the valley and on the hilltop, respectively.}
\label{TablePar}
\end{table}

\begin{figure}[h!]
	\centering
	\begin{tabular}{cc}
		\includegraphics[width=0.455\linewidth]{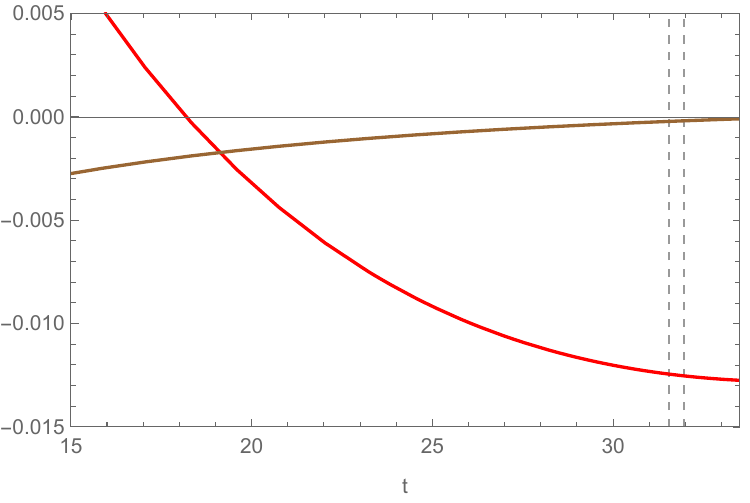}&
		\includegraphics[width=0.455\linewidth]{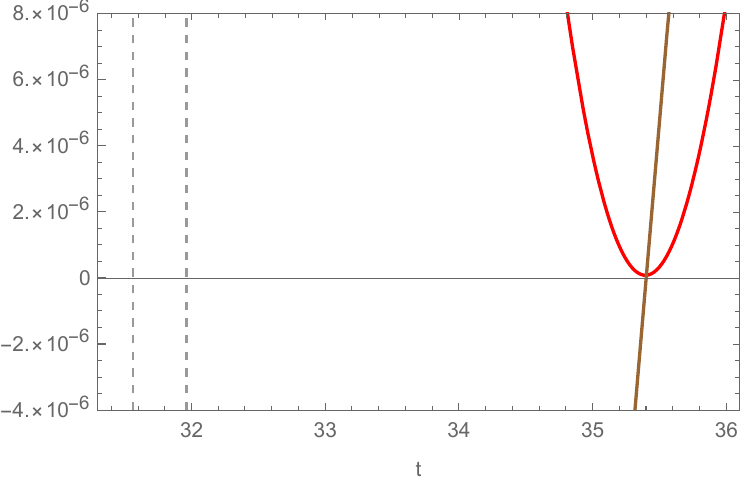}
	\end{tabular}
	\caption{Left: The two-loop RG flow of the quartic Higgs self-coupling $\lambda$ (red line) and its beta function $\beta_{\lambda}$ (brown line) for the central Higgs mass \eqref{IntervalsMh} and Yukawa top-quark mass \eqref{IntervalsMt}. The value of the quartic self-coupling $\lambda\approx10^{-2}$ within the inflationary RG interval $\Delta t_{\mathrm{inf}}$, marked by the two vertical black dashed lines, shows for these masses that condition \ref{cond1} is not satisfied by the SM RG flow. Right: The two-loop RG flow of the quartic Higgs self-coupling $\lambda$ (red line) and its beta function $\beta_{\lambda}$ (brown line) for $M_{\mathrm{H}}=125.1\;\mathrm{GeV}$ and $M_{\mathrm{H}}=170.9\;\mathrm{GeV}$, tuned such that $\lambda(t)$ has a minimum $\lambda_0\approx 0$. However, the RG scale $t_0$ where this minimum occurs is outside the inflationary RG interval bounded by the two vertical black dashed obtained from the values provided in the first row of Table \ref{TablePar}.}
	\label{SMLambdaBetaLambda}
\end{figure}

\begin{figure}[h!]
	\centering
	\begin{tabular}{cc}
		\includegraphics[width=0.455\linewidth]{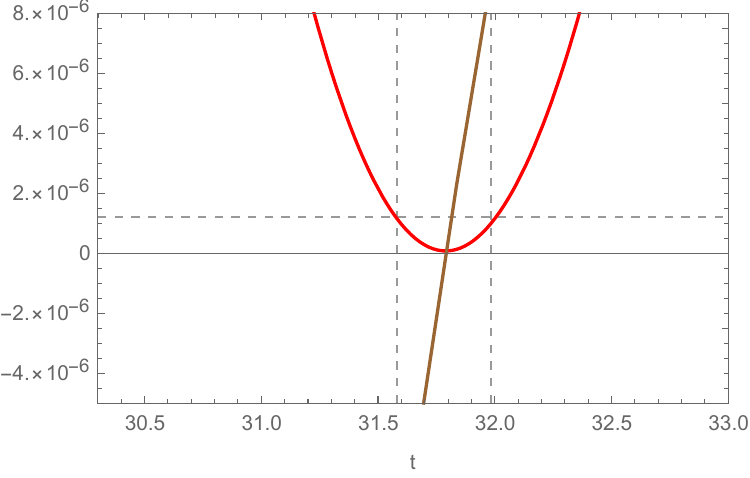}&
		\includegraphics[width=0.455\linewidth]{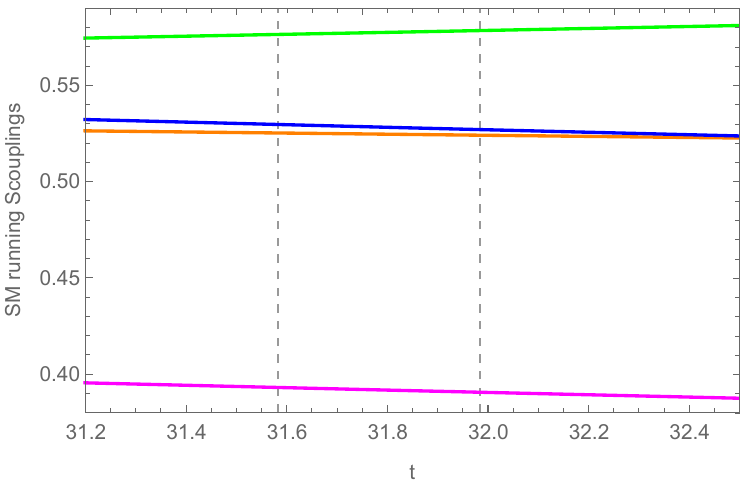}
	\end{tabular}
	\caption{Left:The two-loop RG flow of the quartic Higgs self-coupling $\lambda$ (red line) and its beta function $\beta_{\lambda}$ (brown line) for a Higgs mass $M_{\mathrm{H}}=117.5\; \mathrm{GeV}$ and a Yukawa top-quark mass $M_{\mathrm{t}}=166.95\;\mathrm{GeV}$ chosen such that $\lambda(t)$ has a minimum $\lambda_0=0$ at $t_0$ lying within the inflationary RG interval bounded by the two vertical black dashed lines. Right: The two-loop RG flow of the gauge couplings $g_1$ (green line), $g_2$ (orange line) and $g_3$ (blue line) as well as the Yukawa top-quark coupling $y_{\mathrm{t}}$ (Magenta line) for $M_{\mathrm{H}}=117.5\; \mathrm{GeV}$ and $M_{\mathrm{t}}=166.95\;\mathrm{GeV}$ inside the RG scale interval $\Delta t_{\mathrm{inf}}$ bounded by the two vertical black dashed lines.}
	\label{SMLambdaBetaGauge}
	\end{figure}

In order to investigate whether the inflationary scenario with $\lambda\lesssim10^{-11}$ can be dynamically realized by the RG flow of the SM, we numerically integrate the RG equations with the two-loop beta functions for the SM couplings extracted from the expressions provided in the Appendix B of \cite{Buttazzo2013}. The RG flow is highly sensitive to variations in the initial conditions for $M_{\mathrm{H}}$ and $M_{\mathrm{t}}$ set at the electroweak scale. The recent experimental bounds on these values is found from \cite{Tanabashi2018},
\begin{align}
M_{\mathrm{H}}={}&125.10\pm 0.14\;\mathrm{GeV},\label{IntervalsMh}\\
M_{\mathrm{t}}={}&172.9\pm 0.4\;\mathrm{GeV}.\label{IntervalsMt}
\end{align}
Thus, we can check explicitly whether there are RG trajectories which satisfy the conditions \ref{cond1} and \ref{cond2} at the RG scales corresponding to the inflationary energy scales provided in Table \eqref{TablePar} for the range of observationally allowed values of the Higgs mass and the top mass provided in \eqref{IntervalsMh} and \eqref{IntervalsMt}. 
The RG flow of $\lambda$ and its beta function $\beta_{\lambda}$ for the central values given in \eqref{IntervalsMh} and \eqref{IntervalsMt} is depicted in the left plot of Fig. \ref{SMLambdaBetaLambda}, where the black dashed lines mark the bounds of the inflationary interval. Clearly, for these values of $M_\mathrm{H}$ and $M_{\mathrm{t}}$, the first (and second) conditions are not satisfied.  
Nevertheless, the running of the self-coupling $\lambda$ can in principle be made arbitrarily small but non-negative by tuning its minimum to be $\lambda_0=0$ at a particular RG point $t_0$. This is depicted in the right plot of Fig. \ref{SMLambdaBetaLambda} for $M_{\mathrm{H}}=125.1\;\mathrm{GeV}$ and $M_{\mathrm{t}}=170.9\;\mathrm{GeV}$ where $\lambda_0=0$ at $t_0=35.25$. However, the energy scale $E_0$, corresponding to $t_0$, clearly does not lie inside the inflationary interval bounded by the two black dashed lines. Therefore the conditions \ref{cond1} and \ref{cond2} can again not be satisfied.
Finally, we could tune $M_{\mathrm{H}}$ and $M_{\mathrm{t}}$, such that and $\lambda_0=0$ for $t_{0}\in[t_{\mathrm{in}},t_{\mathrm{end}}]$ as depicted in the left plot of Fig. \ref{SMLambdaBetaGauge}. 
 While in this case condition \ref{cond1} could be satisfied, the required values for $M_{\mathrm{H}}=117.5\;\mathrm{GeV}$ and $M_{\mathrm{t}}=166.95\;\mathrm{GeV}$ would be in strong conflict with the observational bounds \eqref{IntervalsMh} and \eqref{IntervalsMt}.

Therefore, we conclude that condition \ref{cond1} cannot be satisfied by the pure RG flow of the SM. Similar results were obtained in the context of non-minimal Higgs inflation \cite{Hamada2014, Bezrukov2014, Hamada2015}. There, for a non-minimal coupling as small as $\xi\approx 10$, values as small as $\lambda_0\approx 10^{-6}$ could be arranged to lie within the inflationary interval (but not values as small as $\lambda_0\approx 10^{-11}$ required for the scenarios considered in Sec. \ref{Sec:EffectiveSingleFieldu} and Sec. \ref{Wiggles}).

Moreover, even if values of $\lambda_0\lesssim10^{-11}$ would be attainable at a given RG point $t_0$ within the inflationary interval $[t_{\mathrm{in}},t_{\mathrm{end}}]$ by a finetuned combination of $M_{\mathrm{H}}$ and $M_{\mathrm{t}}$, such a small value of $\lambda_0$ could not be sustained over the whole course of the inflationary interval $\Delta t_{\mathrm{inf}}$ within the pure RG flow of the Standard Model. In order to illustrate this, let us look again at the (already excluded) example shown in the left plot of Fig. \ref{SMLambdaBetaGauge}. The intersection point of the horizontal and vertical black dashed lines with the running of $\lambda$ (red line) indicates that the smallest value of $\lambda$ which could be sustained over the whole interval $\Delta t_{\mathrm{inf}}$ is $\lambda\approx 10^{-6}$.

The value of $\lambda\approx 10^{-6}$ can as well be consistently obtained from a rough estimate of the dominant one-loop contribution in $\beta_{\lambda}$. Let us assume that the first condition is satisfied. Let  $\lambda_{0}\approx10^{-11}$ at the center $t_0\approx(t_{\mathrm{in}}+t_{\mathrm{end}})/2$ of the inflationary RG interval $[t_{\mathrm{in}},t_{\mathrm{end}}]$.
Since in both scenarios (``valley'' and ``hill''), the inflationary RG interval $\Delta t_{\mathrm{inf}}\simeq0.44$ is very short (about $1.4\%$ of the total RG interval), the change of $\lambda$ around $t_0$ during inflation can be estimated by the linear approximation 
\begin{align}
\Delta \lambda_{\mathrm{inf}}={}&2|\lambda(t_{0}+\Delta t_{\mathrm{inf}}/2)-\lambda(t_{0})|\nonumber\\
\approx{}&\Delta t_{\mathrm{inf}}\,|\beta_{\lambda}(t_{0})|,
\end{align}
In this way, the condition \ref{cond2} on $\lambda$ translates into a condition on its beta function $\beta_{\lambda}(t_{0})$ at the point $t_0$ and the size of the inflationary RG interval $\Delta t_{\mathrm{inf}}$,
\begin{align}
\Delta t_{\mathrm{inf}}\,|\beta_{\lambda}(t_{0})|\lesssim 10^{-11}.
\end{align}
By assumption, $\lambda_0\approx10^{-11}$, such that we can estimate the dominant contribution to the one-loop beta function $\beta_{\lambda}(t_0)$ by neglecting terms proportional to $\lambda_0$ and $\lambda^2_0$. Moreover, we can safely neglect the contributions from the bottom quark and tau quark and only keep the dominant contributions form the heavy SM particles (the $W^{\pm}$ bosons, the $Z$ boson and the Yukawa top quark $q_{\mathrm{t}}$) for which \cite{Buttazzo2013},
\begin{align}
\beta_{\lambda}(t_{0})\approx \frac{1}{(4\pi)^2}\left[
\frac{27}{200}g_1^4+\frac{9}{20}g_2^2g_1^2+\frac{9}{8}g_2^4-6y^4_{\mathrm{t}}\right].\label{domoneloop}
\end{align} 
All couplings on the right-hand-side of \eqref{domoneloop} are to be evaluated at $t_{0}$. For the example shown in Fig. \ref{SMLambdaBetaLambda}, the values the gauge couplings ${g_1(t_0)\approx0.58}$, ${g_2(t_0)\approx0.52}$ and the Yukawa top quark ${y_{\mathrm{t}}(t_0)\approx0.39}$ at $t_0\approx31.8$ can be read off from the right plot. Inserting these values in \eqref{domoneloop} and using the value of inflationary RG interval $\Delta t_{\mathrm{inf}}=0.44$ from Table \ref{TablePar}, we find that $\lambda$ changes during the inflationary interval by an amount of  $\Delta \lambda_{\mathrm{inf}}\approx\Delta t_{\mathrm{inf}}|\beta_{\lambda}(t_0)|\approx10^{-6}\gg10^{-11}$.
Therefore, even if the condition \ref{cond1} would be satisfied, the condition \ref{cond2} would fail.

Although the preceding analysis suggests that the identification of $\varphi$ with the SM Higgs boson for values ${\lambda\lesssim 10^{-6}}$ does not seem to be appropriate, we emphasize that the RG analysis was based on the pure SM running, taking into account neither the RG running of $\xi$ and $\alpha$, nor their influence on the SM beta functions. Therefore, it remains to be seen if the full RG evolution of the scalaron-Higgs model can lead to significant changes, which ultimately would justify the identification of $\varphi$ with the SM Higgs boson and would allow for a consistent realization of the inflationary scenario with $\lambda\approx 10^{-11}$ within the scalaron-Higgs model.

Note that throughout this article we have restricted ourselves to a tree-level analysis. Only in this section, we invoke the RG analysis in order to investigate whether values as small as $\lambda\lesssim 10^{-11}$ could be dynamically attained at inflationary energy scales from $\lambda\approx 10^{-1}$ at the EW scale. On the basis of the pure SM RG flow, we find that the two conditions \ref{cond2} are not satisfied and $\varphi$ cannot be identified with the SM Higgs boson for $\lambda\lesssim10^{-11}$. However, had we found that the conditions \ref{cond2} were satisfied and $\varphi$ could be identified with the SM Higgs boson, in addition to the modified values of the running couplings at the inflationary energy scale, the RG improvement would also lead to a modification of the equations for the slow-roll parameters \eqref{SlowRollPot} and the spectral observables \eqref{PowerLawPS}. These equations would receive additional correction terms due to the derivatives of the running couplings. Even if for $\lambda\lesssim 10^{-11}$ no consistent RG improved scalaron-Higgs scenario is possible, it would be interesting to study the RG improvement and the impact of the additional correction terms for values of $\lambda$ for which the conditions \eqref{cond2} are satisfied. We leave this for future work. 
\subsection{Stabilization of the electroweak vacuum}
Independent of whether the inflationary scenario with $\lambda\lesssim 10^{-11}$ can be dynamically realized in the scalaron-Higgs model by a running $\lambda(t)$, the dynamics of the pure Standard Model RG flow drives the Higgs self-coupling $\lambda$ to very small values at high energies. As can be seen from the right plot in Fig. \ref{SMrunning}, depending on the values of $M_{\mathrm{H}}$ and $M_{\mathrm{t}}$ at the EW scale, $\lambda(t)$ can even become negative during its RG evaluation. In particular this can happen at a RG scale $t_{\mathrm{inst}}$ (defined as the moment where $\lambda$ crosses zero $\lambda(t_{\mathrm{inst}})=0$) corresponding to energy scales much lower than that of inflation $t_{\mathrm{inst}}\ll t_{\mathrm{inf}}$. 

The resulting instability of the effective Higgs potential must be considered a serious problem. The scalaron-Higgs model changes the stability analysis of the RG improved potential in at least two aspects compared to the Standard Model RG analysis:

First, by adding the two marginal operators $\xi\varphi^2 R$ and $\alpha R^2$ to the Standard Model, the impact of the two additional couplings $\xi(t)$ and $\alpha(t)$ on the RG flow must be taken into account. Consequently, the RG system of the SM must be extended by the beta functions $\beta_{\xi}$ and $\beta_{\alpha}$ and in addition the SM beta functions must be modified by $\xi$ and $\alpha$ dependent contributions. Depending on how $\alpha$ and $\xi$ affect the RG system, this might change the RG flow of $\lambda$ in such a way that $\lambda$ is stabilized, i.e. $\lambda>0$, $\forall t\in[0,t_{\mathrm{in}}]$. Based on results of \cite{Avramidi1986}, it was reported in \cite{Gorbunov2018} that the one-loop contributions to the beta function of $\lambda$ from $\xi$ and $\alpha$ is positive and therefore could prevent $\lambda$ from becoming negative at high energies.
However, in order to make precise quantitative statements, a refined numerical RG analysis would be required.

Second, the structure of the effective scalaron-Higgs potential itself might provide a mechanism which could stabilize the electroweak vacuum. While a detailed analysis would again require knowledge about the RG improved two-field effective scalaron-Higgs potential and therefore numerical access to the full scalaron-Higgs RG system, one stabilization mechanism can be illustrated analytically within the effective single field approximation. Let us consider again the parameter region $\beta^2\ll1$, which leads to the effective single field scenario considered in Sec.~\ref{Sec:EffectiveSingleFieldu}. Even if the scalaron-Higgs RG flow cannot drive the running $\beta^2(t)$ into the region $\beta^2(t_{\mathrm{inf}})\ll1$  at $t_{\mathrm{inf}}$, this region might be accessed for energy scales ${t_{\mathrm{inst}}< t_{\mathrm{inf}}}$ well below the energy scale of inflation $\beta^2(t_{\mathrm{inst}})\ll1$.
Moreover, at $t_{\mathrm{inst}}<t_{\mathrm{inf}}$ the CMB constraint \eqref{constraintLambda} need not be satisfied and therefore the condition on $\lambda\lesssim 10^{-11}$ might be relaxed (only $\beta^2\ll1$ is required).

In the $\beta^2\ll1$ effective single-field scenario with the potential \eqref{Vsingleu}, the background dynamics predominantly takes place along the $\varphi$ direction. According to \eqref{uphi}, we have $u\approx\varphi$ for $u/M\ll1$. In this approximation, the functional form of the potential \eqref{Vsingleu} acquires the $\varphi^4$ shape of the Higgs potential,
\begin{align}
\hat{V}(u)\approx\frac{\bar{\lambda}}{4}\varphi^4,\label{ufourpot}
\end{align}
but with a modified quartic coupling constant $\bar{\lambda}$,
\begin{align}
\bar{\lambda}:=\lambda\left(1+\delta\lambda\right),\qquad \delta\lambda:=4\alpha\frac{\lambda}{\xi^2}.\label{barlambda}
\end{align}
Thus, the only remnant of the original two-field model is the non-negative contribution $4\alpha(\lambda/\xi)^2$ to the effective quartic self-coupling $\bar{\lambda}$.\footnote{We assume that $\alpha(t)$ is positive over the whole RG evolution.} 
On the one hand, for the stability of the effective Higgs potential, we must have \mbox{$\bar{\lambda}\geq0$} at energy scales $t\geq t_{\mathrm{inst}}$, irrespectively of whether $\lambda$ itself remains positive under the RG evolution. On the other hand, we should have $|\delta\lambda(t)|\ll1$ close to the electroweak scale \mbox{$t_{\mathrm{EW}}\approx 0$} to prevent conflicts with particle physics measurements. 
If the full RG dynamics (including $\xi$ and $\alpha$) would drive $\lambda(t)$ to negative values $\lambda(t_{\mathrm{inst}})<0$ for $t\geq t_{\mathrm{inst}}$, we must ensure that $|\delta\lambda(t)|>1$ for $t\geq t_{\mathrm{inst}}$ in order for $\bar{\lambda}(t)>0$. Thus, in order to stabilize the effective Higgs potential \eqref{ufourpot}, a turn-over among the two terms in $\bar{\lambda}$ between $t_{\mathrm{EW}}$ and $t_{\mathrm{inst}}$ should happen, such that the $\delta\lambda (t)$ contribution in \eqref{barlambda} changes from negligible small to dominant
\begin{align}
|\delta\lambda(t)|
\begin{cases}
\ll1{}& \;\mathrm{for}\;t\ll t_{\mathrm{inst}},\\
>1{}& \;\mathrm{for}\;t\geq t_{\mathrm{inst}}.
\end{cases}
\end{align}

Independent of this example, which showed how the stability of the effective Higgs potential can be modified within the single field approximation, full access to the RG flow of the scalaron-Higgs model, including the effects of non-zero $\xi$ and $\alpha$, would allow to perform a complete numerical stability analysis. We hope to address this in future work. 
%
%----------------------------------------------------------
\section{Conclusions and outlook}\label{Sec:Conclusion}
%----------------------------------------------------------
%
The model of scalaron-Higgs inflation provides a natural and elegant framework for a unified description of particle physics and cosmology which only assumes Einstein gravity and the Standard Model of particle physics together with two additional marginal operators: the non-minimal coupling of the Higgs field to gravity and an additional curvature invariant given by the the Ricci scalar squared. Since the scalaron-Higgs model features an approximate scale invariance for large curvatures and large values of the Higgs field, it naturally provides a quasi de Sitter stage as well as a graceful exit due to the Einstein-Hilbert term which breaks the scale invariance at lower energies. The higher derivatives of the quadratic curvature invariant lead to an additional propagating scalar degree of freedom -- the scalaron.

We formulated the scalaron-Higgs model as a scalar-tensor theory of two minimally coupled scalar fields with a curved scalar field space.
In view of the latter, we applied a field covariant formalism and derived the inflationary two-field dynamics on a flat FLRW background as well as the linear perturbations propagating on this background. 

In contrast to single-field models of inflation, the inflationary trajectory is no longer unique but depends on the initial conditions for the dynamical background equations, whose solutions enter the dynamical equations for the cosmological perturbations. In this way the dependence on the initial conditions for the background dynamics propagates into the inflationary observables. 

For a broad range of parameters, the scalaron-Higgs potential features two prominent valleys which serve as natural attractor solutions for the inflationary trajectory. We identified four classes of qualitatively distinct trajectories on the basis of their initial conditions and discussed the observational consequences for each class. For a quartic Higgs coupling at the electroweak scale $\lambda=M_{\mathrm{H}}^2/2v^2\approx10^{-1}$, we found that all classes inevitably reduce to an effective single-field model for which the inflationary dynamics takes place predominantly along the $\hat{\chi}$ direction inside one of the two valleys. For the resulting effective single-field model we derived analytical expressions of the inflationary slow-roll observables and found that they are indistinguishable from the model of non-minimal Higgs inflation and Starobinsky's model of inflation. 
Due to the universal predictions for the spectral index and the tensor-to-scalar ratio, which are independent of any model parameters, only the CMB normalization condition constrains the parameter combination $\lambda/(\xi^2+4\lambda\alpha)$ and imposes an upper bound on the non-minimal coupling $\xi\lesssim 10^4$ and the scalaron coupling $\alpha\lesssim 10^9$. If the latter is dominated by the $1/4\alpha$ term, the situation with the strong coupling $\xi$, present in the model of non-minimal Higgs inflation, is relaxed in a natural way, as $\xi$ can take on small values. Likewise, if the CMB constraint is instead dominated by the combination $\lambda/\xi^2$, the parameter $\alpha$ can be varied.

We also provided analytical expressions for the effective adiabatic and isocurvature masses and showed that inflation inside the valley does not lead to any sizable isocurvature effects until the end of inflation.
In addition, we discussed the limit of a vanishing non-minimal coupling $\xi\to0$, for which the two valleys degenerate to a single valley and showed that the inflationary predictions are those of Starobinsky's model. In particular, we showed that even for very small $\xi\ll1$, no significant isocurvature effects arise until the end of inflation. The new results for small $\xi$ extend the results of \cite{He2018} and are important as they relax the situation with a strong non-minimal coupling present in Higgs inflation.
 
Another interesting region in the parameter space is that of extremely small values $\lambda\lesssim10^{-11}$, for which the scalar field $\varphi$ cannot be identified with the SM Higgs boson a priori. 
We showed that, depending on the initial conditions for the inflationary background trajectories, different scenarios are possible. The main effect of an extremely small $\lambda$ is that the multifield potential is stretched and flattened in the $\varphi$ direction. In this case, inflation inside the valley takes place predominantly in $\varphi$ direction and again leads to an effective single-field model, which however yields different observational predictions than that of non-minimal Higgs inflation or that of Starobinsky's model of inflation. In particular, depending on the precise values of the parameters, it predicts a larger tensor-to-scalar ratio. 

For extremely small $\lambda$, also the plateau on the hilltop of the two-field model is flattened in $\varphi$ direction. Therefore, apart from effective single-field inflation inside the valley, a true multifield scenario becomes possible for which the inflationary trajectory starts on the hilltop and stays there for a sufficient number of efolds before it falls into one of the valleys during the observable number of efolds. In this scenario we found observable multifield effects which manifest themselves in the form of oscillatory features in the power spectrum and might even provide a theoretical explanation of the anomalies observed at large scales in the CMB temperature anisotropy spectrum.

Motivated by the RG flow of the SM, which dynamically drives the self-coupling $\lambda$ to very small values at high energies, we investigated the conditions under which the $\lambda\lesssim 10^{-11}$ scenario can be realised within scalaron-Higgs inflation. While the preliminary RG analysis of the pure SM suggests that only values as small as $\lambda\lesssim 10^{-6}$ can be attained (and retained) during inflation, a conclusive statement can only be made by an exact numerical evaluation of the full RG improved scalaron-Higgs model, which takes into account the effects of $\xi$ and $\alpha$ on the RG evolution.   

Finally, we found that the scalaron-Higgs model might help to stabilize the electroweak vacuum. While the RG flow of the scalaron-Higgs model, which includes the additional beta functions for $\xi$ and $\alpha$ as well as the influence of $\xi$ and $\alpha$ on the SM beta functions, might itself prevent $\lambda$ from running to negative values at high energy scales, in addition another stabilization mechanism becomes possible in certain parameter regions were an effective single field description is adequate. In this case, the scalaron-Higgs potential acquires the structure of the SM Higgs potential but with an effective self-coupling $\bar{\lambda}$, which is the sum of the SM Higgs coupling $\lambda$ and an additional positive contribution.

A complete stability analysis requires a full numerical study of the RG improved scalaron-Higgs model, which we plan to address in a forthcoming work.    
% 
%------------------------------------------------------------------------------

\begin{acknowledgments} 
We thank Claus Kiefer for valuable discussions and for supporting this collaboration. A.G thanks the Bonn-Cologne Graduate School of Physics and Astronomy (GSC 260) for the continued financial support received during his stay in Cologne. We are grateful to Alexei A. Starobinsky for valuable comments.
\end{acknowledgments}
%
%------------------------------------------------------------------------------
\appendix
%------------------------------------------------------------------------------
%
%------------------------------------------------------------------------------
\section{Scalaron-Higgs field space geometry}
\label{App:FieldSpaceGeometry}
%------------------------------------------------------------------------------
%
In terms of the field parametrizations $\Phi^{I}=(\hat{\chi},\varphi)$,
the field space metric $G_{IJ}$ for the scalaron-Higgs model \eqref{ActScal} is given explicitly by
\begin{equation}
\begin{aligned}
G_{IJ}={}&\text{diag}\left(G_{\hat{\chi}\hat{\chi}},G_{\varphi\varphi}\right),\\ G_{\hat{\chi}\hat{\chi}}={}&1,\qquad G_{\varphi\varphi}=\exp{\left(-\gamma\frac{\hat{\chi}}{M_{\mathrm{P}}}\right)}.
\end{aligned}
\end{equation}
The inverse is trivially calculated as
\begin{equation}
\begin{aligned}
G^{IJ}={}&\text{diag}\left(G^{\hat{\chi}\hat{\chi}},G^{\varphi\varphi}\right),\\ G^{\hat{\chi}\hat{\chi}} ={}& 1,\qquad G^{\varphi\varphi} ={} \exp{\left(\gamma\frac{\hat{\chi}}{M_{\mathrm{P}}}\right)}.
\end{aligned}
\end{equation}
The non-vanishing Christoffel components are given by
\begin{equation}
\begin{aligned}
{\Gamma^{\hat{\chi}}}_{\varphi\varphi} ={}& \frac{\gamma}{2M_\mathrm{P}}\exp{\left(-\gamma\frac{\hat{\chi}}{M_{\mathrm{P}}}\right)},\\{\Gamma^{\varphi}}_{\hat{\chi}\varphi} ={}& {\Gamma^{\varphi}}_{\varphi\hat{\chi}} = -\frac{\gamma}{2M_{\mathrm{P}}}.\label{GeoGam}
\end{aligned}
\end{equation}
Since the field space is two dimensional, the Riemann tensor is determined in terms of the constant scalar curvature $R_0$,
\begin{equation}
\begin{aligned}
R_{IJKL}={}&\frac{R_0}{2}\left(G_{IK}G_{JL}-G_{IL}G_{JK}\right),\\ R_{JL}={}&G^{IK}R_{IJKL}=\frac{R_0}{2}G_{JL}.
\end{aligned}
\end{equation}
The Einstein tensor vanishes identically $G_{IJ}\equiv0$.
The non vanishing components of the Riemann tensor, the Ricci tensor and the Ricci scalar are given explicitly as
\begin{align}
R_{\hat{\chi}\varphi\varphi\hat{\chi}} ={}& -R_{\hat{\chi}\varphi\hat{\chi}\varphi} =R_{\varphi\hat{\chi}\hat{\chi}\varphi}=-R_{\varphi\hat{\chi}\varphi\hat{\chi}}\nonumber\\={}& \frac{\gamma^2}{4M_\mathrm{P}^2}\exp{\left(-\gamma\frac{\hat{\chi}}{M_\mathrm{P}}\right)},\label{GeoCurv}\\
R_{\hat{\chi}\hat{\chi}} ={}& -\frac{\gamma^2}{4M^2_{\mathrm{P}}},\\
R_{\varphi\varphi} ={}& -\frac{\gamma^2}{4M^2_{\mathrm{P}}}\exp{\left(-\gamma\frac{\hat{\chi}}{M_{\mathrm{P}}}\right)},\\
 R_0 ={}& -\frac{\gamma ^2}{2M^2_{\mathrm{P}}}.
\end{align}
In particular, a negative $R_0$ implies that the field space is hyperbolic.
%
%------------------------------------------------------------------------------
\bibliography{ScalaronHiggsInflationArXivV3}{}
%------------------------------------------------------------------------------
\end{document}